\documentstyle[12pt,a4]{article}
\newcommand{\be}{\begin{equation}}
\newcommand{\ee}{\end{equation}}
\newcommand{\sect}[1]{\setcounter{equation}{0}\section{#1}}

\newcommand{\vs}[1]{\rule[- #1 mm]{0mm}{#1 mm}}
\newcommand{\hs}[1]{\hspace{#1mm}}
\newcommand{\mb}[1]{\hs{5}\mbox{#1}\hs{5}}

\newcommand{\bea}{\begin{eqnarray}}
\newcommand{\ena}{\end{eqnarray}}
\newcommand{\shalf}{\sm{1}{2}}
\newcommand{\half}{\frac{1}{2}}
\newcommand{\wt}[1]{\widetilde{#1}}
\newcommand{\und}[1]{\underline{#1}}
\newcommand{\ov}[1]{\overline{#1}}
\newcommand{\sm}[2]{\frac{\mbox{\footnotesize #1}\vs{-2}}
                   {\vs{-2}\mbox{\footnotesize #2}}}
\newcommand{\prt}{\partial}

\newcommand{\Z}{Z\hspace{-2mm}Z}

\newcommand{\po}{{\mbox{\small{$\ast$}}}}

\newcommand{\cg}{{\cal G}}
\newcommand{\cd}{{\cal D}}
\newcommand{\ch}{{\cal H}}

\newcommand{\ci}{{\cal I}}

\newcommand{\crr}{{\cal R}}
\newcommand{\r}[1]{{\cal R}_{#1}}
\newcommand{\rpi}[1]{{\cal R}_{#1}^{\pi}}

\newcommand{\NP}[1]{Nucl.\ Phys.\ {\bf #1}}
\newcommand{\PL}[1]{Phys.\ Lett.\ {\bf #1}}

\newcommand{\CMP}[1]{Comm.\ Math.\ Phys.\ {\bf #1}}

\newcommand{\MPL}[1]{Mod.\ Phys.\ Lett.\ {\bf #1}}

\newcommand{\IJMP}[1]{Int.\ Jour.\ of\ Mod.\ Phys.\ {\bf #1}}

\begin{document}
\renewcommand{\thefootnote}{\fnsymbol{footnote}}

\newpage
\setcounter{page}{0}
\pagestyle{empty}

\vs{30}

\begin{center}

{\LARGE {\bf W-algebras and superalgebras }}\\[.5cm]
{\LARGE {\bf from constrained WZW models:}}\\[.5cm]
{\LARGE {\bf a group theoretical classification.}}\\[1cm]

\vs{10}

{\large
L. Frappat\footnote{Groupe
d'Annecy, LAPP Chemin de Bellevue BP 110, F-74941 Annecy-le-Vieux Cedex,
France}, E. Ragoucy${}^{\po}$
and P. Sorba${}^{\po}$\footnote{Groupe de Lyon, ENS Lyon 46 all\'ee
d'Italie, F-69364 Lyon Cedex 07, France}}\\[.5cm]
{\em Laboratoire de Physique Th\'eorique }
{\small E}N{\large S}{\Large L}{\large A}P{\small P}
\footnote{URA 14-36 du
CNRS, associ\'ee \`a l'E.N.S. de Lyon, et au L.A.P.P. (IN2P3-CNRS)}\\
\end{center}
\vs{20}

\centerline{ {\bf Abstract}}

\indent

We present a classification of $W$ algebras and superalgebras arising in
Abelian as well as non Abelian Toda theories. Each model, obtained from a
constrained WZW action, is related with an
$Sl(2)$ subalgebra (resp. $OSp(1|2)$ superalgebra) of a simple Lie algebra
(resp. superalgebra) $\cg$. However, the determination of an $U(1)_Y$ factor,
commuting with $Sl(2)$ (resp. $OSp(1|2)$),
appears, when it exists, particularly useful to characterize
the corresponding $W$ algebra. The (super) conformal spin contents
of each $W$ (super)algebra is performed. The class of all the superconformal
algebras (i.e. with conformal spins $s\leq2$) is easily obtained as a byproduct
of our general results.
\vfill
\rightline{{\small E}N{\large S}{\Large L}{\large A}P{\small P}-AL-391/92}
\rightline{ July 1992}

\newpage
\pagestyle{plain}
\tableofcontents
\newpage
\renewcommand{\thefootnote}{\arabic{footnote}}
\setcounter{footnote}{0}

\sect{Introduction}

\indent

Lots of efforts have been done these recent years to detect and understand
the infinite dimensional symmetries which underly two dimensional field
theories. A particular role is played by Toda theories, since each of them
possesses a $W$ symmetry \cite{1,2}. More recently, it has been shown that in
fact Toda
models can be seen as constrained WZW models \cite{4}. One can say that such
a property
reenforces the fundamental role of WZW models in the realm of conformal field
theories. It also provides a natural framework to compute explicitly the $W$
algebras which then appear.

\indent

In order to reduce a WZW model to a Toda one, some of the
conserved current components have to be set to constants or zero.
It can be realized that, from a given simple Lie algebra (or superalgebra)
$\cg$, different choices of constraints can be proposed, each of them
giving rise to a different Toda model, to which will be associated a $W$
(super)algebra. Actually, to each such a Toda model corresponds a (integral or
half-integral) grading \cite{3} of $\cg$ specified by a Cartan element
$H\in\cg$. In
other words, $\cg$, which is chosen maximally non-compact, admits a vector
space decomposition:
\be
\cg=\bigoplus_{h\in\half{\bf Z}} \cg_h
\mb{with}
{[H,X_h]}=hX_h \mb{for any} X_h\in\cg_h \label{grad}
\ee
As an example, the usual or Abelian Toda model associated to $\cg$ is obtained
by taking $H$ as the Cartan generator of the principal $Sl(2)$ in the algebra
(or superprincipal $OSp(1|2)$ in the superalgebra) $\cg$.

\indent

For each such a
grading $H$ can be defined either an $Sl(2)$ \cite{3,7,8,14} or an
$Sl(2)\oplus U(1)$ \cite{faux}
(resp. $OSp(1|2)$ \cite{12,13} or $OSp(1|2)\oplus U(1)$)
sub(super)algebra of $\cg$
generated by $\{ M_0,M_\pm\}\oplus\{Y\}$ (resp.
$\{ M_0,M_\pm,F_\pm\}\oplus\{Y\}$)
and such that $H=M_0+Y$. More precisely, even when the $U(1)$ part is not
zero, the $Sl(2)$ (resp. $OSp(1|2)$) subalgebra is sufficient to
characterize the $W$ algebra: one can then say that the different Toda models
in $\cg$ are classified by the different $Sl(2)$ (resp. $OSp(1|2)$)
subalgebras of $\cg$. However, interesting informations on the structure of the
corresponding $W$ algebra can be obtained when the $Y$ generator exists. As
will be shown below, a conserved hypercharge can be associated to it, which may
greatly simplify the Poisson bracket computation of the different
primary fields constituting the $W$ algebra.

\indent

The determination of the semi-simple subalgebras of a simple Lie algebra has
been considered by Dynkin \cite{11} and explicited for algebras of rank up
to 6 by Lorente and Gruber \cite{LG}.
Hereafter, we have reconsidered in some details this
classification in order to study the $Sl(2)$ and $Sl(2)\oplus U(1)$
subalgebras. The reduction of the fundamental and adjoint representations of a
(semi-)simple Lie algebra is achieved in the general case, allowing to
determine, by working in the so-called highest weight Drinfeld-Sokolov gauge
\cite{5,4},
the conformal spin content of the $W$ algebras. Tables relative to algebras
of low rank ($r\leq4$) are provided. The usefulness of the conserved
hypercharge $Y$ is illustrated to calculate the PB of the algebra of spins
2,$\sm{3}{2},\sm{3}{2}$,1 first considered in \cite{Poly,Bersh}.

\indent

As already mentioned, in the supersymmetric case when $\cg$ is a simple Lie
superalgebra, the $Sl(2)$ algebra is replaced by its supersymmetric "extension"
$OSp(1|2)$ \cite{12,13}.
It is therefore the classification of $OSp(1|2)\oplus U(1)$
subsuperalgebras in $\cg$ which is now of interest. The perfect knowledge of
the
algebra case helps a lot for this supersymmetric generalization. Note that a
first attempt in this direction can be found in \cite{12}. Hereafter, we
explicitly achieve this classification in a way which, we believe, is clear and
allows a direct use. As in the algebra case, general formulae for the
decomposition of the fundamental and adjoint representations of a simple
Lie superalgebra with respect to $OSp(1|2)\oplus U(1)$ subsuperalgebras are
given, and the (super)conformal spin content of the super $W$ algebras
determined. In order to illustrate these results, and mainly to allow a
comparison with the extended superconformal algebras \cite{quasic},
tables are constructed
for superalgebras of rank up to 4.

\sect{$W$ algebras and (half-)integral gradings}

\subsection{$W$ algebras in Toda theories \label{sec2.1}}

\indent

It has been elegantly shown that, starting from a WZW model, the action of
which is $S(g)$ and the fields $g(x)$ (resp. superfields $g(x,\theta)$)
belong to
the group (resp. supergroup) $G$, and imposing some of the components of the
conserved (super) currents to be constant or zero leads to a Toda model.

Let us, at this point, briefly fix some notations.

\indent

As far as $G$ is a group,
the WZW conserved currents read:
\be
J_+ = g^{-1} \prt_+ g \ \ \ \ J_- = (\prt_-g)g^{-1}
\ee
with
\be
\prt_- J_+ = \prt_+ J_- =0.
\ee

\indent

When considering a supersymmetric WZW model \cite{13}, a supergroup element
will
locally be defined as:
\be
g(x,\theta) = exp (\varphi^i B_i + \psi^j F_j)
\ee
where the $\varphi^i$ (resp. $\psi^j$) are bosonic
(resp. fermionic) superfields,
and the $B_i$ (resp. $F_j$) commuting (resp. anticommuting) generators in the
considered finite dimensional superalgebra $\cg$. Then the corresponding
supercurrents are:
\be
J_+ = \hat{g}^{-1} D_+ g \ \ \ \ J_- = (D_-g) g^{-1}
\ee
where $\hat{g}$ differs from $g$ by the change of sign on its fermionic
generator part, the bosonic ones staying unchanged. We note that the fermionic
character of $D_{\pm} = \theta_{\pm} \prt_{x_{\pm}} + \prt_{\theta_{\pm}}$
implies the supercurrents to develop as:
\be
J = \Psi^i B_i + \Phi^j F_j
\ee
the $\Psi^i$ being fermionic and the $\Phi^j$ bosonic superfields.

The choice of the $J$ components which are constrained to be constant with
respect to those which are put to zero naturally defines a grading
(see \ref{grad}) on the
(super)algebra $\cg$. The simplest and most known example is
the Abelian Toda model relative to $\cg$. In this case the J components
associated to the opposite of the simple roots have constant values while those
relative to the other negative roots are put to zero. The grading is ruled by
the generator $H$, sum of the Cartan generators in the Cartan Weyl basis. The
$\cg$ subalgebra ${\cg}_0$ is exactly the Cartan subalgebra of $\cg$ in this
basis, the simple root generators $E_{+\alpha}$ form the $\cg$ subspace
$\cg_{+1}$, and their partners $E_{-\alpha}$ the subspace ${\cg}_{-1}$;
finally $\cg_+$ is constructed from the positive roots and $\cg_-$ from the
negative ones.

As could be expected, imposing a set of constraints reduces the huge symmetry
provided by the Kac-Moody current algebra to a subset of quantities,
polynomials in the current components and their derivatives, which will
constitute a $W$-algebra. For example, the original
conformal symmetry of the WZW
model itself is broken when constraints corresponding to the grading $H$ are
imposed, and in order to construct the Virasoro symmetry for this Toda model a
$H$ dependent correction term has to be added to the former one.

More precisely, the stress energy tensor reads \cite{4}:
\be
T_H = \shalf  Tr J^2 - Tr H \prt J \label{eq:2}
\ee
when $\cg$ is an algebra, and \cite{13}:
\be
T_H =  Str (\sm{1}{3} J\hat{J}J + \shalf JDJ) - Str(H. D^2 J) \label{eq:2b}
\ee
when $\cg$ is a superalgebra.

\indent

The determination of the other generators of the $W$ algebra can be achieved as
follows.

\indent

If $\cg$ is an algebra, one selects in $\cg_{-1}$ a (constant) element $M_-$
such that \cite{4}
\be
Ker (ad M_-) \cap {\cg}_+ = \{0\}.\label{eq:1}
\ee
Then one expresses $J$ as:
\be
J=M_- + J_{>-1}
\ee
where the variable dependent part $J_{>-1}$ belongs to $\oplus_h\,{\cg}_h$
with $h> -1$.

\indent

If $\cg$ is a superalgebra, then one picks up in ${\cg}_{-1/2}$ a
fermionic (constant) element $F_-$ with $\{F_-,F_-\} = M_- \neq 0$ such
that:
\be
Ker (ad F_-) \cap \cg_+ = \{ 0 \}
\ee
and one expresses $J$ as:
\be
J=F_- + J_{>-\half}.
\ee

\indent

Finally one has just to use the gauge transformations:
\be
J \rightarrow g J g^{-1} + (\prt g) g^{-1}
\ee
where $g$ belongs to the local Lie groups generated by $\cg_+$, or:
\be
J \rightarrow \hat{g} J g^{-1} + (D_-g) g^{-1}
\ee
in the supersymmetric case, to transform $J$ into:
\be
J' = \mu_- + \sum_h W_{h+1} (J) X_h \mb{with} \mu_- = M_-\
(\mbox{resp.}\ F_-) \label{eqJg}
\ee
where the $W_{h+1}(J)$ are gauge invariant polynomials
generating the $W$ algebra
associated to the Toda theory. The PB among $W$ generators will be calculated
from the PB:
\be
\{ J^a(x), J^b (y') \}_{PB} = i f^{ab}_c \delta(x-x') J^c (x') + k^{ab}
\prt_x \delta(x-x').
\ee
when $\cg$ is a Lie algebra and:
\be
\{ J^a(X), J^b (X') \}_{PB} = i(-1)^{[a](1+[b])} f^{ab}_c \delta(X-X')
J^c (X') + k\eta^{ab} D_x \delta(X-X').
\ee
when $\cg$ is a superalgebra. $f^{ab}_c$ are the structure constants,
$\eta^{ab}$ the scalar product and
$k$ the central extension parameter of the Kac Moody (super)algebra; by $[a]$
is
expressed the $\Z_2$ grading of the generator $T^a$: $[a] =0$ (resp. 1) if
$T^a$ is
a commuting (resp. anticommuting) generator (see \cite{13} for more details).

\indent

Using $(\ref{eq:2b})$ one understands that $W_{h+1} (J)$ has
a (super) conformal weight
$1+h$ under $T_H$.

\indent

Note that the condition (\ref{eq:1}) expresses the non degeneracy
for $h>0$, of the
operator:
\be
ad M_-\ :\ \cg_h \rightarrow \cg_{h-1}.
\ee

Then Drinfeld-Sokolov (D.S.) gauges can be used to determine a complete set of
gauge invariant quantities $W_{h+1} (J)$. In the highest weight D.S. gauge,
each $W_{h+1} (J)$ is "carried" by the highest weight of a given $Sl(2)$
subalgebra built from $M_-$.

\subsection{Properties of (half) integral gradations \label{zut}}

\indent

We have presented in \cite{faux} three propositions establishing a
correspondence between (integral and half integral) gradings of a simple Lie
algebra $\cg$ which specify Toda theories, and $Sl(2) \oplus U(1)$
subalgebras of
$\cg$. The generalisation to the superalgebra case is straightforward,
replacing the $Sl(2)$ part by its "supersymmetric extension" $OSp(1|2)$.
Therefore, we limit ourselves to enounce hereafter these properties.

\indent

Let $H$ be a grading operator of a (super)algebra $\cg$. Then:

\indent

\underline{Proposition 1}:

$i)$ $\cg$ being an algebra, any
element $M_-\in\cg_-$ can be embedded in one of its
$Sl(2)$ subalgebra.

$ii)$ $\cg$ being a superalgebra, any fermionic
element $F_-\in\cg_-$ with $\{F_-,F_-\}=M_-\neq0$ can be embedded in one of its
$OSp(1|2)$ subalgebra.

\indent

\underline{Proposition 2}:

Let $M_-\in\cg_{-1}$ (resp. $F_-\in\cg_{-1/2}$).
Then, it is always possible to
write $H$ as:
\be
H=M_0+Y
\ee
with $M_0$ being the Cartan part of an $Sl(2)$ algebra constructed from
$M_-$ (resp. an $OSp(1|2)$ superalgebra built on $F_-$), and the
generator $Y$ commuting, when non zero, with this three (resp. five)
dimensional subalgebra.

Moreover, the $Sl(2)$ part constructed from $M_-$ (resp. $OSp(1|2)$
superalgebra built
on $F_-$) is unique up to a
conjugation by group elements generated from the subalgebra
$\hat{\cg}_0=Ker(adM_-)\cap\cg_0$.

\indent

\underline{Proposition 3}:

$i)$ Let $M_-,M_0,M_+$ and $Y$ generate an
$Sl(2)\oplus U(1)$ subalgebra of $\cg$ with $M_-\in\cg_{-1}$ and
$M_0+Y=H$. Decompose $\cg$, considered as a vector space, into irreducible
representations $\cd_{j_i}(y_i)$ of this algebra,
where $y_i$ denotes the eigenvalue
of $Y$ on the $Sl(2)$ representation $\cd_{j_i}$. Then
\be
Ker(adM_-)\cap\cg_+=\{0\} \mb{iff} |y_i|\leq j_i \mb{for any} \cd_{j_i}(y_i)
\mb{in} \cg. \label{Ndeg}
\ee

$ii)$ Let $M_-,F_-,M_0,F_+,M_+$ and $Y$ generate an
$OSp(1|2)\oplus U(1)$ subsuperalgebra of $\cg$ with $F_-\in\cg_{-1/2}$ and
$M_0+Y=H$. Decompose $\cg$, considered as a vector space, into irreducible
representations $\crr_{j_i}(y_i)$ of this algebra,
where $y_i$ denotes the eigenvalue
of $Y$ on the $OSp(1|2)$ representation
$\crr_{j_i}=\cd_{j_i}\oplus\cd_{j_i-1/2}$. Then
\be
Ker(adF_-)\cap\cg_+=\{0\} \mb{iff} |y_i|\leq j_i \mb{for any} \crr_{j_i}(y_i)
\mb{in} \cg. \label{Ndeg2}
\ee
in the following, we will call the condition (\ref{Ndeg}) (resp. \ref{Ndeg2})
a {\em non degeneracy condition} for $adM_-$ (resp. for $adF_-$).
Of course, as the grades satisfy $h_i=j_i+y_i$, one must impose
$h_i\in\half\Z$ in the $\cg$ adjoint representation to have (half)integral
grading.

\indent

These three propositions have to be completed by:

\indent

\und{Proposition 4:}

The gradations $H=M_0+Y$ and $M_0$ lead to the same $W$ algebra.

\indent

This last proposition has been proven in \cite{14}. From
the point of view of the decomposition under $Sl(2)\oplus U(1)$, note
that (\ref{Ndeg}) ensures that the highest weight of the $Sl(2)$ subalgebra are
in the $\cg_{\geq0}$ part of $\cg$ for both $H=M_0$ and $H=M_0+Y$ gradations.
This is in agreement with
the "halving" used in \cite{14}.

\subsection{Primary fields of $W$ algebras}

\indent

The spin of the $W$ generators corresponding to a given gradation $H$ are
obtained from the highest weights of the $Sl(2)\oplus U(1)$ (resp.
$OSp(1|2)\oplus U(1)$ ) decomposition of
the $\cg$-adjoint representation (DS gauge).
Now, we have to know whether the $W$ generators
are (super) primary fields under $T_H$. The (super) primary fields satisfy
the following Poisson bracket:
\bea
\{T_H(x),W_{h+1}(x')\}_{PB} &=& (h+1)W_{h+1}(x')\prt_x\delta(x-x')
+\prt W_{h+1}(x')\delta(x-x') \label{gol1}\\
\{T_H(X),W_{h+1/2}(X')\}_{PB} &=& (h+\half)\prt_x\delta(X-X')W_{h+1}(X')
+\delta(X-X')\prt W_{h+1}(X')+ \nonumber \\
&& -\half D_X\delta(X-X')W_{h+1/2}(X') \label{gol2}
\ena
where we have used for the supersymmetric case the conventions
\be
X=(x,\theta)\mb{and} \delta(X-X')=(\theta-\theta')\delta(x-x')
\ee
Note that (\ref{gol1}) corresponds to PB between fields
and (\ref{gol2}) between superfields.
We will say, in the former case, that $W_{h+1}$ has {\em spin} $h+1$,
whereas, in the latter case, $W_{h+1/2}$ carries a {\em
superspin}\footnote{Note that the two components of the superfield $W_{h+1/2}$
are of conformal spins $h+\half$ and $h+1$.}
$h+\half$.
In fact, it is clear from the expression of
$T_H$ that the only generators $W_{h+1}$ (resp. $W_{h+1/2}$)
which are not primary are those which
satisfy $<H,X_h>\neq 0$, where $<,>$ is the $\cg$ non degenerated scalar
product and $X_h$ is the
generator of $\cg$ carrying $W_{h+1}$ (resp. $W_{h+1/2}$) in (\ref{eqJg}).
This implies that $X_h$
is a Cartan generator, so that $h=0$ and $W_{h+1}\equiv W_1$
(resp. $W_{h+1/2}\equiv W_{1/2}$) forms a singlet
representation of $Sl(2)$ (resp. $OSp(1|2)$).
Actually, by linear combinations, one can always
eliminate these non primary generators, but one.
Since for $H=M_0$ all
the $W$ generators are primary (except $T_{M_0}$ of course), we can think at
the non primary generator as carried by $Y$ itself \cite{14}. This is ensured
by
the equality
\be
<H,Y>=<M_0+Y,Y>=<Y,Y>\neq0 \mb{iff} Y\neq0
\ee
We will call this (super) generator $W_1^Y$ (resp. $W_{1/2}^Y$). Note that
because of its spin 1 (resp. superspin $\half$), the
PB of $T_H$ with $W^Y_1$ ($W_{1/2}^Y$) differs from the PB of $T_H$ with a
(super)
primary field only by a central extension term, corresponding to a second order
derivative (resp. fermionic derivative $D$) of a (super)delta distribution.

\indent

{\em Thus, all the $W$ generators are primary with respect to $T_H$, except
$T_H$ itself and, when $Y\neq0$, a spin 1 generator $W^Y_1$
(resp. a superspin $\half$ generator $W_{1/2}^Y$) carried by $Y$.
In that case, $W^Y_1$ ($W_{1/2}^Y$) differs from a primary
field (resp. superfield) by a central extension term.}

\subsection{Classification of constrained WZW models. \label{class}}

\indent

The above properties suggest a way to determine all the different
(super)Toda models associated with (half-)integral gradings of a simple Lie
(super)algebra $\cg$, and their corresponding (super) $W$ algebras, namely:

\indent

$i)$ Classify all the $Sl(2)$ (resp. $OSp(1|2)$) sub(super)algebras of $\cg$

\indent

$ii)$ Add to each of these simple sub(super)algebras a commuting
$U(1)$ factor such that in the decomposition of the $\cg$ adjoint
representation
into $Sl(2)\oplus U(1)$ representations $\cd_{j_i}(y_i)$ (resp. $OSp(1|2)\oplus
U(1)$ representations $\crr_{j_i}(y_i)$),
the following conditions hold:
\bea
&&
\vert y_{i}\vert \leq j_i \hs{5} i=1,\cdots, n \label{eq:4} \\
&& j_i+y_i\in \Z \mb{(integral grading)} \ j_i+y_i\in \half\Z
\mb{(half-integral grading)}\label{eq:5}
\ena
Note that the $y_i$ values are naturally restricted when calculating the
$Sl(2)\oplus U(1)$ (resp. $OSp(1|2) \oplus U(1)$) decomposition of the adjoint
representation of $\cg$ coming from the product of fundamental representations
already decomposed into $Sl(2)\oplus U(1)$ (resp. $OSp(1|2)\oplus U(1)$)
representations: this will be explicited in the following.

\indent

$iii)$ Then to each such an $Sl(2)\oplus U(1)$ (resp. $OSp(1|2)\oplus
U(1)$) sub(super)algebra of $\cg$
satisfying (\ref{eq:4}) and (\ref{eq:5}) there will correspond
a classical (i.e. PB) $W$ algebra
generated by the $n$ elements $W_{{h_1}+1}$, $\ldots$, $W_{{h_n}+1}$
(resp. $W_{{h_1}+1/2}$, $\ldots$, $W_{{h_n}+1/2}$)
of conformal (super)spin under the (super)Virasoro algebra defined in
(\ref{eq:2},\ref{eq:2b}) $h_1+1$, $\ldots$, $h_n+1$
(resp. ${h_1}+\half$, $\ldots$, ${h_n}+\half$) with $h_i$ given by
\be
h_i=y_{i}+j_i
\ee
as a consequence of a Drinfeld-Sokolov highest weight gauge \cite{4,5}.

\indent

$iv)$ Reconstruct the grading $H$ from the $Sl(2)\oplus U(1)$
(resp. $OSp(1|2)\oplus U(1)$) decomposition.
Varying $Y$ for a fixed $Sl(2)$ ($OSp(1|2)$ super)algebra will give all
the isomorphic gradations.

\indent

$v)$ Deduce informations of the PB from the $Sl(2)\oplus U(1)$
(resp. $OSp(1|2)\oplus U(1)$) reduction.

\indent

These five steps will be explicited in the following. In the part
\ref{part1}, we will focus on the algebras case, while in the part
\ref{part2} the previous results will be used to state
the superalgebras case.

\newpage

\part{$W$ algebras built on Lie algebras \label{part1}}

\sect{The different $Sl(2)$ subalgebras in a simple Lie algebra $\cg$
\label{sect3}}

\subsection{$Sl(2)$ associated to regular subalgebras \label{gy}}

\indent

The classification of $Sl(2)$ subalgebras of a simple Lie algebra $\cg$
has been achieved by Dynkin \cite{11}. His technics can be summarized as
follows:

\indent

{\em Any $Sl(2)$ subalgebra in $\cg$ can be seen, up to a few exceptions
occuring in $D_n$ and $E_{6,7,8}$ algebras (which will be discussed in a
moment), as the principal $Sl(2)$ algebra of a regular $\cg$ subalgebra.}

\indent

Let us remind that the principal $Sl(2)$ subalgebra of a (semi) simple algebra
$\cg$ admits as a positive vector root, the generator:
\be
E_+= \sum_{\alpha\in\Phi_+} E_\alpha
\ee
if $\Phi_+$ denotes a system of simple roots in $\cg$.

\indent

In order to define a regular subalgebra of $\cg$, let us consider, once given a
Cartan subalgebra $\ch$, the canonical vector space decomposition:
\be
\cg= \left(\bigoplus_{\alpha\in\Delta}\ \cg_\alpha \right) \oplus\ch
\ee
where $\Delta$ is the root system of $\cg$, and $\cg_\alpha$ the one
dimensional $\cg$ subspace associated to the root $\alpha$. Then, a subalgebra
$\cg'$ of $\cg$ is said regular if it can be seen, up to a conjugation of
$\cg$, as:
\be
\cg'= \left(\bigoplus_{\beta\in\Delta'}\ \cg_\beta\right)
\oplus\ch' \label{luc4}
\ee
with $\Delta'\subset\Delta$ and $\ch'\subset\ch$. Obviously, the principal
$Sl(2)$ of $\cg$ is not regular (except if $\cg=Sl(2)$). When a subalgebra is
not regular, it is said singular.

\indent

Discarding the possible $U(1)$ factors, one can assert that a semi-simple
regular subalgebra of $\cg$ admits as Dynkin diagram (DD) a subdiagram of the
DD of $\cg$. By subdiagram of $\cg$, we understand a DD obtained in the
following way.

\indent
$i)$ First extend the $\cg-$DD by adjunction of the lowest root of $\cg$, and
consider the different DD's obtained by removing one dot.

\indent
$ii)$ For these diagrams repeat the procedure $i)$ to each of their connected
parts until no new (connected or not connected) diagram with
$r=$rank($\cg$) dots is obtained.

\indent
$iii)$ Finally, from each such a "maximal" subdiagram, remove $m$ dots with
$0\leq m<r$. One gets in this way all the $\cg$ subdiagrams.

\indent

Of course, by this construction, the same subdiagram may appear several times.
Such identical diagrams correspond in general to conjugate subalgebras, with
three exceptions as long as we discard the $E$ series: see below.

\indent

When the $\cg-$DD possesses roots of different length (the long
ones represented by
\begin{picture}(15,12)
\put(8,1){\circle{11}}
\end{picture},
and the short ones by
\begin{picture}(15,12)
\put(8,1){\circle*{11}}
\end{picture})
it is understood (and rather obvious) that the two subdiagrams
\begin{picture}(15,12)
\put(8,1){\circle{11}}
\end{picture}\
and
\begin{picture}(15,12)
\put(8,1){\circle*{11}}
\end{picture}\
correspond to two non-conjugate $Sl(2)$ algebras.

\subsection{Exceptions}

\indent

We have explicited in section \ref{gy} how to get the regular subalgebras
of a given algebra $\cg$ using the method of subdiagrams. In that way,
we get most of the $Sl(2)$ embeddings of $\cg$, the other ones being obtained
as follows:

\indent

1) Regular subalgebras of the type $A_{n_1}+A_{n_2}+\cdots+A_{n_k}$, with
$n_1+n_2+\cdots+n_k=2n-k$ and $n_i$'s all odd, show up twice in $D_{2n}$,
$n>2$ (three times in $D_4$), each of them being related to the other by an
outer automorphism of $D_{2n}$. As an example, there are two regular
subalgebras of the type $3A_1$, $A_1+A_3$ and $A_5$ in $D_6$.

\indent

2) For $\cg=B_n$ and $D_n$, $n>3$, the diagram
$\bigcirc\hspace{-.3mm}\rule[1mm]{5mm}{.1mm}\hspace{-1.1mm}\bigcirc
\hspace{-1.1mm}\rule[1mm]{5mm}{.1mm}\hspace{-.3mm}\bigcirc $
must be considered twice, one been related to an "algebra $A_3$", and the other
one to "$D_3$". Indeed, the $\cg$ subdiagram

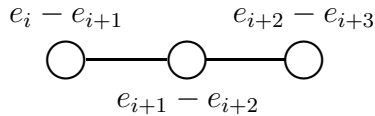
\begin{picture}(400,60)
\thicklines
\put(165,30){\circle{14}}
\put(165,46){\makebox(0.4444,0.6667){\tenrm
$e_i-e_{i+1}$}}
\put(173,30){\line( 1, 0){ 30}}
\put(211,30){\circle{14}}
\put(211,14){\makebox(0.4444,0.6667){\tenrm
$e_{i+1}-e_{i+2}$}}
\put(218,30){\line( 1, 0){ 30}}
\put(255,30){\circle{14}}
\put(255,46){\makebox(0.4444,0.6667){\tenrm
$e_{i+2}-e_{i+3}$}}
\end{picture}

defines a system of simple roots for "$A_3$", while the subdiagram

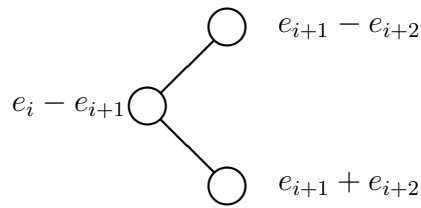
\begin{picture}(400,90)
\thicklines
\put(221,40){\circle{14}}
\put(191,40){\makebox(0.4444,0.6667){\tenrm
$e_{i}-e_{i+1}$}}
\put(226,45){\line( 1, 1){ 20}}\put(226,35){\line( 1,-1){ 20}}
\put(251,70){\circle{14}}
\put(297,70){\makebox(0.4444,0.6667){\tenrm
$e_{i+1}-e_{i+2}$}}
\put(251,10){\circle{14}}
\put(297,10){\makebox(0.4444,0.6667){\tenrm
$e_{i+1}+e_{i+2}$}}
\end{picture}

provides a system of simple roots of "$D_3$". In order to convince the reader,
we remark that the fundamental representation of $D_n$ reduces with respect to
$A_3$ as $\und{2n}=\und{4}+\ov{{4}}+(2n-8)\und{1}$, and with respect to
$D_3$ as $\und{2n}=\und{6}+(2n-6)\und{1}$.

\indent

3) Again $B_n$ and $D_n$ admit two different types of $2A_1$ subalgebras
associated to the diagrams

\begin{picture}(200,40)
\thicklines
\put(110,8){\circle{14}}
\put(110,22){\makebox(0.4444,0.6667){\tenrm
$e_{1}-e_{2}$}}
\put(160,8){\circle{14}}
\put(160,22){\makebox(0.4444,0.6667){\tenrm
$e_{3}-e_{4}$}}
\end{picture}
and
\begin{picture}(110,40)
\thicklines
\put(30,8){\circle{14}}
\put(30,22){\makebox(0.4444,0.6667){\tenrm
$e_{1}-e_{2}$}}
\put(80,8){\circle{14}}
\put(80,22){\makebox(0.4444,0.6667){\tenrm
$e_{1}+e_{2}$}}
\end{picture}

The fundamental of $D_n$ reduces with respect to the first
algebra as
$\und{2n}=(\und{2}+\ov{{2}},\und{0})+(\und{0},\und{2}+\ov{{2}})
+(2n-8)(\und{0},\und{0})$ and with respect to the second as
$\und{2n}=(\und{2},\und{2})+(2n-4)(\und{0},\und{0})$. We can note that as well
as in case 1), it is the bifurcation appearing in the (extended) DD
of $(B_n)$ $D_n$ which is responsible for these doublings.

\indent

4) As mentioned at the beginning of the previous paragraph, there exist
also $Sl(2)$
subalgebras which cannot be seen as principal $Sl(2)$ part of a {\em regular}
subalgebra. This concerns only the $D_n$ and $E_{6,7,8}$ algebras. In the
$D_n$ case, one has to add $[\sm{$n$-2}{2}]$ $Sl(2)$ subalgebras, each of them
being a principal subalgebra of the singular ones:
\be
B_i\oplus B_j \mb{with} i+j=n-1 \mb{and} i\neq j
\ee
For example, in $D_4\equiv SO(8)$ and $D_5\equiv SO(10)$, there exists one
algebra of this type, associated to $SO(3)\oplus SO(5)$ and $SO(3)\oplus SO(7)$
respectively. In $D_6\equiv SO(12)$, two new $Sl(2)$ algebras,
principal with respect to
$SO(3)\oplus SO(9)$ and $SO(5)\oplus SO(7)$ have to be considered.

\subsection{Example: subalgebras of $Sp(6)$}

\indent

We illustrate the method on the algebra\footnote{We remind the notations:
$A_n\equiv Sl(n+1)$, $B_n\equiv SO(2n+1)$, $C_n\equiv Sp(2n)$ and
$D_n\equiv SO(2n)$.} $C_3$. Its DD and extended DD are
respectively:

\begin{picture}(400,60)
\thicklines
\put(15,30){\circle*{14}}
\put(23,30){\line( 1, 0){ 30}}
\put(61,30){\circle*{14}}
\put(69,32){\line( 1, 0){ 30}}\put(69,28){\line( 1, 0){ 30}}
\put(107,30){\circle{14}}
\put(147,30){\makebox(1,1){\tenrm and}}
\put(200,30){\circle{14}}
\put(208,32){\line( 1, 0){ 30}}\put(208,28){\line( 1, 0){ 30}}
\put(239,30){\circle*{14}}
\put(247,30){\line( 1, 0){ 30}}
\put(278,30){\circle*{14}}
\put(286,32){\line( 1, 0){ 30}}\put(286,28){\line( 1, 0){ 30}}
\put(324,30){\circle{14}}
\end{picture}

Deleting one dot, we get at the first step:

\begin{picture}(400,60)
\thicklines
\put(15,30){\circle*{14}}
\put(23,30){\line( 1, 0){ 30}}
\put(61,30){\circle*{14}}
\put(69,32){\line( 1, 0){ 30}}\put(69,28){\line( 1, 0){ 30}}
\put(107,30){\circle{14}}
\put(147,30){\makebox(0.4444,0.6667){\tenrm and}}
\put(240,30){\circle{14}}
\put(278,30){\circle*{14}}
\put(286,32){\line( 1, 0){ 30}}\put(286,28){\line( 1, 0){ 30}}
\put(324,30){\circle{14}}
\end{picture}

which correspond to the subalgebras $C_3$ and $A_1\oplus C_2$ respectively.
Then, extending
once more this last diagram, we get a new three dots DD:

\begin{picture}(400,60)
\thicklines
\put(15,30){\circle{14}}
\put(61,30){\circle{14}}
\put(69,32){\line( 1, 0){ 30}}\put(69,28){\line( 1, 0){ 30}}
\put(107,30){\circle*{14}}
\put(115,32){\line( 1, 0){ 30}}\put(115,28){\line( 1, 0){ 30}}
\put(153,30){\circle{14}}
\put(180,30){\line( 1, 0){30}}
\put(210,30){\line( -2, 1){10}}\put(210,30){\line( -2, -1){10}}
\put(235,30){\circle{14}}
\put(265,30){\circle{14}}
\put(295,30){\circle{14}}
\end{picture}

There is no new possibility to get three dots DD by the extension-deletion
procedure. A direct chirurgy on them leads to the DD:

\begin{picture}(500,60)
\thicklines
\put(1,30){\circle*{14}}
\put(9,30){\line( 1, 0){ 26}}
\put(43,30){\circle*{14}}
\put(70,30){\makebox(0.4444,0.6667){\tenrm ;}}
\put(102,30){\circle*{14}}
\put(110,32){\line( 1, 0){ 30}}\put(110,28){\line( 1, 0){ 30}}
\put(148,30){\circle{14}}
\put(169,30){\makebox(0.4444,0.6667){\tenrm ;}}
\put(202,30){\circle*{14}}
\put(242,30){\circle{14}}
\put(269,30){\makebox(0.4444,0.6667){\tenrm ;}}
\put(302,30){\circle{14}}
\put(332,30){\circle{14}}
\put(359,30){\makebox(0.4444,0.6667){\tenrm ;}}
\put(389,30){\circle*{14}}
\put(415,30){\makebox(0.4444,0.6667){\tenrm ;}}
\put(446,30){\circle{14}}
\end{picture}

Gathering all these DD's, we can list the regular (semi-)simple
subalgebras of $C_3$:
\be
\begin{array}{ll}
\mbox{ Rank 3 subalgebras } & C_3\ ;\ C_2\oplus A_1\ ;\ 3A_1 \\
\mbox{ Rank 2 subalgebras } & C_2\ ;\ A_2^2\ ;\ 2A_1\ ;\ A_1\oplus A_1^2 \\
\mbox{ Rank 1 subalgebras } & A_1\ ;\ A_1^2
\end{array}
\ee
where the $Sl(2)$ and $Sl(3)$ subalgebras
built on short roots are marked with the superscript (Dynkin index) 2.
To each regular subalgebra of $C_3$ will be
associated a principal $Sl(2)$ subalgebra, and this construction will exhaust
the set of $Sl(2)$ subalgebras.

\sect{$Sl(2)$ decompositions of simple Lie algebras}

\indent

Given any $Sl(2)$ subalgebra of a Lie algebra $\cg$ in the $A,B,C,D$ series, we
need to know the decomposition of the adjoint representation of $\cg$ with
respect to this three dimensional subalgebra. For such a purpose, we will first
compute the $Sl(2)$ decomposition of the $(1,0,0,...,0)$ fundamental
representation of $\cg$. We will deduce the $Sl(2)$ decomposition of the $\cg$
adjoint representation by computing the product of the fundamental
representation by its
contragredient one: for the $A_n$ series, the adjoint representation is
given by this product, once
throwing away a trivial representation, while in the $B_n$ and $D_n$ (resp.
$C_n$) cases, one has to select the antisymmetric (resp. symmetric) part.

\subsection{The $\cg$ fundamental representation with respect to a $Sl(2)$
subalgebra}

\subsubsection{$Sl(n)$ case}

\indent

Any $Sl(2)$ subalgebra is the principal subalgebra of a (sum of) $Sl(p)$
subalgebra(s) in $Sl(n)$. For each $Sl(p)$ subalgebra will correspond a
$\cd_{(p-1)/2}$ representation of $Sl(2)$ in the $\und{n}$
of $Sl(n)$, which will be completed with singlets.

\indent

For instance, if we look at the $Sl(2)$
principal subalgebra of $Sl(p)\oplus Sl(q)$
in $Sl(n)$, we will have
\be
\und{n}= \cd_{(p-1)/2}\oplus \cd_{(q-1)/2}\oplus (n-p-q)\cd_0
\ee

\subsubsection{$Sp(2n)$ case}

\indent

An $Sl(2)$ subalgebra is the principal subalgebra of a (sum of) $Sp(2p)$
subalgebra(s), $Sl(2)^1$ subalgebra(s), or $Sl(q)^2$ subalgebra(s). The
superscript refers to the Dynkin index of the $Sl(m)$ subalgebra considered:
it is 1 when the $Sl(2)$ subalgebra is
constructed on a long root, and 2 in the other cases. The $Sp(2p)$ subalgebra
contributes to the fundamental representation via a $\cd_{p-(1/2)}$ $Sl(2)$
representation, while the $Sl(q)^2$ (resp. $Sl(2)^1$) yields the
$\cd_{(q-1)/2}+{{\ov{\cd}}_{(q-1)/2}}$ (resp. $\cd_{1/2}$) representations. The
$\und{2n}$ representation is then completed by singlets. For example, for the
decomposition of $Sp(2n)$ under the principal $Sl(2)$ of $Sp(2p)\oplus
Sl(q)^2\oplus r Sl(2)^1$, we have:
\be
\und{2n}= \cd_{p-(1/2)}\oplus (\cd_{(q-1)/2}\oplus{{\ov{\cd}}_{(q-1)/2}})
\oplus r\cd_{1/2}\oplus (2n-2p-2q-2r)\cd_0
\ee

\subsubsection{$SO(n)$ case}

\indent

When $Sl(2)$ is principal subalgebra of either an $SO(2p+1)$ or an $SO(2p+2)$
one, the $\und{n}$ fundamental of $SO(n)$ contains a $\cd_p$ representation. In
the case of an $Sl(q)$, $q\neq 2$, subalgebra, then it is the sum
$\cd_{(q-1)/2}\oplus {{\ov{\cd}}_{(q-1)/2}}$ which shows up. For $q=2$, one
must
distinguish the case $Sl(2)^1$ (long root) which leads to
$\cd_{1/2}\oplus {{\ov{\cd}}_{1/2}}$ from the case $Sl(2)^2$ (short root)
leading to $\cd_1$.

\indent

Finally, we have mentionned in Section \ref{sect3} the existence
of two $Sl(2)\oplus Sl(2)$ and two $Sl(4)\equiv SO(6)$ algebras. The
corresponding decompositions are:
\bea
&&Sl(2)\oplus Sl(2) \
\left\{
\begin{array}{l}
\und{n}=2(\cd_{1/2}\oplus {{\ov{\cd}}_{1/2}})\oplus(n-8)\cd_0\\
\und{n}=\cd_1\oplus (n-3)\cd_0
\end{array}
\right.\\
&&Sl(4) \
\left\{
\begin{array}{l}
\und{n}=\cd_{3/2}\oplus {{\ov{\cd}}_{3/2}}\oplus(n-8)\cd_0\\
\und{n}=\cd_2\oplus (n-5)\cd_0
\end{array}
\right.
\ena
We remind that for each $SO(2n)$ subalgebras, there exist $Sl(2)$ algebras
related to the singular embeddings $SO(2k+1)\oplus SO(2n-2k-1)$, $0<2k<n$.

\subsection{The $\cg$ adjoint representation with respect to $Sl(2)$
subalgebras}

\indent

To achieve the $Sl(2)$ reduction of the adjoint representation for any
simple Lie algebra $\cg$ from the knowledge of the fundamental
representation, the
following formulae are especially convenient:
\bea
\left( \cd_n\times \cd_n\right)_A &=& \cd_{2n-1}\oplus \cd_{2n-3}\oplus \cdots
\oplus \cd_1 \hs{5} n\in \Z \label{pi}\\
\left( \cd_{n-(1/2)}\times \cd_{n-(1/2)}\right)_A &=&
\cd_{2n-2}\oplus \cd_{2n-4}\oplus \cdots\oplus \cd_0 \hs{5} n\in \Z \\
\left( \cd_n\times \cd_n\right)_S &=&
\cd_{2n}\oplus \cd_{2n-2}\oplus \cdots\oplus \cd_0 \hs{5} n\in \Z \\
\left( \cd_{n-(1/2)}\times \cd_{n-(1/2)}\right)_S &=&
\cd_{2n-1}\oplus \cd_{2n-3}\oplus \cdots
\oplus \cd_1 \hs{5} n\in \Z \label{pf}
\ena
the subscript (A) S standing for (Anti-)Symmetric part of the product. We have
also, for $m,p\in\Z$ and $j,k\in\half\Z$:
\bea
\left\{\rule{.0mm}{4mm} (m\cd_j)\times (m\cd_j)\right\}_A &=&
\sm{$m(m+1)$}{2} \left( \cd_j\times
\cd_j\right)_A \oplus\sm{$m(m-1)$}{2} \left( \cd_j\times
\cd_j\right)_S \nonumber\\
&=& m \left( \cd_j\times \cd_j\right)_A
\oplus\sm{$m(m-1)$}{2} \left( \cd_j\times
\cd_j\right) \label{pi2}\\
\left\{\rule{.0mm}{4mm} (m\cd_j)\times (m\cd_j)\right\}_S &=&
\sm{$m(m+1)$}{2} \left( \cd_j\times
\cd_j\right)_S \oplus\sm{$m(m-1)$}{2} \left( \cd_j\times
\cd_j\right)_A \nonumber\\
&=& m \left( \cd_j\times \cd_j\right)_S \oplus\sm{$m(m-1)$}{2}
\left( \cd_j\times\cd_j \right) \label{pm2}\\
\left\{\rule{.0mm}{4mm} (m\cd_j)\times(p\cd_k)\oplus
(p\cd_k)\times(m\cd_j)\right\}_A
&=& \left\{\rule{.0mm}{4mm} (m\cd_j)\times(p\cd_k)\oplus (p\cd_k)\times(m\cd_j)
\right\}_S \nonumber\\
&=& mp\left( \cd_j\times \cd_k\right) \label{pf2}
\ena
where $m\cd_j$ stands for the direct sum of $m$ representations $\cd_j$.

\subsection{Example: case of $SO(8)$\label{SO8}}

\indent

We illustrate our methods on $SO(8)$. First, we give the decompositions of its
fundamental and adjoint representations with respect to the $Sl(2)$ subalgebras
associated to {\em regular} subalgebras:
\be
\begin{array}{lll}
SO(8) & \und{8}=\cd_3\oplus \cd_0 & (8\times8)_A=\cd_5\oplus 2\cd_3
\oplus \cd_1\\
Sl(4) & \und{8}=\cd_{3/2}\oplus {{\ov{\cd}}_{3/2}} &
(8\times8)_A=\cd_3\oplus 3\cd_2\oplus \cd_1\oplus 3\cd_0 \\
Sl(4)' & \und{8}=\cd_{2}\oplus 3\cd_0 &
(8\times8)_A=\cd_3\oplus 3\cd_2\oplus \cd_1\oplus 3\cd_0 \\
Sl(3) & \und{8}=2\cd_1\oplus 2\cd_0 &
(8\times8)_A=\cd_2\oplus 7\cd_1\oplus 2\cd_0 \\
Sl(2) & \und{8}=\cd_{1/2}\oplus{{\ov{\cd}}_{1/2}}\oplus 4\cd_0 &
(8\times8)_A=\cd_1\oplus 8\cd_{1/2}\oplus 9\cd_0 \\
4Sl(2) & \und{8}=2\cd_{1}\oplus 2\cd_0 &
(8\times8)_A=\cd_2\oplus 7\cd_{1}\oplus 2\cd_0 \\
3Sl(2) & \und{8}=\cd_{1}\oplus 2{\cd_{1/2}}\oplus \cd_0 &
(8\times8)_A=2\cd_{3/2}\oplus 3\cd_1\oplus 4\cd_{1/2}\oplus 3\cd_0 \\
2Sl(2) & \und{8}=\cd_{1}\oplus 5\cd_0 &
(8\times8)_A=6\cd_1\oplus 10\cd_0 \\
(2Sl(2))' & \und{8}=2(\cd_{1/2}\oplus{{\ov{\cd}}_{1/2}})&
(8\times8)_A=6\cd_1\oplus 10\cd_0
\end{array}
\nonumber
\ee
One remarks that the reductions associated to the two $Sl(4)$ on the one hand,
and the two $2Sl(2)$ on the other hand, give the same decomposition of the
adjoint: this special case has to be related to the triality in $SO(8)$.

\indent

Finally, one has to add the {\em singular} embedding:
\be
SO(3)\oplus SO(5)\hs{5} \und{8}=\cd_1\oplus \cd_2\hs{5}
(8\times 8)_A= 2\cd_3\oplus \cd_2 \oplus 3\cd_1
\ee

\sect{$Sl(2)\oplus U(1)_Y$ decompositions of simple Lie algebras}

\subsection{$Sl(n)$ algebras}

\indent

We start by considering the case $\cg=Sl(n)$, which has already been studied in
some detail in \cite{faux}. Let us recall that, for such an algebra, all the
$Sl(2)$ representations of equal dimension $\cd_j$ have the same $U(1)_Y$
eigenvalue $y_j$ in the $\und{n}$ fundamental representation, so that a general
decomposition reads
\be
\und{n}= \bigoplus_{j} n_j\, \cd_j(y_j) \mb{with} j's \mb{all different}
\ee
where $n_j$ is the multiplicity of $\cd_j$.
One will have to impose to the product $\und{n}\times\ov{{n}}-\cd_0(0)$,
the non degeneracy
condition $|y|\leq j$ for any representation $\cd_j(y)$ in the $\cg$ adjoint
representation. Note that the condition $y\in\half\Z$, which ensures a
(half-)integral gradation, has to be imposed only in the adjoint
representation, and \und{not} in the fundamental.

\indent

As an example, consider the $Sl(2)$ which is principal with respect to $A_n$ in
$A_{n+2}$. Then
\bea
\und{n+3} &=& \cd_{n/2}(y) \oplus2\cd_0(-\sm{$n+1$}{2}y)\\
\ov{{n+3}} &=& \cd_{n/2}(-y) \oplus2\cd_0(\sm{$n+1$}{2}y)
\ena
where we have, of course, imposed the traceless condition for $Y$. It follows:
\be
\und{n+3}\times\ov{{n+3}}-\cd_0(0)=
\left(\oplus_{j=1}^n \cd_j(0)\right) \oplus 2\cd_{n/2}(\sm{$n+3$}{2}y) \oplus
2\cd_{n/2}(-\sm{$n+3$}{2}y) \oplus 4\cd_0(0)
\ee
with the condition $|\frac{n+3}{2}y|\leq\frac{n}{2}$ and
$y'=\frac{n+3}{2}y\in\half\Z$.

\subsection{$SO(n)$ algebras}

\indent

Now, let us turn to the $\cg=B_n$ or $D_n$ case. These algebras have a real
fundamental representation, so that if $\cd_j(y)$, $y\neq0$, appears in the
decomposition, then $\cd_j(-y)$ must also be present with the same
multiplicity.
To get the adjoint representation, we have to improve the formulae
(\ref{pi}$-$\ref{pf2}) by specifying the $U(1)$ dependence.
Using the reality of the adjoint representation, one is led to
\bea
&&\left\{ \rule{.0mm}{5mm} [n\cd_j(y)\oplus n\cd_j(-y)]\times
[n\cd_j(y)\oplus n\cd_j(-y)] \right\}_A
= n^2 \left( \rule{.0mm}{5mm} \cd_j\times \cd_j\right)(0)\oplus \\
&& \hs{10} \oplus\left(\rule{.0mm}{5mm}
(n\cd_j)\times (n\cd_j)\right)_A(2y) \oplus
\left(\rule{.0mm}{5mm} (n\cd_j)\times (n\cd_j)\right)_A(-2y)
\mb{for} j\in\half\Z\mb{and}n\in\Z \nonumber
\ena
where $((n\cd_j)\times (n\cd_j))_A$ is computed via (\ref{pi2}).
This formula shows
that from a term $n\cd_j(y)$ in the fundamental, we will always get a term
$\cd_0(2y)$ in the adjoint, except if $n=1$ and $j$ is integer. Moreover, when
$n=1$ and $j$ is integer but non zero, there will always exists a
$\cd_1(\pm2y)$ term in the adjoint representation. The non degeneracy condition
$|y|\leq j$ for $\cd_j(y)$ will then led to set $y=0$, except for $n=1$ and $j$
integer, where, for $j\neq0$, we will have $|2y|\leq1$ and $2y\in\half\Z$,
that is $y=0$, or $y=\pm\sm{1}{4}$, or $y=\pm\half$.

\indent

{\em Thus, for the orthogonal series, the only $Sl(2)$ representation with non
zero $U(1)$ eigenvalues are those which appear in the fundamental
representation as $n(\cd_p(y)\oplus \cd_p(-y))$ with $n=1$ and $p$ integer.
Moreover, for $p\neq0$, we have $|2y|=0$, or $\half$, or 1.}

\indent

Note that these restrictions are necessary but not sufficient conditions on
$y$: we still have to impose the non degeneracy condition in the $\cg$ adjoint.
To be complete, let us add the formula:
\bea
&&\left\{ \rule{.0mm}{5mm} 2[n\cd_j(y)\oplus n\cd_j(-y)]\times
[p\cd_k(y')\oplus p\cd_k(-y')] \right\}_A=
(\cd_n\times \cd_p)(y+y')\oplus
\nonumber \\
&&\hs{14} \oplus(\cd_n\times \cd_p)(-(y+y'))\oplus
 (\cd_n\times \cd_p)(y-y')\oplus (\cd_n\times \cd_p)(-(y-y'))
\ena

\indent

As an example, we look at the principal $Sl(2)$ of $SO(2n-1)$ in $SO(2n+1)$:
\bea
&&\und{2n+1}= \cd_{n-1}(0)\oplus \cd_0(y)\oplus \cd_0(-y)= \ov{{2n+1}}\\
&&(\und{2n+1}\times\und{2n+1})_A =(\cd_{2n-3}\oplus
\cd_{2n-1}\oplus\cdots\oplus
\cd_1 \cd_0)(0)\oplus \cd_{n-1}(y)\oplus \cd_{n-1}(-y) \nonumber
\ena
with the condition $|y|\leq n-1$.

\subsection{$Sp(2n)$ algebras}

\indent

Finally, let us study the case $\cg=C_n$. From the $SO(n)$ case, it is easy to
deduce the rule:
\bea
&&\left\{ \rule{.0mm}{5mm} [n\cd_j(y)\oplus n\cd_j(-y)]\times
[n\cd_j(y)\oplus n\cd_j(-y)] \right\}_S
= n^2 (\cd_j\times \cd_j)(0)\oplus \\
&& \hs{10} \oplus\left( \rule{.0mm}{5mm} (n\cd_j)\times (n\cd_j)
\right)_S(2y) \oplus
\left( \rule{.0mm}{5mm} (n\cd_j)\times (n\cd_j)\right)_S(-2y)
\mb{for} j\in\half\Z\mb{and}n\in\Z \nonumber
\ena
where $((n\cd_j)\times (n\cd_j))_A$ is computed via (\ref{pm2}).

\indent

Then, the $Sl(2)\oplus U(1)$ decomposition of $C_n$ is deduced from the
$B_n$ one by exchanging integer and half-integer:

\indent

{\em For the symplectic series, the only $Sl(2)$ representations with non
zero $U(1)$ eigenvalues are those which appear in the fundamental
representation as $n(\cd_{p+\half}(y)\oplus \cd_{p+\half}(-y))$
with $n=1$ and $p$ integer.
Moreover, the allowed eigenvalues for the $U(1)$ generator
$Y$ are $|2y|=0$, or $\half$, or 1.}

\indent

We illustrate these results on the decomposition under the $Sl(2)$ of
$Sl(2)^2\oplus Sp(2n-2)$ in $Sp(2n+2)$:
\bea
\und{2n+2}&=& \cd_{n-(3/2)}(0) \oplus\cd_\half(y) \oplus\cd_\half(-y) \\
\ov{{2n+2}}&=& \cd_{n-(3/2)}(0) \oplus\cd_\half(-y) \oplus\cd_\half(y) \\
(\und{2n+2}\times\ov{2n+2})_S &=& (\cd_{2n-3}\oplus \cd_{2n-5} \oplus\cdots
\oplus\cd_1)(0)\oplus (\cd_1\oplus\cd_0)(0)\oplus \\
&& \oplus\cd_1(2y)\oplus\cd_1(-2y)
\oplus (\cd_{n-1}\oplus \cd_{n-2})(y)\oplus
(\cd_{n-1}\oplus \cd_{n-2})(-y) \nonumber
\ena
with $|2y|\leq 1$.

\subsection{The algebra $G_2$\label{luc3}}

\indent

This (rank 2) algebra admits the system of roots:
\be
\pm(e_i\pm e_j) \mb{;} \pm(2e_i-e_j-e_k)
\mb{with} i,j,k=1,2,3 \mb{all different}
\ee
The fundamental representation of $G_2$ is seven-dimensional, and its adjoint
has the dimension 14. These representations are real.
To simplify the discussion about $Sl(2)\oplus U(1)$ decomposition, we remark
that $G_2$ can be embedded in $SO(7)$ (in a singular way). As a consequence,
its
adjoint representation will be present in the antisymmetric part of the
product $\und{7}\times\und{7}$.
Indeed, we have \cite{slan}:
\be
(\und{7}\times\und{7})_A = \und{7}\oplus \und{14}
\ee
Thus, we can obtain the adjoint representation from the fundamental by
$\und{14}=(\und{7}\times\und{7})_A - \und{7}$. It is then sufficient to know
the decomposition of the fundamental. This is done with the
same rules as for the $SO(n)$ algebras (because of the embedding $G_2\subset
SO(7)$). Note that none of the $Sl(2)$ subalgebras of $G_2$ can be extended to
a $Sl(2)\oplus U(1)$ subalgebra in such a way that (\ref{Ndeg}) is still
satisfied. The results are presented in table \ref{t0}. The defining vector is
given in the Cartan basis of $SO(7)$, the Cartan generators of
$G_2$ being given by $H_1-H_2$ and $2H_2-H_1-H_3$ (see section \ref{mach}).

\subsection{The algebra $F_4$}

\indent

This exceptional algebra has rank 4 and dimension 52. Its fundamental
representation has dimension 26, and $F_4$ can be (irregularily) embedded in
$SO(26)$. However, one cannot directly obtains the adjoint representation from
the fundamental one, since a new representation appears in the antisymmetric
part of the product:
\be
(26\times26)_A= 52+273
\ee
Thus, the only information on the $U(1)$ dependence comes from the non
degeneracy condition (\ref{Ndeg}). The $Sl(2)$ algebras
has already been studied in
\cite{11}, where the decomposition of the fundamental representation was given:
we recall in table \ref{tabf4} this decomposition giving the conformal spin
content.

\sect{Classification of (half-)integral gradings.\label{mach}}

\indent

The decomposition of the adjoint of a simple Lie algebra $\cg$ in terms of
$Sl(2)\oplus U(1)$ representations gives an exhaustive classification of the
different constrained WZW theory arising from a (half-)integral grading.
Moreover, the different values of $Y$ (at fixed $Sl(2)$ subalgebra) leads to
the equivalent theories \cite{14}.
Thus, if we know how to reconstruct the gradation
$H$ from this decomposition, we will be able to give an explicit classification
of gradations. This is the aim of this section.

\subsection{Defining vectors \label{defV}}

\indent

An $Sl(2)$ algebra in a simple Lie algebra $\cg$ is specified \cite{11}
by its defining vector $(f_1,...,f_r)$, itself defined from the relation
\be
M_0=\sum_{i=1}^r f_iH_i \mb{} f_i\mb{rational} \label{eqf}
\ee
where $M_0$ denotes the Cartan part of $Sl(2)$ and $\{H_1,...,H_r\}$ a
Cartan subalgebra of $\cg$. For the A,B,C,D algebras of rank up to 6, a
defining vector for all $Sl(2)$ subalgebras has been explicitely computed in
\cite{LG}, and we will use the same normalization here, up to a global factor
2. Let us precise the Cartan basis for these algebras.

\indent

For $B_n$ ($D_n$) algebras, where the roots are $\pm e_i\pm e_j$ and
$\pm e_i$ ($\pm e_i\pm e_j$)
$i,j=1,...,n$. A basis of simple roots is given by $e_i-e_{i+1}$
($i=1,...,n-1)$ and
$e_n$ ($e_{n-1}+e_n$) for $B_n$ ($D_n$).
The Cartan generators satisfy
\be
{[H_i,E_{e_j}]}=\delta_{ij}E_{e_j}\ \ \forall\ i,j=1,...,n
\ee

\indent

For $C_n$ algebras, the roots are $\pm e_i\pm e_j$ and $\pm 2e_i$
$i,j=1,...,n$, a basis of simple roots being $e_i-e_{i+1}$ $i=1,..n-1$ and
$2e_n$.
The Cartan generators satisfy
\be
{[H_i,E_{2e_j}]}=2\delta_{ij}E_{2e_j}\ \ \forall\ i,j=1,...,n
\ee

\indent

For $A_n$ algebras, the $H_i$ generators (with $i=1,...,n+1$) satisfy
\be
{[H_i,E_{e_j-e_k}]}=(\delta_{ij}-\delta_{ik})E_{e_j-e_k}\ \ \forall\
i,j,k=1,...,n+1 \mb{with} j\neq k
\ee
where the roots are $\pm (e_i-e_j)$ $i,j=1,...,n+1$ and the Cartan generators
take the form $H_i-H_j,\ i\neq j$. Note that here, the definition of
the defining vector is slightly modified, since it has a dimension which is no
longer the rank of $A_n$, but the rank plus one. The tracelessness condition
is then used to recover the rank.

\indent

The defining vectors can always be chosen such that
\be
f_n\leq f_{n-1} \leq\cdots\leq f_1 \mb{if} f=(f_1,f_2,...,f_n) \label{ordre}
\ee
Moreover, for $\cg=B_n,C_n,D_n$, we have
\be
f_i\geq0\ \ \forall i \label{truc}
\ee
For $\cg=A_n$, (\ref{truc}) becomes the tracelessness condition
\be
\sum_{i=1}^{n+1} f_i=0
\ee
When adopting this convention for the defining vector, the gradation is such
that the simple roots introduced above have a positive grade.

\indent

If $f'=(f'_1,f'_2,...,f'_p,0,..,0)$ and
$f''=(f''_1,f''_2,...,f''_q,0,..,0)$ are the defining vectors
associated to the two principal $Sl(2)$'s
of two subalgebras $\cg'$ and $\cg''$ of
$\cg$, and if $\cg'\oplus \cg''$ is also a subalgebra of $\cg$, then a defining
vector of the principal $Sl(2)$ of $\cg'\oplus\cg''$ is $f=(f_1,...,f_{p+q})$,
where the set $\{ f_1,...,f_{p+q}\}$ is the union of the sets
$\{f'_1,...,f'_p\}$ and $\{f''_1,...,f''_q\}$, and can be arranged according to
(\ref{ordre}).

\subsection{Calculation of the defining vector from the $Sl(2)$ decomposition}

\indent

We set for a while $Y=0$, and look at the gradation produced by $M_0$, Cartan
generator of a given $Sl(2)$ subalgebra of $\cg$. This $Sl(2)$ subalgebra can
always be seen as the principal embedding of a (regular or singular) subalgebra
of $\cg$. From the properties mentionned above, we just have to
specify the defining vector associated to simple subalgebras (except the case
$2A_1$ in $SO(n)$, see below). Due to the Cartan basis we have adopted, the
defining vectors are directly related to the eigenvalues of $M_0$ in the
fundamental representation of $\cg$.

\indent

First consider the case $\cg=A_n$. The defining vector components are
just the eigenvalues of $M_0$, since one can always
diagonalize $M_0$ with hermitian matrices. Then, we have the rules:
\bea
A_{2p}\subset A_n &\longrightarrow& f=(p,p-1,...,1,
\underbrace{0,...,0}_{n+1-2p},-1,-2,...,-p) \\
A_{2p+1}\subset A_n &\longrightarrow& f=(p+\sm{1}{2},p-\sm{1}{2},...,\sm{1}{2},
\underbrace{0,...,0}_{n-2p-1},\sm{-1}{2},\sm{-3}{2},...,-p-\sm{1}{2})
\ena
For example, the defining vector of $A_2$ (resp. $A_1$) in $A_4$ is
$(1,0,0,0,-1)$ (resp. $(\sm{1}{2},0,0,0,\sm{-1}{2})$).
The defining vector of $A_2\oplus A_1$ is $(1,\sm{1}{2},0,-\sm{1}{2},-1)$.

\indent

Let us now turn to the $SO(n)$ case. Because of the antisymmetry of the
matrices in the fundamental representation, the Cartan generators cannot be
diagonal. In fact, they are constructed with $\sigma_2$ matrices on the
diagonal. Each $\sigma_2$ matrix possesses $+1$ and $-1$ as eigenvalues, so
that one has only to specify the positive $M_0-$eigenvalues in the defining
vector. The general rules are:
\bea
B_p \mb{or} D_{p+1}\subset B_n &\longrightarrow& f=(p,p-1,...,1,0,...0) \\
D_{p+1}\subset D_n &\longrightarrow& f=(p,p-1,...,1,0,...0) \\
A_{2p} \subset B_n\mb{or}D_n &\longrightarrow&
f=(p,p,p-1,p-1,...,1,1,0,...0) \\
A_{2p+1} \subset B_n\mb{or}D_n &\longrightarrow&
f=(p+\sm{1}{2},p+\sm{1}{2},p-\sm{1}{2},p-\sm{1}{2},...,\sm{1}{2},\sm{1}{2}
,0,...0)
\ena
As there are some exceptional embeddings
of $Sl(2)$ algebras in orthogonal ones, there will be also exceptions for the
defining vectors.
For $A_3\equiv D_3$, they are two different defining vectors, one associated to
"$A_3$", and the other one to "$D_3$":
\bea
"A_3"\subset B_n\mb{or} D_n &\longrightarrow& f=(\sm{3}{2},\sm{3}{2},
\sm{1}{2},\sm{1}{2},0,...0) \\
"D_3"\subset B_n\mb{or} D_n &\longrightarrow& f=(2,1,0,...0)
\ena
They are also two defining vectors for $2A_1\subset SO(m)$
\bea
"2A_1"\subset B_n\mb{or} D_n &\longrightarrow& (\sm{1}{2},\sm{1}{2},\sm{1}{2}
,\sm{1}{2},0,...,0) \\
"2A_1"\subset B_n\mb{or} D_n &\longrightarrow& (1,0,...,0)
\ena
Finally, for the short root of $B_n$, we have
\be
A_1^2\subset B_n \longrightarrow (1,0,...,0)
\ee
The defining vectors associated to the singular embeddings $(B_i\oplus
B_j)\subset D_n$ (with $i+j=n-1, i\neq j$) are computed with the
above rules.

\indent

Finally, we study the case of $Sp(2n)$ algebras. The rules are similar to those
of $SO(n)$ algebras:
\bea
A_1^1 \subset C_n &\longrightarrow& f=(\sm{1}{2},0,...,0) \\
A^2_{2p} \subset C_n &\longrightarrow&
f=(p,p,p-1,p-1,...,1,1,0,...0) \\
A^2_{2p+1} \subset C_n &\longrightarrow&
f=(p+\sm{1}{2},p+\sm{1}{2},p-\sm{1}{2},p-\sm{1}{2},...
,\sm{1}{2},\sm{1}{2},0,...0) \\
C_p \subset C_n &\longrightarrow&
f=(p+\sm{1}{2},p-\sm{1}{2},...,\sm{1}{2},0,...0)
\ena

\subsection{Case of $Sl(2)\oplus U(1)$ decomposition}

\indent

When $H\neq M_0$, we can no longer speak about defining vector for $H$,
since $H$
cannot be embedded in an $Sl(2)$ algebra. However,
it is still possible to compute a vector $f=(f_1,...,f_n)$ that defines $H$,
using the relation (\ref{eqf}).
We give hereafter the rules to compute this vector associated to $H$.

\indent

Let us first look at the $SO(n)$ case, where $Y$ appears, in the fundamental
representation,
only in combinations
$\cd_m(y)\oplus \cd_m(-y)$ with $m$ integer. The rule is then
\bea
&&\cd_m(y)\oplus \cd_m(-y)\ \mbox{in Fund}^{\mbox{l}}\ (m\in\Z_+) \\
&&\longrightarrow\
f=(m+y,m-y,m-1+y,m-1-y,...,1+y,1-y,y,0,...,0) \nonumber
\ena
before ordering following (\ref{ordre}).
For example, for $A_4\subset D_6$, we have
\bea
&&\und{12}=\cd_2(y_2)\oplus\ov{\cd}_2(-y_2) \oplus
\cd_0(y_0)\oplus\ov{\cd}_0(-y_0) \\
&& f=(2+y_2,2-y_2,1+y_2,1-y_2,y_2,y_0)
\nonumber
\ena

\indent

For the case $\cg=A_n$, the defining vector can be read directly in the
fundamental decomposition: the piece corresponding to a representation
$\cd_i(y_i)$ in the fundamental is $(i+y_i,i-1+y_i,...,-i+y_i)$.
Note that the
different eigenvalues $y_i$ are related by a traceless condition:
\be
\sum_i\ m_i(2i+1)y_i=0 \mb{for} \und{n+1}= \bigoplus_i\ m_i\cd_i(y_i)
\ee
They are determined in the adjoint representation, by the usual condition
$|y|\leq j$ and $y\in\half\Z$ for any representation $\cd_j(y)$ in the adjoint.

\indent

For example, for the reduction of $A_2$ with respect to its regular $A_1$
algebra, we have
\bea
&&\und{3}= \cd_\half(y)\oplus \cd_0(-2y) \mb{thus} f=(\sm{1}{2}+y,
-\sm{1}{2}+y,-2y)\\
&&\und{8}= (\cd_1\oplus\cd_0)(0) \oplus \cd_\half(3y) \oplus \cd_\half(-3y)\\
&& |\pm3y|\leq\half \mb{and} \pm3y\in\half\Z \ \ \Rightarrow\ \
y=0,\pm\frac{1}{6}
\ena

\indent

Finally, for the symplectic algebras, the rules are analogous to those of the
$B_n$ case, that is:
\bea
&&\cd_{m+\half}(y)\oplus \cd_{m+\half}(-y)\ \mbox{in Fund}^{\mbox{l}}\
(m\in\Z_+) \\
&&\longrightarrow\
f=(m+\sm{1}{2}+y,m+\sm{1}{2}-y,m-\sm{1}{2}+y,m-\sm{1}{2}-y,...,
\sm{1}{2}+y,\sm{1}{2}-y,0,...,0) \nonumber
\ena

\sect{Poisson brackets of $W$ algebras}

\subsection{Generalities}

\indent

The knowledge of the spin contents of a $W$ algebra with the use of a
$Sl(2)\oplus U_Y(1)$ decomposition, together with the Proposition 4 of section
\ref{zut}, allows us to determine many of the PB of this algebra,
when $Y$ exists. Indeed, let $W_{I}$ be the $W$ generators,
$I\in\ci$ indexing the generators. The theory possesses a grading
operator $H$, and we suppose here that $H\neq M_0$. The spin content associated
to the stress energy tensor $T_H$ is then
given by $s_I=1+j_I+y_I$. It is conserved through the PB, so that starting from
the general form:
\bea
&& \{ W_{I}(x) ,W_{J}(x') \}_{PB} = \sigma_0(I,J)
\prt^{j_I+y_I+j_J+y_J+1}\delta(x-x')+ \nonumber\\
&&\mb{} +\sum_K \sum_{p,q} \sigma_1(I,J,K,p,q)
\left(\prt^p W_{K}(x')\right) \left(\prt^q\delta(x-x')\right)+
\nonumber\\
&& \mb{} +\sum_{K,L} \sum_{p,q,r} \sigma_2(I,J,K,L,p,q,r)
\left(\prt^p W_{K}(x')\right) \left(\prt^r W_{L}(x')\right)
\left(\prt^r\delta(x-x')\right)+ \nonumber\\
&& \mb{} \vdots \nonumber
\ena
where the $\sigma_n(..)$ are coefficients, the conformal invariance
imposes the sums to satisfy the equalities
\bea
&& p,q,K \mb{such that} p+j_K+y_K+q=j_I+y_I+j_J+y_J \nonumber\\
&& p,q,r,K,L \mb{such that} p+j_K+y_K+q+j_L+y_L+r+1=j_I+y_I+j_J+y_J
\nonumber\\
&& \vdots \label{spinH}
\ena

But Proposition 4 ensures that this algebra is the same as the one obtained
from the grading operator\footnote{This can be guessed if one remarks that
the $Sl(2)$ highest weights are the same for $H=M_0$ and $H=M_0+Y$.} $M_0$.
The main
change between these two algebras is the stress energy tensor ($T_H$ or
$T_{M_0}$). Then, the conformal invariance of the PB when the gradation is
given
by $M_0$ imposes:
\bea
&& p+j_K+q=j_I+j_J \nonumber\\
&& p+j_K+q+j_L+r+1=j_I+j_J \label{spinM0}\\
&& \vdots \nonumber
\ena
Gathering (\ref{spinH}) and (\ref{spinM0}) leads to:
\bea
&& p+j_K+q=j_I+j_J \mb{and} y_K=y_I+y_J \nonumber\\
&& p+j_K+q+j_L+r+1=j_I+j_J \mb{and} y_K+y_L=y_I+y_J \nonumber\\
&& \vdots \nonumber
\ena
For each line, the second equality shows that the charge
associated to the $U(1)_Y$ generator is conserved. This
severely limits the number of
allowed fields in the r.h.s. of the PB, since not only the $T_{M_0}$-conformal
spin
(associated to $Sl(2)$) but also the "hypercharge" associated to $Y$ is
conserved. Note that in this context, $T_{M_0}$ has a zero $U(1)_Y$ value.

\indent

Finally, let us add that there may exist several independent Cartan
generators $Y_i$
which can be added to $M_0$ in such a way that $H_i=M_0+Y_i$ is a non
degenerate
gradation, the corresponding $Sl(2)$ subalgebra of which is still
$(M_\pm,M_0)$.
For example, in the decomposition of $SO(8)$ with respect to $Sl(3)$
(see section \ref{SO8}), we have
\bea
&&\und{8}= \cd_1(y_1)\oplus\cd_1(-y_1)\oplus\cd_0(y_0)\oplus\cd_0(-y_0)
\nonumber\\
&& (8\times 8)_A= (\cd_2\oplus\cd_1\oplus2\cd_0)(0)
\oplus\cd_1(2y_1)\oplus\cd_1(-2y_1)\oplus \nonumber\\
&&\mb{} \oplus\cd_1(y_1+y_0)\oplus\cd_1(-(y_1+y_0))
\oplus\cd_1(y_1-y_0)\oplus\cd_1(-(y_1-y_0))
\ena
In the above decomposition of the adjoint representation, one sees that
$y_0$ and $y_1$ can take the values $0,\half$, independently from one another,
without violating the non degeneracy condition.
So, we can decompose $Y$ in $Y_0+Y_1$, $Y_0$ and $Y_1$ being defined by
the vectors
$f_0=(\frac{3}{2},\half,\half,0)$ and $f_1=(1,1,\half,0)$.

\indent

Thus, we will now write the $W$ generators as
\be
W_{j+y+1}\equiv W_{j+1}^{(\vec{y})}
\ee
$j+1$ being the conformal spin in the basis where all the fields (but $T$) are
primary, and $\vec{y}$ being the set of "hypercharges" associated to the
different possible $U(1)_Y$.

\indent

For instance, in the case of $SO(8)$ reduced by $Sl(3)$, we will
have as $W$ generators:
\bea
&& T^{(0,0)}_{M_0} \mb{,} W_1^{Y1} \mb{,} W_1^{Y2} \nonumber\\
&& W_3^{(0,0)} \mb{,} W_2^{(1,0)} \mb{,} W_2^{(-1,0)} \nonumber\\
&& W_2^{(1/2,1/2)} \mb{,} W_2^{(-1/2,-1/2)} \mb{,}
W_2^{(1/2,-1/2)} \mb{,} W_2^{(-1/2,1/2)} \nonumber
\ena
where the doublet superscript indicates the hypercharges of the $W$ generator
with respect to $W_1^{Y1}$ and $W_1^{Y2}$.

\subsection{Use of the stress-energy tensor}

\indent

We know that the theory associated to $H$ contains a stress-energy
tensor $T_H$, and that all the fields but $W_1^Y$ are primary.
Moreover, from the
equation (\ref{eq:2}), it is clear that
\be
T_H=T_{M_0}+\prt W^Y_1 \mb{for} H=M_0+Y
\ee
Then, a generator $W_j^{(y)}$ being primary (we omit $T$ and $W^Y_1$) with
respect to $T_H$ and $T_{M_0}$, we will have
\be
\{ \prt_x W^Y_1(x), W_j^{(y)}(x') \}_{PB} = yW^{(y)}_j(x') \prt_x\delta(x-x')
\ee
Note that although $T_{M_0}$ is not an eigenvector of $W^1_Y$, we associated to
it an "eigenvalue" 0.
Of course, if there are several $U(1)$, each of them will satisfy this
property.

\indent

{\em Thus, the generator $W^Y_1$ associated to $Y=H-M_0$ generates a conserved
"hypercharge", and all the $W$ generators except $T$ are
$W_1^{Y}-$eigenvectors:
\be
\{ W^Y_1(x), W_j^{(y)}(x') \}_{PB} = yW^{(y)}_j(x') \delta(x-x')
\ee
$T$ possesses a zero hypercharge, but the PB reads:
\be
\{ T(x),W^Y_1(x') \}_{PB} = \prt W^Y_1(x') \delta(x-x')
+W^Y_1(x') \prt\delta(x-x')
\ee
}

\indent

Finally, let us remark that the set of spin 1 generators must be closed,
because of the conservation of the conformal spin. This shows that we will have
a KM algebra, corresponding to the part of $\cg$ which has not been used for
the definition of the $Sl(2)$ algebra.

\subsection{Example \label{Wber}}

\indent

As an example, let us look at the $W$ algebra coming from non Abelian Toda on
$Sl(3)$. The $W$ generators are
\be
W_2,\ W_{3/2+y},\ W_{3/2-y},\ W_1 \mb{with} y=0\mb{or}\half
\ee
Applying the above procedure to the PB of this algebra, we can determine their
structure. As a notation, we will write $\prt$ for $\prt_x$ and
$\prt'$ for $\prt_{x'}$.
\bea
\{W_2(x), W_2(x')\}_{PB} &=& \left( a_1\prt' W_2(x')
+a_3\prt'^2 W_1(x') +a_4 \prt'(W_1W_1)(x') \right.\nonumber\\
&&\left. +a_2 W_{3/2+y}W_{3/2-y}(x')
\right)\delta(x-x') \nonumber\\
&& +\left( a_5 W_2(x') +a_6 W_1W_1(x')
+a_7 W_1(x') \right)
\prt\delta(x-x') \nonumber\\
&& +a_8 W_1(x') \prt^2\delta(x-x')
+a_9 \prt^3\delta(x-x') \\
\{W_2(x), W_{3/2\pm y}(x')\}_{PB} &=& a_{10}\prt' W_{3/2\pm y}(x') \delta(x-x')
+a_{11} W_{3/2\pm y}(x') \prt\delta(x-x') \\
\{W_2(x), W_1(x')\}_{PB} &=& \left( a_{12}\prt' W_1(x')
+a_{13}\prt' W_2(x') \right)\delta(x-x') + \nonumber \\
&& +a_{14} W_1(x') \prt\delta(x-x')
+a_{15} \prt^2\delta(x-x') \\
\{W_{3/2\pm y}(x), W_{3/2\pm y}(x')\}_{PB} &=& 0 \\
\{W_{3/2 +y}(x), W_{3/2 -y}(x')\}_{PB} &=& \left( a_{16}\prt' W_1(x') +
a_{17} W_1W_1(x') +a_{18} W_2(x') \right)\delta(x-x')+
\nonumber \\
&& +a_{19} W_1(x') \prt\delta(x-x') +
a_{20} \prt^2\delta(x-x') \\
\{W_1(x), W_{3/2\pm y}(x')\}_{PB} &=& a^\pm_{21} W_{3/2\pm y}(x')
\delta(x-x')\\
\{W_1(x), W_1(x')\}_{PB} &=& a_{22} \prt\delta(x-x')
\ena
Now, assuming that $Y=0$, replacing $W_2$ by $T$ the Virasoro tensor, and
recognizing in $W_1$ the $W_1^Y$ generator,
we are led to the constraints:
\bea
&& a_1=1\ ,\ a_5=2\ ,\
a_2=a_3=a_4=a_6=a_7=a_8=0 \\
&& a_{10}=1\ ,\ a_{11}=\frac{3}{2} \\
&& a_{12}=1\ ,\ a_{14}=1\ ,\ a_{13}=a_{15}=0 \\
&& a_{21}^\pm=\pm1
\ena
Thus, the $W$ algebra associated to the regular $Sl(2)$ in $Sl(3)$ must
satisfy:
\bea
\{T(x), T(x')\}_{PB} &=& \prt' T(x') \delta(x-x')
+2T(x') \prt\delta(x-x') +c \prt^3\delta(x-x') \\
\{T(x), W_{3/2}^\pm(x')\}_{PB} &=& \prt' W_{3/2}^\pm(x') \delta(x-x')
+\frac{3}{2} W_{3/2}^\pm(x') \prt\delta(x-x') \\
\{T(x), W_1(x')\}_{PB} &=& \prt' W_1(x') \delta(x-x') +
W_1(x') \prt\delta(x-x') \\
\{W_{3/2}^\pm(x), W_{3/2}^\pm(x')\}_{PB} &=& 0 \\
\{W_{3/2}^+(x), W_{3/2}^-(x')\}_{PB} &=& \left( a_{16}\prt' W_1(x') +
a_{17} W_1W_1(x') +a_{18} T(x') \right)\delta(x-x')+
\nonumber \\
&& +a_{19} W_1(x') \prt\delta(x-x') +
a_{20} \prt^2\delta(x-x') \\
\{W_1(x), W_{3/2}^\pm(x')\}_{PB} &=& \pm W_{3/2}^\pm(x')
\delta(x-x')\\
\{W_1(x), W_1(x')\}_{PB} &=& k\prt\delta(x-x')
\ena
which has to be compared with the $W$ algebra explicited in \cite{Bersh}.
Note that the Jacobi identities for the PB of the $W$ algebra will also
constrain the remaining structure constants.

\sect{$W$ algebras from Lie algebras of rank up to 4}

\indent

As an application of the above formulation, we present here an exhaustive
classification of $W$ algebras arising from constrained WZW models based on
classical algebras of rank up to 4. For such a purpose, we follow the point of
view
developped in section \ref{class}, using the results presented in sections
\ref{sect3}$-$\ref{mach}. Although the algebras $B_2$ and $C_2$ on the one
hand,
and $A_3$ and $D_3$ on the other hand are isomorphic, we have separately
considered these four algebras to show the differences in the calculations.
The classification is listed in tables \ref{t1}-\ref{tabf4}, where the
decomposition of the fundamental
of $\cg$ with respect to $Sl(2)\oplus U(1)$ is given. We give the minimal (i.e.
the lowest dimensional) regular subalgebras containing the $Sl(2)$, when they
exist. For the singular embedding associated to $D_4$, we mention the
$SO(3)\oplus SO(5)$ subalgebra.
Then, we give the conformal spin content $s=j+1$, with
the convention: $n\ast s$ means that the spin $s$ appears $n$ times.
In the same column, we give under the spin $s$ the hypercharge(s) $y$ when it
exists. Finally, we write the different gradations that lead to this $W$
algebra.

\begin{table}
\begin{tabular}{|l|c|l|l|l|} \hline
$\cg$ & Subalg. & $Sl(2)\oplus U(1)$ decompos.
& Spin contents & Gradation\\
& &(fundamental rep.) & (Hypercharge) & \\ \hline
&&&& \\
$A_1$ & $A_1$ & $\cd_{1/2}$ & 2 & $(\sm{1}{2},\sm{-1}{2})$ \\
&&&& \\
\hline
&&&& \\
$A_2$ & $A_1$ & $\cd_{1/2}(y)\oplus \cd_0(-2y)$ & $2,\sm{3}{2},\sm{3}{2},
1$ & $(\sm{1}{2},0,\sm{-1}{2})$ \\
&& & $(0,3y,-3y,0)$ & $(\sm{2}{3},\sm{-1}{3},\sm{-1}{3})$ \\
&&&& \\
& $A_2$ & $\cd_1$ & $3,2$ & (1,0,-1) \\
&&&& \\
\hline
&&&& \\
$B_2$ & $A_1$
& $2\cd_{1/2} \oplus \cd_0$ & $2,2\po \sm{3}{2},3\po 1$
& $(\sm{1}{2},\sm{1}{2})$
\\
&&&& \\
& $\left.\begin{array}{c} A^2_1 \\ 2A_1 \end{array} \right\}$
& $\cd_1\oplus \cd_0(y)\oplus \cd_0(-y)$
&2,2,2,1
& $(1,0)$\\
&&& $(0,y,-y,0)$ & $(1,\sm{1}{2})$ \\
&&& & $(1,1)$\\
&&&& \\
&$B_2$& $\cd_2 $
& $4,2$ & $(2,1)$\\
&&&& \\
\hline
&&&& \\
$C_2$ & $A_1$
& $\cd_{1/2} \oplus 2\cd_0$ & $2,2\po \sm{3}{2},3\po 1$ & $(\sm{1}{2},0)$
\\
&&&& \\
& $\left.\begin{array}{c} 2A_1 \\ A^2_1 \end{array} \right\}$
& $\cd_{1/2}(y)\oplus \cd_{1/2}(-y) $&
2,2,2,1
& $(\sm{1}{2},\sm{1}{2})$\\
&&& $(0,2y,-2y,0)$ & $(\sm{3}{4},\sm{1}{4})$\\
&&& & $(1,0)$\\
&&&& \\
&$C_2$& $\cd_{3/2} $
& $4,2$ & $(\sm{3}{2},\sm{1}{2})$\\
&&&& \\
\hline
\end{tabular}
\caption{$W$ algebras for Lie algebras of rank 1 and 2\label{t1}.}
\end{table}

\clearpage

\begin{table}
\begin{tabular}{|l|c|l|l|l|} \hline
$\cg$ & Subalg. & $Sl(2)\oplus U(1)$ decompos.
& Spin contents & Gradation\\
& &(fundamental rep.) & (Hypercharge) & \\ \hline
&&&& \\
$A_3$ & $A_1$ & $\cd_{1/2}(y)\oplus 2\cd_0(-y)$ &
$2,4\po \sm{3}{2},4\po 1$ & $(\sm{1}{2},0,0,\sm{-1}{2})$ \\
&& & $(0,2y,2y,-2y,-2y,4\po 0)$
& $(\sm{3}{4},\sm{-1}{4},\sm{-1}{4},\sm{-1}{4})$ \\
&&&& \\
& $2A_1$ & $2\cd_{1/2}$ & $4\po 2,3\po 1$ & $(\sm{1}{2},\sm{1}{2}
,\sm{-1}{2},\sm{-1}{2})$ \\
&&&& \\
& $A_2$ & $\cd_1(y)\oplus \cd_0(-3y)$ & $3,2,2,2,1$ & (1,0,0,-1) \\
&& &$(0,2y,-2y,0,0)$ & $(\sm{5}{4},\sm{1}{4},\sm{-3}{4},\sm{-3}{4})$ \\
&& && $(\sm{9}{8},\sm{1}{8},\sm{-3}{8},\sm{-7}{8})$ \\
&&&& \\
& $A_3$ & $\cd_{3/2}$ & $4,3,2$ & $(\sm{3}{2},\sm{1}{2}
,\sm{-1}{2},\sm{-3}{2})$\\
&&&& \\ \hline
&&&& \\
$D_3$ & $A_1$ & $2\cd_{1/2}\oplus\cd_0(y)\oplus \cd_0(-y)$ &
$2,4\po \sm{3}{2},4\po 1$ & $(\sm{1}{2},\sm{1}{2},0)$ \\
&& & $(0,y,y,-y,-y,4\po 0)$
& $(\sm{1}{2},\sm{1}{2},\sm{1}{2})$ \\
&&&& \\
& $2A_1$ & $\cd_1\oplus3\cd_0$ & $4\po 2,3\po 1$ &
$(1,0,0)$ \\
&&&& \\
& $A_2$ & $\cd_1(y)\oplus \cd_1(-y)$ & $3,2,2,2,1$ & (1,1,0) \\
&& &$(0,2y,-2y,0,0)$ & $(\sm{5}{4},\sm{3}{4},\sm{1}{4})$ \\
&& && $(\sm{3}{2},\sm{1}{2},\sm{1}{2})$ \\
&&&& \\
& $D_3$ & $\cd_2\oplus\cd_0$ & $4,3,2$ & $(2,1,0)$\\
&&&& \\ \hline
\end{tabular}
\caption{$W$ algebras for $A_3\equiv D_3$\label{t2}.}
\end{table}

\clearpage

\begin{table}
\begin{tabular}{|l|c|l|l|l|} \hline
$\cg$ & Subalg. & $Sl(2)\oplus U(1)$ decompos.
& Spin contents & Gradation\\
& &(fundamental rep.) & (Hypercharge) & \\ \hline
&&&& \\
$B_3$ &$A_1$
& $2\cd_{1/2} \oplus 3 \cd_0$ & $2,6\po \sm{3}{2},6\po 1$ &
$(\sm{1}{2},\sm{1}{2},0)$ \\
&&&& \\
& $\left.\begin{array}{c} A_1^2 \\ 2A_1 \end{array}\right\}$
&$\cd_1\oplus 4\cd_0$ & $5\po 2,6\po 1$ & $(1,0,0)$ \\
&&&& \\
&$A_1\oplus A_1^2$ & $\cd_1\oplus2\cd_{1/2}$ &
$\sm{5}{2},\sm{5}{2},2,2,\sm{3}{2},\sm{3}{2},1,1,1$ &
$(1,\sm{1}{2},\sm{1}{2})$ \\
&&&& \\
& $\left.\begin{array}{c} A_2 \\ 2A_1 \oplus A^2_1 \end{array}\right\}$
& $\cd_1(y) \oplus\cd_1(-y) \oplus\cd_0(0)$
& $3,5\po 2,1$ & $(1,1,0)$ \\
& & &$(0,2y,y,-y,-2y,0,0)$ & $(\sm{3}{2},\sm{1}{2},\sm{1}{2})$ \\
&&&& \\
&$\left.\begin{array}{c} A_3 \\ B_2 \end{array}\right\}$
& $\cd_2(0)\oplus \cd_0(y) \oplus \cd_0 (-y)$ & $4,3,3,2,1$ &
$(2,1,0)$ \\
&&&$(0,y,-y,0,0)$ & $(2,1,\sm{1}{2})$ \\
&&& & $(2,1,1)$ \\
&&& & $(2,\sm{3}{2},1)$ \\
&&& & $(2,2,1)$ \\
&&&& \\
&$B_3$ & $ \cd_3$ & $6,4,2$ & $(3,2,1)$\\
&&&& \\
\hline
&&&& \\
$C_3$ & $A_1$ & $\cd_{1/2}\oplus 4\cd_0$ &
$2,4\po \sm{3}{2},10\po 1$ & $(\sm{1}{2},0,0)$ \\
&&&& \\
& $\left.\begin{array}{c} A_1^2 \\ 2A_1 \end{array}\right\}$
& $\cd_{1/2}(y)\oplus \cd_{1/2}(-y)\oplus 2\cd_0(0)$
& $3\po 2,4\po \sm{3}{2},4\po 1$ & $(\sm{1}{2},\sm{1}{2},0)$ \\
&& &$(0,2y,-2y,2\po y,2\po (-y),4\po 0)$ & $(1,0,0)$ \\
&&&& \\
& $A_2^2$ & $2\cd_1$ & $3\po 3, 2,3\po 1$ &
$(1,1,0)$ \\
&&&& \\
& $C_2$ & $\cd_{3/2}\oplus 2\cd_0$ & $4,\sm{5}{2},\sm{5}{2},2,3\po 1$
& $(\sm{3}{2},\sm{1}{2},0)$ \\
&&&& \\
& $\left.\begin{array}{c} A_1\oplus A_1^2 \\ 3A_1 \end{array}\right\}$
& $3\cd_{1/2}$ & $6\po 2,3\po 1$
& $(\sm{1}{2},\sm{1}{2},\sm{1}{2})$ \\
&&&& \\
& $C_2\oplus A_1$ & $\cd_{3/2}\oplus \cd_{1/2}$ & $4,3,3\po 2$
& $(\sm{3}{2},\sm{1}{2},\sm{1}{2})$ \\
&&&& \\
& $C_3$ & $\cd_{5/2}$ & $6,4,2$ & $(\sm{5}{2},\sm{3}{2},\sm{1}{2})$ \\
&&&& \\
\hline
\end{tabular}
\caption{$W$ algebras for $B_3$ and $C_3$\label{t2bis}.}
\end{table}

\clearpage

\begin{table}
\begin{tabular}{|c|l|l|l|} \hline
Subalg. & $Sl(2)\oplus U(1)$ decompos.
& Spin contents & Gradation\\
&(fundamental rep.) & (Hypercharge) & \\ \hline
&&& \\
$A_1$
& $\cd_{1/2}(y) \oplus 3\cd_0(\sm{-2$y$}{3})$
& $2,6\po \sm{3}{2},9\po 1$ &
$(\sm{1}{2},0,0,0,\sm{-1}{2})$ \\
&& $(0,3\po \sm{5$y$}{3},3\po \sm{-5$y$}{3},9\po 0)$ &
$(\sm{4}{5},\sm{-1}{5},\sm{-1}{5},\sm{-1}{5},\sm{-1}{5})$ \\
&&& \\
$2A_1$ &$2\cd_{1/2}(y)\oplus\cd_0(-4y)$
& $4\po 2,4\po \sm{3}{2},4\po 1$
& $(\sm{1}{2},\sm{1}{2},0,\sm{-1}{2},\sm{-1}{2})$ \\
&& $(4\po0,2\po 5y,2\po (-5y),4\po 0)$ &
$(\sm{3}{5},\sm{3}{5},\sm{-2}{5},\sm{-2}{5},\sm{-2}{5})$ \\
&&& \\
$ A_2$ & $\cd_{1}(y)\oplus 2\cd_{0}(\sm{-3$y$}{2})$
& $3,5\po 2,4\po 1$ & $(1,0,0,0,-1)$ \\
&& $(0,2\po \sm{5$y$}{2},0,2\po \sm{-5$y$}{2},4\po 0)$ &
$(\sm{6}{5},\sm{1}{5},\sm{-3}{10},\sm{-3}{10},\sm{-4}{5})$ \\
&&& $(\sm{7}{5},\sm{2}{5},\sm{-3}{5},\sm{-3}{5},\sm{-3}{5})$ \\
&&& \\
$A_2\oplus A_1$ & $\cd_1(y)\oplus \cd_{1/2}(\sm{-3$y$}{2})$
& $3,2\po \sm{5}{2},2\po 2,2\po \sm{3}{2},1$ &
$(1,\sm{1}{2},0,\sm{-1}{2},-1)$ \\
&& $(0,\sm{5$y$}{3},\sm{-5$y$}{3},0,0,\sm{5$y$}{3},\sm{-5$y$}{3},0)$ &
$(\sm{13}{20},\sm{1}{20},\sm{3}{10},\sm{-19}{20},\sm{-7}{10})$ \\
&&&\\
$A_3$ & $\cd_{3/2}(y)\oplus \cd_0(-4y)$ & $4,3,2\po \sm{5}{2},2,1$ &
$(\sm{3}{2},\sm{1}{2},0,\sm{-1}{2},\sm{-3}{2})$ \\
&& $(0,0,5y,-5y,2\po 0)$ &
$(\sm{9}{5},\sm{4}{5},\sm{-1}{5},\sm{-6}{5},\sm{-6}{5})$ \\
&&& $(\sm{8}{5},\sm{3}{5},\sm{-2}{5},\sm{-2}{5},\sm{-7}{5})$ \\
&&& $(\sm{17}{10},\sm{7}{10},\sm{-3}{10},\sm{-4}{5},\sm{-13}{10})$ \\
&&&\\
$A_4$ & $\cd_2$ & $5,4,3,2$ & $(2,1,0,-1,-2)$ \\
&&& \\ \hline
\end{tabular}
\caption{$W$ algebras for $A_4$\label{t6}.}
\end{table}

\clearpage

\begin{table}
\begin{tabular}{|c|l|l|l|} \hline
Subalg. & $Sl(2)\oplus U(1)$ decompos.
& Spin contents & Gradation\\
&(fundamental rep.) & (Hypercharge) & \\ \hline
&&& \\
$A_1$
& $2\cd_{1/2} \oplus 5\cd_0$ & $2,10\po \sm{3}{2},13\po 1$ &
$(\sm{1}{2},\sm{1}{2},0,0)$ \\
&&& \\
$\left.\begin{array}{c} A_1^2 \\ 2A_1 \end{array}\right\}$
&$\cd_1\oplus 6\cd_0$
& $7\po 2,15\po 1$ & $(1,0,0,0)$ \\
&&& \\
$(2A_1)'$ &$4\cd_{1/2}\oplus \cd_0$
& $6\po 2,4\po \sm{3}{2},10\po 1$ &
$(\sm{1}{2},\sm{1}{2},\sm{1}{2},\sm{1}{2})$ \\
&&& \\
$\left.\begin{array}{c} A_1\oplus A_1^2 \\ 3A_1 \end{array}\right\}$
& $\cd_1\oplus 2\cd_{1/2}\oplus \cd_0(y)\oplus \cd_0(-y)$
& $\sm{5}{2},\sm{5}{2},4\po 2,6\po \sm{3}{2},6\po 1$&
$(1,\sm{1}{2},\sm{1}{2},0)$ \\
&& $(0,0,y,-y,0,0,y,y,-y,-y,4\po 0)$ & $(1,\sm{1}{2},\sm{1}{2},\sm{1}{2})$ \\
&&& \\
$\left.\begin{array}{c} A_2 \\ 4A_1 \\ 2A_1 \oplus A_1^2 \end{array}\right\}$
& $\cd_1(y)\oplus \cd_1(-y)\oplus 3\cd_0(0)$
& $3,9\po 2,4\po 1$&
$(1,1,0,0)$ \\
&& $(0,3\po y,3\po(-y),2y,-2y,5\po 0)$
& $(\sm{3}{2},\sm{1}{2},\sm{1}{2},0)$ \\
&&& \\
$ A_2\oplus A^2_1$
& $3\cd_{1}$ & $3\po 3,6\po 2,3\po 1$ & $(1,1,1,0)$ \\
&&& \\
$A_3$& $2\cd_{3/2}\oplus \cd_0$
& $4,3\po 3,2\po \sm{5}{2},2,3\po 1$ &
$(\sm{3}{2},\sm{3}{2},\sm{1}{2},\sm{1}{2})$ \\
&&&\\
$\left.\begin{array}{c} B_2 \\ A_3 \end{array}\right\}$
& $\cd_2\oplus 4\cd_0$ & $4,4\po 3,2,6\po 1$ & $(2,1,0,0)$ \\
&&&\\
$B_2\oplus A_1$ & $\cd_2\oplus 2\cd_{1/2}$
& $4,2\po \sm{7}{2},2\po \sm{5}{2},2,2,3\po 1$ &
$(2,1,\sm{1}{2},\sm{1}{2})$ \\
&&&\\
$\left.\begin{array}{c} B_2\oplus 2A_1 \\ A_3 \oplus A_1^2 \end{array}\right\}$
&$\cd_{2}\oplus \cd_{1} \oplus \cd_0$ & $4,4,3,3,4\po 2$ &
$(2,1,1,0)$ \\
&&&\\
$\left.\begin{array}{c} B_3 \\ D_4 \end{array}\right\}$
& $\cd_{3}(0)\oplus \cd_0(y) \oplus \cd_0(-y)$
& $6,3\po 4,2,1$ & $(3,2,1,0)$ \\
&&$(0,y,-y,3\po 0)$ & $(3,2,1,y)$ \\
&&& \\
$B_4$ & $\cd_4$ & $8,6,4,2$ & $(4,3,2,1)$ \\
&&& \\ \hline
\end{tabular}
\caption{$W$ algebras for $B_4$\label{t4}.}
\end{table}

\clearpage

\begin{table}
\begin{tabular}{|c|l|l|l|} \hline
Subalg. & $Sl(2)\oplus U(1)$ decompos.
& Spin contents & Gradation\\
 &(fundamental rep.) & (Hypercharge) & \\ \hline
&&& \\
$A_1$
& $\cd_{1/2} \oplus 6\cd_0$ & $2,6\po \sm{3}{2},21\po 1$ &
$(\sm{1}{2},0,0,0)$ \\
&&& \\
$\left.\begin{array}{c} A_1^2 \\ 2A_1 \end{array}\right\}$
&$\cd_{1/2}(y)\oplus \cd_{1/2}(-y)\oplus 4\cd_0(0)$
& $3\po 2,8\po \sm{3}{2},11\po 1$ & $(\sm{1}{2},\sm{1}{2},0,0)$ \\
&& $(0,2y,-2y,4\po y,4\po (-y),11\po 0)$ & $(1,0,0,0)$ \\
&&& \\
$\left.\begin{array}{c} A_1\oplus A_1^2 \\ 3A_1 \end{array}\right\}$
& $3\cd_{1/2}\oplus 2\cd_0$ & $6\po 2,6\po \sm{3}{2},6\po 1$&
$(\sm{1}{2},\sm{1}{2},\sm{1}{2},0)$ \\
&&& \\
$\left.\begin{array}{c} 2A^2_1 \\ 4A_1 \\ 2A_1 \oplus A_1^2
\end{array}\right\}$
& $4\cd_{1/2}$ & $10\po 2,6\po 1$
& $(\sm{1}{2},\sm{1}{2},\sm{1}{2},\sm{1}{2})$ \\
&&& \\
$A^2_2$& $2\cd_1\oplus 2\cd_0$ & $3,3,3,5\po 2,6\po 1$ &
$(1,1,0,0)$ \\
&&&\\
$A^2_2\oplus A_1$& $2\cd_1\oplus \cd_{1/2}$
& $3\po 3,2\po \sm{5}{2},2\po 2, 2\po \sm{3}{2},3\po 1$ &
$(1,1,\sm{1}{2},0)$ \\
&&&\\
$C_2$& $\cd_{3/2}\oplus 4\cd_0$ & $4,4\po \sm{5}{2},2,10\po 1$ &
$(\sm{3}{2},\sm{1}{2},0,0)$ \\
&&&\\
$C_2\oplus A_1$& $\cd_{3/2}\oplus \cd_{1/2} \oplus 2\cd_0$
& $4,3,2\po \sm{5}{2},3\po 2, 2\po \sm{3}{2},3\po 1$ &
$(\sm{3}{2},\sm{1}{2},\sm{1}{2},0)$ \\
&&&\\
$\left.\begin{array}{c} C_2\oplus A^2_1 \\ C_2 \oplus 2A_1 \end{array}\right\}$
& $\cd_{3/2}(0)\oplus \cd_{1/2}(y) \oplus \cd_{1/2}(-y)$
& $4,2\po 3,6\po 2,1$ & $(\sm{3}{2},\sm{1}{2},\sm{1}{2},\sm{1}{2})$ \\
&&$(0,y,-y,2y,-2y,y,-y,3\po 0)$ & $(\sm{3}{2},1,\sm{1}{2},0)$ \\
&&& \\
$\left.\begin{array}{c} A_3^2 \\ 2C_2 \end{array}\right\}$
& $\cd_{3/2}(y)\oplus \cd_{3/2}(-y)$ & $3\po 4,3,3\po 2,1$
& $(\sm{3}{2},\sm{3}{2},\sm{1}{2},\sm{1}{2})$ \\
&&$(0,2y,-2y,0,2y,-2y,2\po 0)$ & $(2,1,1,0)$ \\
&&& \\
$C_3$ & $ \cd_{5/2}\oplus 2\cd_0$ & $6,4,\sm{7}{2},\sm{7}{2},2,3\po 1$
& $(\sm{5}{2},\sm{3}{2},\sm{1}{2},0)$ \\
&&& \\
$C_3\oplus A_1$ & $\cd_{5/2}\oplus \cd_{1/2}$ &
$6,4,4,3,2,2$ & $(\sm{5}{2},\sm{3}{2},\sm{1}{2},\sm{1}{2})$ \\
&&& \\
$C_4$ & $\cd_{7/2}$ & $8,6,4,2$ &
$(\sm{7}{2},\sm{5}{2},\sm{3}{2},\sm{1}{2})$ \\
&&& \\ \hline
\end{tabular}
\caption{$W$ algebras for $C_4$\label{t3}.}
\end{table}

\clearpage

\begin{table}[p]
\begin{tabular}{|c|l|l|l|} \hline
Subalg. & $Sl(2)\oplus U(1)$ decompos.
& Spin contents & Gradation\\
&(fundamental rep.) & (Hypercharge) & \\ \hline
&&& \\
$A_1$
& $2\cd_{1/2} \oplus 4\cd_0$ & $2,8\po \sm{3}{2},9\po 1$ &
$(\sm{1}{2},\sm{1}{2},0,0)$ \\
&&& \\
$2A_1$ &$\cd_1\oplus 5\cd_0$
& $6\po 2,10\po 1$ & $(1,0,0,0)$ \\
&&& \\
$(2A_1)'$ &$4\cd_{1/2}$ & $6\po 2,10\po 1$
& $(\sm{1}{2},\sm{1}{2},\sm{1}{2},\sm{1}{2})$ \\
&&& \\
$3A_1$ & $\cd_1\oplus 2\cd_{1/2}\oplus \cd_0$
& $\sm{5}{2},\sm{5}{2},3\po 2,4\po \sm{3}{2},3\po 1$&
$(1,\sm{1}{2},\sm{1}{2},0)$ \\
&&& \\
$\left.\begin{array}{c} A_2 \\ 4A_1 \end{array}\right\}$
& $\cd_{1}(y_1)\oplus \cd_{1}(-y_1)\oplus $
& $3,7\po 2,2\po 1$ & $(1,1,0,0)$ \\
& $\oplus \cd_{0}(y_0)\oplus \cd_{0}(-y_0)$
& $(0,\pm y_1\pm y_0,\pm2y_1,3\po0)$ & $(\sm{3}{2},\sm{1}{2},\sm{1}{2},0)$ \\
&&& $(2,1,0,0)$ \\
&&& $(1,1,\sm{1}{2},0)$ \\
&&& $(1,1,1,0)$ \\
&&& $(\sm{3}{2},\sm{1}{2},\sm{1}{2},\sm{1}{2})$ \\
&&& \\
$A_3$& $2\cd_{3/2}$
& $4,3\po 3,2,3\po 1$ &
$(\sm{3}{2},\sm{3}{2},\sm{1}{2},\sm{1}{2})$ \\
&&&\\
$D_3$& $\cd_2 \oplus 3\cd_0$
& $4,3\po 3,2,3\po 1$ &
$(2,1,0,0)$ \\
&&&\\
$B_2\oplus B_1$& $\cd_2\oplus \cd_1$
& $4,4,3,3\po 2$ & $(2,1,1,0)$ \\
&&&\\
$D_4$ & $\cd_3\oplus \cd_0$ & $6,4,4,2$ & $(3,2,1,0)$ \\
&&& \\ \hline
\end{tabular}
\caption{$W$ algebras for $D_4$\label{t5}.}
\end{table}

\begin{table}[p]
\begin{tabular}{|c|l|l|l|} \hline
Minimal including & $Sl(2)$ decomposition
& Spin contents & Defining \\
Regular Subalgebra &(fundamental rep.) & & Vector \\ \hline
&&& \\
$A_1$
& $2\cd_{1/2} \oplus 3\cd_0$
& $2,4\po\sm{3}{2},1,1,1$ & $(\sm{1}{2},\sm{1}{2},0)$ \\
&&& \\
$A_1^2$ & $\cd_1\oplus2\cd_{1/2}$ & $\sm{5}{2},\sm{5}{2},2,1,1,1$
& $(1,\sm{1}{2},0)$\\
&&& \\
$A_1\oplus A_1^2$ & $2\cd_1 \oplus\cd_0$
& $3,2,2,2$ & $(1,0,0)$\\
&&& \\
$G_2$ &$\cd_3$ & $6,2$ & $(2,\sm{3}{2},\sm{1}{2})$\\
&&& \\
\hline
\end{tabular}
\caption{ Classification for $G_2$. \label{t0}}
\end{table}

\clearpage

\begin{table}[p]
\begin{tabular}{|c|l|l|} \hline
Minimal including & $Sl(2)$ decomposition
& Spin contents \\
Regular Subalgebra &(fundamental rep.) & \\ \hline
&& \\
$A_1$
& $6\cd_{1/2}\oplus14\cd_0$
& $2,14\po\sm{3}{2},21\po1$ \\
&& \\
$\left.\begin{array}{c} A_1^2 \\ 2A_1 \end{array}\right\}$
& $\cd_1\oplus8\cd_{1/2} \oplus 7\cd_0$
& $7\po2,10\po\sm{3}{2},15\po1$ \\
&& \\
$\left.\begin{array}{c} A_1\oplus A_1^2 \\ 3A_1 \end{array} \right\}$
& $3\cd_1\oplus6\cd_{1/2} \oplus5\cd_0$
& $2\po\sm{5}{2},6\po2,10\po\sm{3}{2},6\po1$ \\
&& \\
$\left.\begin{array}{c} 4A_1 \\ 2A_1\oplus A_1^2 \\ A_2 \end{array}\right\}$
& $6\cd_1\oplus 8\cd_0$ & $3,13\po2,8\po1$ \\
&& \\
$A^2_2$ & $\cd_2\oplus 7\cd_1$ & $7\po3,2,14\po1$ \\
&& \\
$A_2\oplus A^2_1$
& $\cd_2\oplus2\cd_{3/2}\oplus3\cd_1\oplus2\cd_{1/2}$
& $2\po4,3\po3,6\po2,2\po\sm{3}{2},1$ \\
&& \\
$A_1\oplus A_2^2$ & $2\cd_{3/2}\oplus3\cd_1\oplus4\cd_{1/2}\oplus\cd_0$
& $3\po3,2\po\sm{5}{2},6\po2,4\po\sm{3}{2},3\po1$ \\
&& \\
$\left.\begin{array}{c} A^2_2\oplus A_2 \\ A_3\oplus A_1^2 \\ B_2\oplus A_1^2
\\ B_2\oplus 2A_1 \end{array}\right\}$ & $3\cd_2\oplus3\cd_1\oplus 2\cd_0$
& $2\po4,4\po3,6\po2$ \\
&& \\
$\left.\begin{array}{c} B_2 \\ A_3 \end{array}\right\}$
& $\cd_2\oplus 4\cd_{3/2}\oplus5\cd_0$ & $4,4\po3,4\po\sm{5}{2},2,6\po1$ \\
&& \\
$B_2\oplus A_1$ &
$2\cd_2\oplus2\cd_{3/2}\oplus \cd_1\oplus2\cd_{1/2}\oplus\cd_0$
& $4,2\po\sm{7}{2},3,4\po\sm{5}{2},3\po2,3\po1$ \\
&& \\
$\left.\begin{array}{c} B_3 \\ D_4 \end{array}\right\}$
& $3\cd_3\oplus5\cd_0$ &
$6,5\po4,2,3\po1$ \\
&& \\
$B_4$ & $\cd_5\oplus\cd_4\oplus\cd_2 \oplus\cd_0$ &
$8,2\po6,4,3,2$ \\
&& \\
$C_3$ & $\cd_4\oplus2\cd_{5/2}\oplus\cd_2$ &
$6,2\po\sm{11}{2},4,2\po\sm{5}{2},2,3\po1$ \\
&& \\
$C_3\oplus A_1$ & $\cd_4\oplus\cd_3\oplus2\cd_2$ &
$2\po6,5,4,3,3\po2$ \\
&& \\
$F_4$ & $\cd_8\oplus\cd_4$ & $12,8,6,2$ \\
&& \\
\hline
\end{tabular}
\caption{ Classification for $F_4$. \label{tabf4}}
\end{table}

\clearpage

\part{Super $W$ algebras built on Lie superalgebras \label{part2}}

\sect{The $OSp(1|2)$ subsuperalgebras of simple Lie superalgebras}

\indent

The determination of the different $OSp(1|2)$ subalgebras in a simple Lie
superalgebra $\cg = \cg_B \oplus \cg_F$ is greatly simplified by the two
following remarks:

\noindent
1) The $Sl(2)$ part of $OSp(1|2)$ is in the (semi)simple bosonic part of the
considered superalgebra. The knowledge of a method to classify the $Sl(2)$
subalgebras of a simple Lie algebra can be obviously generalized to the case
of a direct sum of two (or three, cf. $D(2,1;\alpha)$) simple algebras.

\noindent
2) Any representation of $OSp(1|2)$ is completely irreducible, and any
irreducible $OSp(1|2)$ representation $\crr_j$ ($j$ integer or
half-integer) is the direct sum of two
$Sl(2)$ representations $\cd_j \oplus \cd_{j-1/2}$ with an exception for
the trivial one-dimensional representation $\crr_0 = \cd_0$.
{}From the reduction of the fundamental representation of $\cg$ into $Sl(2)$
ones, it is therefore easy to verify whether the $Sl(2)$ under consideration
can be embedded into an $OSp(1|2)$ superalgebra.

\indent

Now, in the same way that the $Sl(2)$ subalgebras of a simple Lie algebra
$\cg$ are principal subalgebras of the $\cg$ regular subalgebras (up to
exceptions arising in the $D_n$ case, see Section \ref{sect3}), it is rather
clear that the $OSp(1|2)$ subsuperalgebras of a simple Lie superalgebra
$\cg$ are superprincipal in the $\cg$ regular subsuperalgebras (up to
exceptions arising in the $D(m,n)$ case). One recalls that the definition of
a regular subsuperalgebra (SSA)
is a direct generalization of that of an algebra
(see \ref{luc4}), and such SSA can be obtained from the
extended Dynkin diagrams for superalgebras, as for simple algebras
\cite{FSS}. Of course, since several Dynkin diagrams can be in general
associated to the same superalgebra, one has to apply the method to each
allowed Dynkin diagram specifying the superalgebra.
A SSA of $\cg$ which is not regular is called singular.
An example of singular SSA of $\cg$ is the superprincipal
$OSp(1|2)$, when it exists. It is defined as
\bea
&& F_+ = \sum_{\alpha \in \Delta} E_{\alpha} \mbox{\ \ \ and \ \ \ }
F_- = \sum_{\alpha \in \Delta} E_{-\alpha} \\
&& E_+ = \{F_+,F_+\} \mbox{\ \ \ , \ \ \ } E_- = \{F_-,F_-\}
\mbox{\ \ \ and \ \ \ } H = \{F_+,F_-\}
\ena
where $\Delta$ is a simple root system of $\cg$.

\noindent
Not all the simple Lie superalgebras admit a superprincipal embedding.
Actually, it is clear from the expression of the $OSp(1|2)$ generators, that
a superprincipal embedding can be defined only if the superalgebra under
consideration has a completely fermionic simple root system $\Delta$ (which
corresponds to a Dynkin diagram with only grey or/and black dots). Notice that
this condition is necessary but not sufficient (the superalgebra $PSl(n|n)$
does not admit a superprincipal embedding although it has a completely
fermionic simple root system). The simple superalgebras admitting
a superprincipal $OSp(1|2)$ are the following:
$Sl(n+1|n)$, $Sl(n|n+1)$, $OSp(2n\pm 1|2n)$, $OSp(2n|2n)$, $OSp(2n+2|2n)$
with $n \ge 1$ and $D(2,1;\alpha)$ with $\alpha \ne 0, \pm 1$.

\indent

Finally, the method for classifying the $OSp(1|2)$ SSAs in a simple
Lie superalgebra $\cg$ can be summarized as follows:

\indent

{\em
Any $OSp(1|2)$ SSA in a simple Lie superalgebra $\cg$ can be
considered as the superprincipal $OSp(1|2)$ SSA of a regular
SSA $\wt{\cg}$ of $\cg$, up to the following exceptions:

i) For $\cg = OSp(2n \pm 2|2n)$ with $n \ge 2$, besides the superprincipal
$OSp(1|2)$ SSAs described above, there exist $OSp(1|2)$
SSAs associated to the singular embeddings
$OSp(2k \pm 1|2k) \oplus OSp(2n-2k \pm 1|2n-2k)$
with $1 \le k \le n-1$.

ii) For $\cg = OSp(2n|2n)$ with $n \ge 2$, besides the $OSp(1|2)$
superprincipal
embedding, there exist $OSp(1|2)$
SSAs associated to the singular embeddings
$OSp(2k \pm 1|2k) \oplus OSp(2n-2k \mp 1|2n-2k) \subset OSp(2n|2n)$
with $1 \le k \le n-1$.}

\sect{$OSp(1|2)$ decompositions of simple Lie superalgebras}

\indent

Following the general method explained above, once the possible
$OSp(1|2)$ embeddings are determined in the simple Lie superalgebra $\cg$, one
has to reduce the adjoint representation of $\cg$ into $OSp(1|2)$
supermultiplets. Consider an $OSp(1|2)$
SSA of $\cg$, and let $\wt{\cg}$ be the minimal including
regular SSA of $\cg$ having this $OSp(1|2)$ as superprincipal
embedding.
We will show on the example of $Sl(m|n)$ how to obtain
the decomposition of a simple
Lie superalgebra starting from the decompositions of its bosonic and fermionic
parts with respect to the bosonic $Sl(2)$ subalgebra of the
$OSp(1|2)$ under consideration. Moreover, we will see that such a
decomposition can be obtained in a systematic way from the decomposition of the
fundamental representation of the superalgebra with respect to the $OSp(1|2)$.

\subsection{The unitary superalgebras $Sl(m|n)$}

\indent

The bosonic part of $\cg = Sl(m|n)$ with $m \ne n$
is $\cg_B = Sl(m) \oplus Sl(n) \oplus U(1)$ and the fermionic
part $\cg_F$ is the $(\und{m},\ov{n}) \oplus (\ov{m},\und{n})$ representation
of $Sl(m)\oplus Sl(n)$. The regular SSAs of $Sl(m|n)$ which admit a
superprincipal embedding are of the $Sl(p+1|p)$ or $Sl(p|p+1)$ type.

\indent

Consider an $OSp(1|2)$ SSA of $\cg$ such that the minimal including
regular SSA in $\cg$ is $\wt{\cg} = Sl(p+1|p)$ with $p \le
\inf(m-1,n)$. Under $Sl(2)$ (of $OSp(1|2)$),
the representations $\und{m}$ and $\und{n}$
of $Sl(m)$ and $Sl(n)$ decompose as
\bea
&&\und{m} = \cd_{p/2} \oplus (m-p-1) \cd_0 \nonumber \\
&&\und{n} = \cd_{(p-1)/2} \oplus (n-p) \cd_0
\label{luc1}
\ena
Therefore the fermionic part $\cg_F$ reduces to
\bea
(\und{m},\ov{n}) \oplus (\ov{m},\und{n})
&=& 2 \left(\cd_{p/2} \oplus (m-p-1) \cd_0 \right) \times \left(\cd_{(p-1)/2}
\oplus (n-p) \cd_0 \right) \nonumber \\
&=& 2\cd_{p-1/2} \oplus 2\cd_{p-3/2} \oplus ... \oplus 2\cd_{1/2} \oplus
2(m-p-1) \cd_{(p-1)/2} \nonumber \\
&&\oplus 2(n-p) \cd_{p/2} \oplus 2(m-p-1)(n-p) \cd_0
\ena
The bosonic part $\cg_B$ is decomposed as
\bea
\cg_B &=& Sl(m) \oplus Sl(n) \oplus U(1) \nonumber \\
&=& \left(\cd_{p/2} \oplus (m-p-1) \cd_0 \right) \times
\left(\cd_{p/2} \oplus (m-p-1) \cd_0 \right) \nonumber \\
&&\oplus \left(\cd_{(p-1)/2} \oplus (n-p) \cd_0 \right) \times
\left(\cd_{(p-1)/2} \oplus (n-p) \cd_0 \right) - \cd_0 \nonumber \\
&=& \cd_p \oplus 2\cd_{p-1} \oplus ... \oplus 2\cd_1 \oplus 2(m-p-1) \cd_{p/2}
\nonumber \\
&&\oplus 2(n-p) \cd_{(p-1)/2} \oplus [(m-p-1)^2 + (n-p)^2 + 1] \cd_0
\ena
Gathering the $Sl(2)$ representations $\cd_j$ into $OSp(1|2)$ irreducible
representations $\crr_j$, one finds that the adjoint representation of
$Sl(m|n)$ decomposes under the superprincipal $OSp(1|2)$ of $Sl(p+1|p) \subset
Sl(m|n)$ as\footnote{In the following, we will use $\frac{{\bf \footnotesize
Ad}[\cg]}{\wt{\cg}}$ to denote the decomposition of the adjoint representation
of $\cg$ with respect to the superprincipal $OSp(1|2)$ of
$\wt{\cg}\subset\cg$.}:
\bea
\frac{{\bf Ad}[Sl(m|n)]}{Sl(p+1|p)} &=&
\crr_p \oplus \crr_{p-1/2} \oplus \crr_{p-1} \oplus
... \oplus \crr_{1/2}
\oplus 2(n-p) \crr_{p/2} \oplus 2(m-p-1) \crr'_{p/2} \nonumber \\
&&\oplus [(m-p-1)^2 + (n-p)^2] \crr_0 \oplus 2(m-p-1)(n-p) \crr'_0
\label{luc2}
\ena
Notice that the $W_{j+1/2}$ superfield corresponding to the
representation $\crr_j =
\cd_j\oplus\cd_{j-1/2}$ has two component fields $w_{j+1}$ and $w_{j+1/2}$
of spins $j+1$ and $j+1/2$ respectively. If the
representation $\cd_j$ comes from the bosonic (resp. fermionic) part,
$w_{j+1}$ is commuting (resp. anticommuting), whereas
$w_{j+1/2}$ is anticommuting (resp. commuting). Therefore, if $j$ is integer,
the generators $w_{j+1}$ and $w_{j+1/2}$ have the "right" statistics, whereas
they have the "wrong" statistics if $j$ is half-integer. The representations
$\crr_j$ denoted with a prime are used in the case of $W$
superfields obeying to the "wrong" statistics.

\indent

Actually, this decomposition (which was obtained
above in a rather heavy way) can
be derived directly from the decomposition of the fundamental representation
of the superalgebra $Sl(m|n)$ with respect to the $OSp(1|2)$ under
consideration. From (\ref{luc1}), the fundamental representation of
$Sl(m|n)$, of dimension $m+n$, decomposes as
\be
\und{m+n} = \r{p/2} \oplus (m-p-1) \r{0} \oplus (n-p) \rpi{0}
\ee
where we have introduced two kinds of $OSp(1|2)$ representations.
An $OSp(1|2)$ representation is denoted
$\r{j}$ if the representation $D_j$ comes from the decomposition of the
fundamental of $Sl(m)$ and $\rpi{j}$ if $D_j$ comes from the decomposition
of the fundamental of $Sl(n)$.

\noindent
Then the adjoint representation of $Sl(m|n)$ is obtained from the fundamental
one by
\be
{\bf Ad}[Sl(m|n)] = (\und{m+n}) \times (\ov{m+n}) - \und{1}
\label{luc5}
\ee
Using the general formula giving the product of two $OSp(1|2)$ representations
$\crr_{q_1}$ and $\crr_{q_2}$:
\be
\crr_{q_1} \times \crr_{q_2} = \bigoplus_{q=|q_1-q_2|}^{q=q_1+q_2} \crr_q
\mbox{\ \ \ with } q \mbox{\ integer and half-integer,}
\label{luc8}
\ee
one recovers the decomposition of the adjoint representation of $Sl(m|n)$
under the superprincipal $OSp(1|2)$ of $Sl(p+1|p)$ given by (\ref{luc2}).

\indent

Now, we consider the $OSp(1|2)$ superprincipal embedding of $Sl(p|p+1)$
in $\cg$
with $p \le \inf(m,n-1)$. Then the decompositions of
the representations $\und{m}$ and $\und{n}$ of $Sl(m)$ and $Sl(n)$ are:
\bea
&&\und{m} = \cd_{(p-1)/2} \oplus (m-p) \cd_0 \nonumber \\
&&\und{n} = \cd_{p/2} \oplus (n-p-1) \cd_0
\ena
leading to the following decomposition of the fundamental representation
$\und{m+n}$ of $Sl(m|n)$:
\be
\und{m+n} = \rpi{p/2} \oplus (m-p) \r{0} \oplus (n-p-1) \rpi{0}
\ee
Therefore, the decomposition of the adjoint representation reads
\bea
\frac{{\bf Ad}[Sl(m|n)]}{Sl(p|p+1)} &=&
\crr_p \oplus \crr_{p-1/2} \oplus \crr_{p-1} \oplus
... \oplus \crr_{1/2}
\oplus 2(m-p) \crr_{p/2} \oplus 2(n-p-1) \crr'_{p/2} \nonumber \\
&&\oplus [(m-p)^2 + (n-p-1)^2] \crr_0 \oplus 2(m-p)(n-p-1) \crr'_0
\ena

\indent

More generally, if $\wt{\cg}$ is a sum of SSAs of
$Sl(p+1|p)$ or $Sl(p|p+1)$ type,
each factor $Sl(p+1|p)$ gives rise to an $OSp(1|2)$
representation $\r{p/2}$ and each factor $Sl(p|p+1)$ to an $OSp(1|2)$
representation $\rpi{p/2}$ in the decomposition of the fundamental
$\und{m+n}$ of $Sl(m|n)$, completed eventually by singlets $\r{0}$ or
$\rpi{0}$. Then the decomposition of the adjoint representation of $Sl(m|n)$
is obtained by applying (\ref{luc5}).

\indent

Finally, let us consider the case of the superalgebra $PSl(n|n)$ whose
bosonic part is $Sl(n) \oplus Sl(n)$ and its fermionic part is the
$(\und{n},\ov{n}) \oplus (\ov{n},\und{n})$ representation of the bosonic
subalgebra. If the minimal including regular SSA is $Sl(p+1|p)$
with $p \le n-1$, the fundamental representation of $PSl(n|n)$ decomposes as
\be
\und{2n} = \r{p/2} \oplus (n-p-1)\r{0} \oplus (n-p)\rpi{0}
\ee
and the adjoint representation of $PSl(n|n)$ is given by
\be
{\bf Ad}[PSl(n|n)] = (\und{2n}) \times (\ov{2n}) - 2 \ \und{1}
\ee
One finds therefore
\bea
\frac{{\bf Ad}[PSl(n|n)]}{Sl(p+1|p)} &=&
\crr_p \oplus \crr_{p-1/2} \oplus \crr_{p-1} \oplus
... \oplus \crr_{1/2} \oplus 2(n-p) \crr_{p/2} \oplus 2(n-p-1) \crr'_{p/2}
\nonumber \\
&&\oplus [(n-p-1)^2 + (n-p)^2 -1] \crr_0 \oplus 2(n-p-1)(n-p) \crr'_0
\ena
The computation is completely analogous if the minimal including regular
SSA is $Sl(p|p+1)$.

\subsection{The orthosymplectic superalgebras $OSp(M|2n)$}

\subsubsection{Products of $OSp(1|2)$ irreducible representations \label{pouf}}

\indent

Consider an $OSp(1|2)$ SSA of $\cg = OSp(M|2n)$ and let $\wt{\cg}$
be the minimal including SSA in $\cg$. Under the superprincipal
$OSp(1|2)$ of $\wt{\cg}$, the fundamental representation of $\cg$, of dimension
$M+2n$, decomposes in a sum of $OSp(1|2)$ representations, generically denoted
as
\be
\und{M+2n} = (\bigoplus_j \r{j})\oplus( \bigoplus_{j'} \rpi{j'})
\ee
where the representations $\r{j}$ and $\rpi{j'}$ have the same meaning as in
the previous section : a representation $\r{j}$ (resp. $\rpi{j'}$) corresponds
here to an $OSp(1|2)$ representation where the $\cd_j$ comes from the
decomposition of the $SO(M)$ (resp. $Sp(2n)$) part.

\indent

In order to know how to obtain the decomposition of the adjoint representation
of $OSp(M|2n)$ from the decomposition of the fundamental one, we come back for
a while to the Abelian case \cite{13},
specializing for the moment to the superalgebra
$OSp(2m+1|2m)$. In that case, the fundamental representation of $OSp(2m+1|2m)$
of dimension $4m+1$ decomposes under its superprincipal $OSp(1|2)$ as
\be
\und{4m+1} = \r{m}
\ee
and thus the adjoint representation of $OSp(2m+1|2m)$ decomposes as
\be
{\bf Ad}[OSp(2m+1|2m)] = (\cd_m \times \cd_m)_A \oplus
(\cd_{m-1/2} \times \cd_{m-1/2})_S \oplus (\cd_m \times \cd_{m-1/2})
\ee
The two first terms correspond to the adjoint representations of $SO(2m+1)$ and
$Sp(2m)$ respectively, and the last one to the fermionic representation
$(\und{2m+1},\und{2m})$ of the bosonic part. Therefore, one has
\bea
{\bf Ad}[OSp(2m+1|2m)] &=& \left(\bigoplus_{k=1}^m
\cd_{2k-1}\right) \oplus \left(\bigoplus_{k=1}^m \cd_{2k-1}\right) \oplus
\left(\bigoplus_{k=1/2}^{2m-1/2} \cd_k\right) \nonumber \\
&=& \left(\bigoplus_{k=1}^m \cd_{2k-1} \oplus \cd_{2k-3/2}\right)
\oplus \left(\bigoplus_{k=1}^m \cd_{2k-1/2} \oplus \cd_{2k-1}\right)
\nonumber \\
&=& \bigoplus_{k=1}^m \left(\crr_{2k-1} \oplus \crr_{2k-1/2}\right)
\ena
By analogy with the bosonic $SO(2m)$ case (cf. \ref{pi}), we set (with $m$
integer)
\be
\left(\r{m} \times \r{m}\right)_A =
\bigoplus_{k=1}^m \left(\crr_{2k-1} \oplus \crr_{2k-1/2}\right)
\mb{with} k\in\Z
\label{luc6}
\ee

\indent

Now we specialize to the superalgebra $OSp(2m-1|2m)$. In that case, the
fundamental representation of $OSp(2m-1|2m)$ of dimension $4m-1$ decomposes
under its superprincipal $OSp(1|2)$ as
\be
\und{4m-1} = \rpi{m-1/2}
\ee
and thus the adjoint representation of $OSp(2m-1|2m)$ decomposes as
\be
{\bf Ad}[OSp(2m-1|2m)] = (\cd_{m-1} \times \cd_{m-1})_A \oplus
(\cd_{m-1/2} \times \cd_{m-1/2})_S \oplus (\cd_{m-1} \times \cd_{m-1/2})
\ee
Therefore, one has
\bea
{\bf Ad}[OSp(2m-1|2m)] &=& \left(\bigoplus_{k=1}^{m-1}
\cd_{2k-1}\right) \oplus \left(\bigoplus_{k=1}^m \cd_{2k-1}\right) \oplus
\left(\bigoplus_{k=1/2}^{2m-3/2} \cd_k\right) \nonumber \\
&=& \left(\bigoplus_{k=1}^m \cd_{2k-1} \oplus \cd_{2k-3/2}\right)
\oplus \left(\bigoplus_{k=1}^{m-1} \cd_{2k-1/2} \oplus \cd_{2k-1}\right)
\nonumber \\
&=& \bigoplus_{k=1}^{m-1} \left(\crr_{2k-1} \oplus \crr_{2k-1/2}\right)
\oplus \crr_{2m-1}
\ena
By analogy with the bosonic $Sp(2m)$ case (cf. \ref{pf}), we set (with $m$
integer)
\be
\left(\rpi{m-1/2} \times \rpi{m-1/2}\right)_S = \bigoplus_{k=1}^{m-1}
\left(\crr_{2k-1} \oplus \crr_{2k-1/2}\right) \oplus \crr_{2m-1}
\mb{with} k\in\Z
\label{luc7}
\ee
Using the equations (\ref{luc8}), (\ref{luc6}) and (\ref{luc7}),
one obtains also the useful formulae (with $k$ and $m$ integer)
\be
\left(\r{m-1/2} \times \r{m-1/2}\right)_A =
\bigoplus_{k=0}^{m-1} \left(\crr_{2k} \oplus \crr_{2k+1/2}\right)
\ee
and
\be
\left(\rpi{m} \times \rpi{m}\right)_S = \bigoplus_{k=0}^{m-1}
\left(\crr_{2k} \oplus \crr_{2k+1/2}\right) \oplus \crr_{2m}
\ee
The products between $\r{j}$ and $\rpi{j}$ representations are given by
\bea
&& \r{j_1} \times \r{j_2} = \left\{
\begin{array}{ll} \oplus \crr_{j_3} \mbox{\ \ \ if } j_1+j_2
\mbox{ is integer} \\
\oplus \crr'_{j_3} \mbox{\ \ \ if } j_1+j_2 \mbox{ is half-integer} \end{array}
\right. \nonumber \\
&& \rpi{j_1} \times \rpi{j_2} = \left\{
\begin{array}{ll} \oplus \crr_{j_3} \mbox{\ \ \ if } j_1+j_2
\mbox{ is integer} \\
\oplus \crr'_{j_3} \mbox{\ \ \ if } j_1+j_2 \mbox{ is half-integer} \end{array}
\right. \\
&& \r{j_1} \times \rpi{j_2} = \left\{
\begin{array}{ll} \oplus \crr'_{j_3} \mbox{\ \ \ if } j_1+j_2
\mbox{ is integer} \\
\oplus \crr_{j_3} \mbox{\ \ \ if } j_1+j_2 \mbox{ is half-integer} \end{array}
\right.\nonumber
\ena
where the representations $\crr_{j_3}$ and $\crr'_{j_3}$ correspond to $W$
superfields which obey to "right" or "wrong" statistics respectively.

\noindent
Finally, one has
\be
(n\crr_j \times n\crr_j)_A = \frac{n(n+1)}{2} (\crr_j \times \crr_j)_A
\oplus \frac{n(n-1)}{2} (\crr_j \times \crr_j)_S
\ee
\be
(n\crr_j \times n\crr_j)_S = \frac{n(n+1)}{2} (\crr_j \times \crr_j)_S
\oplus \frac{n(n-1)}{2} (\crr_j \times \crr_j)_A
\ee
and
\be
\left((\crr_{j_1} \oplus \crr_{j_2}) \times (\crr_{j_1} \oplus \crr_{j_2})
\right)_A =
(\crr_{j_1} \times \crr_{j_1})_A \oplus (\crr_{j_2} \times \crr_{j_2})_A
\oplus (\crr_{j_1} \times \crr_{j_2})
\ee
\be
\left((\crr_{j_1} \oplus \crr_{j_2}) \times (\crr_{j_1} \oplus \crr_{j_2})
\right)_S =
(\crr_{j_1} \times \crr_{j_1})_A \oplus (\crr_{j_2} \times \crr_{j_2})_S
\oplus (\crr_{j_1} \times \crr_{j_2})
\label{luc9}
\ee
Of course, the same formulae hold for $\rpi{}$ representations.

\indent

It remains to obtain the decompositions of the adjoint representations
of the simple Lie superalgebras from the decompositions of their fundamental
representations for the different possible $OSp(1|2)$ embeddings in order
to classify the super-Toda theories. The following subsections are devoted
to the study of the superalgebras $OSp(2m|2n)$, $OSp(2m+1|2n)$, $OSp(2|2n)$
and to the irregular embeddings.

\subsubsection{The superalgebras $OSp(2m|2n)$}

\indent

The regular SSAs of $\cg = OSp(2m|2n)$ (with $m \ge 2$) which
admit a superprincipal embedding are of the type $OSp(2k|2k)$, $OSp(2k+2|2k)$
and $Sl(p \pm 1|p)$.

\indent

Let $\wt{\cg} = OSp(2k|2k)$ with $1 \le k \le \inf(m,n)$. Under the
superprincipal $OSp(1|2)$ of $\wt{\cg}$, the fundamental representation of
$OSp(2m|2n)$ of dimension $2m+2n$ decomposes as follows:
\be
\und{2m+2n} = \rpi{k-1/2} \oplus (2m-2k+1) \r{0} \oplus (2n-2k) \rpi{0}
\ee
The decomposition of the adjoint representation of $OSp(2m|2n)$ is obtained
from the decomposition of the fundamental representation
by taking the antisymmetric product
of the othogonal part and the symmetric product of the
symplectic part; more precisely,
one has
\bea
\frac{{\bf Ad}[OSp(2m|2n)]}{OSp(2k|2k)} &=&
\left. \left(\rule{.0mm}{5mm} (2m-2k+1) \r{0} \right) \times
\left(\rule{.0mm}{5mm} (2m-2k+1) \r{0}\right)
\right|_A
\nonumber \\
&&\left. \oplus \left(\rpi{k-1/2} \oplus (2n-2k)\rpi{0} \right) \times
\left(\rpi{k-1/2} \oplus (2n-2k)\rpi{0}\right)\right|_S \nonumber \\
&&\oplus \left(\rule{.0mm}{5mm} (2m-2k+1)\r{0}\right) \times
\left(\rpi{k-1/2}
\oplus (2n-2k)\rpi{0}\right)
\ena
Using the formulae (\ref{luc6}) and (\ref{luc7}-\ref{luc9}), one finds
\bea
\frac{{\bf Ad}[OSp(2m|2n)]}{OSp(2k|2k)} &=&
\crr_{2k-1} \oplus \crr_{2k-5/2} \oplus \crr_{2k-3} \oplus \crr_{2k-9/2}
\oplus ... \oplus \crr_{3/2} \oplus \crr_1 \nonumber \\
&\oplus& (2m-2k+1) \crr_{k-1/2} \oplus 2(n-k) \crr'_{k-1/2} \oplus
2(2m-2k+1)(n-k) \crr'_0 \nonumber \\
&\oplus& [(2m-2k+1)(m-k) + (2n-2k+1)(n-k)] \crr_0
\ena

\indent

Now, let $\wt{\cg} = OSp(2k+2|2k)$. Under the superprincipal $OSp(1|2)$ of
$\wt{\cg}$, the fundamental representation of $OSp(2m|2n)$ decomposes as:
\be
\und{2m+2n} = \r{k} \oplus (2m-2k-1) \r{0} \oplus (2n-2k) \rpi{0}
\ee
Therefore, one has
\bea
\frac{{\bf Ad}[OSp(2m|2n)]}{OSp(2k+2|2k)} &=&
\left.\left(\rule{.0mm}{5mm} \r{k} \oplus (2m-2k-1)\r{0} \right) \times
\left(\rule{.0mm}{5mm} \r{k}
\oplus (2m-2k-1)\r{0}\right)\right|_A \nonumber \\
&&\left. \oplus \left(\rule{.0mm}{5mm} (2n-2k) \rpi{0} \right) \times
\left(\rule{.0mm}{5mm} (2n-2k) \rpi{0}
\right)\right|_S \nonumber \\
&&\oplus \left(\rule{.0mm}{5mm} \r{k} \oplus (2m-2k-1)\r{0}\right) \times
\left(\rule{.0mm}{5mm} (2n-2k)\rpi{0}
\right)
\ena
and one obtains in that case
\bea
\frac{{\bf Ad}[OSp(2m|2n)]}{OSp(2k+2|2k)} &=&
\crr_{2k-1/2} \oplus \crr_{2k-1} \oplus \crr_{2k-5/2}
\oplus \crr_{2k-3} \oplus ... \oplus \crr_{3/2} \oplus \crr_1 \nonumber \\
&\oplus& (2m-2k-1) \crr_k \oplus 2(n-k) \crr'_k \oplus 2(2m-2k-1)(n-k)
\crr'_0 \nonumber \\
&\oplus& [(2m-2k-1)(m-k-1) + (2n-2k+1)(n-k)] \crr_0
\ena

\indent

Finally, let us consider the case where $\wt{\cg}$ belongs to the unitary
series. First, we study the case $\wt{\cg} = Sl(2k+1|2k)$ with $4k \le m+n-2$.
The decomposition of the fundamental representation of $OSp(2m|2n)$ under the
superprincipal $OSp(1|2)$ of $\wt{\cg}$ is given by
\be
\und{2m+2n} = 2\r{k} \oplus 2(m-2k-1) \r{0} \oplus 2(n-2k) \rpi{0}
\ee
Therefore, one has
\bea
\frac{{\bf Ad}[OSp(2m|2n)]}{Sl(2k+1|2k)} &=&
\left.\left(\rule{.0mm}{5mm} 2\r{k} \oplus 2(m-2k-1)\r{0} \right) \times
\left(\rule{.0mm}{5mm} 2\r{k}
\oplus 2(m-2k-1)\r{0}\right)\right|_A \nonumber \\
&&\left.\oplus \left(\rule{.0mm}{5mm} 2(n-2k) \rpi{0} \right) \times
\left(\rule{.0mm}{5mm} 2(n-2k) \rpi{0}\right)
\right|_S \nonumber \\
&&\oplus \left(\rule{.0mm}{5mm} 2\r{k} \oplus 2(m-2k-1)\r{0}\right) \times
\left(\rule{.0mm}{5mm} 2(n-2k)\rpi{0}
\right)
\ena
One obtains here
\bea
\frac{{\bf Ad}[OSp(2m|2n)]}{Sl(2k+1|2k)} &=&
\crr_{2k} \oplus 3\crr_{2k-1} \oplus \crr_{2k-2} \oplus ... \oplus \crr_2
\oplus 3\crr_1 \oplus \crr_0 \nonumber \\
&\oplus& 3\crr_{2k-1/2} \oplus \crr_{2k-3/2} \oplus 3\crr_{2k-5/2} \oplus ...
\oplus 3\crr_{3/2} \oplus \crr_{1/2} \nonumber \\
&\oplus& 4(m-2k-1) \crr_{k} \oplus 4(n-2k)\crr'_{k} \oplus
4(m-2k-1)(n-2k) \crr'_0 \nonumber \\
&\oplus& [(2m-4k-3)(m-2k-1) + (2n-4k+1)(n-2k)] \crr_0
\ena
The other cases are similar. One finds easily the following results.
If $\wt{\cg} = Sl(2k-1|2k)$ with $4k \le m+n$, one has
\be
\und{2m+2n} = 2\rpi{k-1/2} \oplus 2(m-2k+1) \r{0} \oplus 2(n-2k) \rpi{0}
\ee
and
\bea
\frac{{\bf Ad}[OSp(2m|2n)]}{Sl(2k-1|2k)} &=&
3\crr_{2k-1} \oplus \crr_{2k-2} \oplus 3\crr_{2k-3} \oplus ... \oplus \crr_2
\oplus 3\crr_1 \oplus \crr_0 \nonumber \\
&\oplus& \crr_{2k-3/2} \oplus 3\crr_{2k-5/2} \oplus \crr_{2k-7/2} \oplus ...
\oplus 3\crr_{3/2} \oplus \crr_{1/2} \nonumber \\
&\oplus& 4(m-2k+1) \crr_{k-1/2} \oplus 4(n-2k)\crr'_{k-1/2} \oplus
4(m-2k+1)(n-2k) \crr'_0 \nonumber \\
&\oplus& [(2m-4k+1)(m-2k+1) + (2n-4k+1)(n-2k)] \crr_0
\ena

\noindent
If $\wt{\cg} = Sl(2k|2k+1)$, one has
\be
\und{2m+2n} = 2\rpi{k} \oplus 2(m-2k) \r{0} \oplus 2(n-2k-1) \rpi{0}
\ee
and
\bea
\frac{{\bf Ad}[OSp(2m|2n)]}{Sl(2k|2k+1)} &=&
3\crr_{2k} \oplus \crr_{2k-1} \oplus 3\crr_{2k-2} \oplus ... \oplus 3\crr_2
\oplus \crr_1 \oplus 3\crr_0 \nonumber \\
&\oplus& \crr_{2k-1/2} \oplus 3\crr_{2k-3/2} \oplus \crr_{2k-5/2} \oplus ...
\oplus \crr_{3/2} \oplus 3\crr_{1/2} \nonumber \\
&\oplus& 4(m-2k) \crr'_{k} \oplus 4(n-2k-1)\crr_{k} \oplus
4(m-2k)(n-2k-1) \crr'_0 \nonumber \\
&\oplus& [(2m-4k-1)(m-2k) + (2n-4k-1)(n-2k-1)] \crr_0
\ena

\noindent
Finally, if $\wt{\cg} = Sl(2k|2k-1)$, one has
\be
\und{2m+2n} = 2\r{k-1/2} \oplus 2(m-2k) \r{0} \oplus 2(n-2k+1) \rpi{0}
\ee
and
\bea
\frac{{\bf Ad}[OSp(2m|2n)]}{Sl(2k|2k-1)} &=&
\crr_{2k-1} \oplus 3\crr_{2k-2} \oplus \crr_{2k-3} \oplus ... \oplus 3\crr_2
\oplus \crr_1 \oplus 3\crr_0 \nonumber \\
&\oplus& 3\crr_{2k-3/2} \oplus \crr_{2k-5/2} \oplus 3\crr_{2k-7/2} \oplus ...
\oplus \crr_{3/2} \oplus 3\crr_{1/2} \nonumber \\
&\oplus& 4(m-2k) \crr'_{k-1/2} \oplus 4(n-2k+1)\crr_{k-1/2} \oplus
4(m-2k)(n-2k+1) \crr'_0 \nonumber \\
&\oplus& [(2m-4k-1)(m-2k) + (2n-4k+3)(n-2k+1)] \crr_0
\ena

\subsubsection{The superalgebras $OSp(2m+1|2n)$}

\indent

The regular SSAs of $\cg = OSp(2m+1|2n)$ which admit a
superprincipal embedding are of the type $OSp(2k|2k)$, $OSp(2k+2|2k)$,
$OSp(2k \pm 1|2k)$ and $Sl(p \pm 1|p)$.

\indent

Let $\wt{\cg} = OSp(2k|2k)$. Under the superprincipal $OSp(1|2)$ of $\wt{\cg}$,
the fundamental representation of $OSp(2m+1|2n)$, of dimension $2m+2n+1$,
decomposes as follows:
\be
\und{2m+2n+1} = \rpi{k-1/2} \oplus (2n-2k) \rpi{0} \oplus (2m-2k+2) \r{0}
\ee
The decomposition of the adjoint representation is then
\bea
\frac{{\bf Ad}[OSp(2m+1|2n)]}{OSp(2k|2k)} &=&
\left.\left(\rule{.0mm}{5mm} (2m-2k+2)\r{0} \right) \times
\left(\rule{.0mm}{5mm} (2m-2k+2)\r{0}\right)
\right|_A \nonumber \\
&&\left. \oplus \left(\rpi{k-1/2} \oplus (2n-2k)\rpi{0} \right) \times
\left(\rpi{k-1/2} \oplus (2n-2k)\rpi{0}\right)\right|_S \nonumber \\
&&\oplus \left(\rule{.0mm}{5mm} (2m-2k+2)\r{0}\right) \times
\left(\rpi{k-1/2} \oplus
(2n-2k)\rpi{0}\right)
\ena
i.e.
\bea
\frac{{\bf Ad}[OSp(2m+1|2n)]}{OSp(2k|2k)} &=&
\crr_{2k-1} \oplus \crr_{2k-5/2} \oplus \crr_{2k-3}
\oplus \crr_{2k-9/2} \oplus ... \oplus \crr_{3/2} \oplus \crr_1 \nonumber \\
&\oplus& 2(m-k+1) \crr_{k-1/2} \oplus 2(n-k) \crr'_{k-1/2} \oplus
4(m-k+1)(n-k) \crr'_0 \nonumber \\
&\oplus& [(2m-2k+1)(m-k+1) + (2n-2k+1)(n-k)] \crr_0
\ena

\indent

Now, let $\wt{\cg} = OSp(2k+2|2k)$. Under the superprincipal $OSp(1|2)$ of
$\wt{\cg}$,
the fundamental representation of $OSp(2m+1|2n)$ decomposes as:
\be
\und{2m+2n+1} = \r{k} \oplus (2m-2k) \r{0} \oplus (2n-2k) \rpi{0}
\ee
Then one obtains
\bea
\frac{{\bf Ad}[OSp(2m+1|2n)]}{OSp(2k+2|2k)} &=&
\left. \left(\rule{.0mm}{5mm} \r{k} \oplus (2m-2k)\r{0} \right) \times
\left(\rule{.0mm}{5mm} \r{k}
\oplus (2m-2k)\r{0}\right)\right|_A \nonumber \\
&&\left. \oplus \left(\rule{.0mm}{5mm} (2n-2k)\rpi{0} \right) \times
\left(\rule{.0mm}{5mm} (2n-2k)\rpi{0} \right)
\right|_S \nonumber \\
&&\oplus \left(\rule{.0mm}{5mm} \r{k} \oplus (2m-2k)\r{0}\right) \times
\left(\rule{.0mm}{5mm} (2n-2k)\rpi{0}
\right)
\ena
i.e.
\bea
\frac{{\bf Ad}[OSp(2m+1|2n)]}{OSp(2k+2|2k)} &=&
\crr_{2k-1/2} \oplus \crr_{2k-1} \oplus
\crr_{2k-5/2}
\oplus \crr_{2k-3} \oplus ... \oplus \crr_{3/2} \oplus \crr_1 \nonumber \\
&\oplus& 2(n-k) \crr'_k \oplus 2(m-k) \crr_k \oplus 4(m-k)(n-k) \crr'_0
\nonumber \\
&\oplus& [(2m-2k-1)(m-k) + (2n-2k+1)(n-k)] \crr_0
\ena

Finally, let $\wt{\cg} = OSp(2k-1|2k)$. Under the superprincipal $OSp(1|2)$ of
$\wt{\cg}$,
the fundamental representation of $OSp(2m+1|2n)$ decomposes as
\be
\und{2m+2n+1}  = \rpi{k-1/2} \oplus (2n-2k) \rpi{0} \oplus (2m-2k+2) \r{0}
\ee
which is the same decomposition as the case $\wt{\cg} = OSp(2k|2k)$. Therefore,
the two SSAs $OSp(2k|2k)$ and $OSp(2k-1|2k)$ (when both can be
embedded in $\cg$) lead to the same
decomposition of the adjoint representation of $\cg$ and consequently to the
same theory. On the same lines, one finds that the two SSAs
$OSp(2k+2|2k)$ and $OSp(2k+1|2k)$ lead to the same theory.

\indent

The last case is $\wt{\cg} = Sl(p \pm 1|p)$. We leave the different
decompositions to the reader. The results are summarized in the table of
Section \ref{soussect}.

\subsubsection{The irregular embeddings}

\indent

We will study now the irregular embeddings, which are present in
$OSp(2n \pm 2|2n)$ and $OSp(2n|2n)$.

\indent

Consider first the superalgebra $\cg = OSp(2n+2|2n)$ and take the $OSp(1|2)$
SSA of $\cg$ such that the minimal including SSA in
$\cg$ (which is now singular) is
$\wt{\cg} = OSp(2k+1|2k) \oplus OSp(2n-2k+1|2n-2k)$
and $1 \le k \le [\frac{n-1}{2}]$. Under the superprincipal $OSp(1|2)$ of
$\wt{\cg}$, the fundamental representation of $\cg$, of dimension $4n+2$,
decomposes as
\be
\und{4n+2} = \r{k} \oplus \r{n-k}
\ee
and we get for the $OSp(2n+2|2n)$ adjoint representation
\be
\frac{{\bf Ad}[OSp(2n+2|2n)]}{OSp(2k+1|2k)\oplus OSp(2n-2k+1|2n-2k)}
 = \left.\left(\rule{.0mm}{5mm} \r{k} \oplus \r{n-k} \right) \times
\left(\rule{.0mm}{5mm} \r{k} \oplus \r{n-k}\right)\right|_A
\ee
which leads to the following decomposition:
\bea
&&\frac{{\bf Ad}[OSp(2n+2|2n)]}{OSp(2k+1|2k)\oplus OSp(2n-2k+1|2n-2k)}
= \crr_{2n-2k-1} \oplus \crr_{2n-2k-3} \oplus ... \oplus \crr_1 \nonumber \\
&&\mb{}\oplus \crr_{2n-2k-1/2} \oplus \crr_{2n-2k-3/2}
\oplus ... \oplus \crr_{3/2}
\oplus \crr_{2k-1} \oplus \crr_{2k-3} \oplus ... \oplus \crr_1 \nonumber \\
&&\mb{} \oplus \crr_{2k-1/2} \oplus \crr_{2k-3/2} \oplus ... \oplus \crr_{3/2}
\oplus \crr_n \oplus \crr_{n-1} \oplus ... \oplus \crr_{n-2k} \nonumber \\
&&\mb{} \oplus \crr_{n-1/2} \oplus \crr_{n-3/2}
\oplus ... \oplus \crr_{n-2k+1/2}
\ena

\indent

Consider then the superalgebra $\cg = OSp(2n-2|2n)$ with $\wt{\cg} =
OSp(2k-1|2k) \oplus OSp(2n-2k-1|2n-2k)$ and $1 \le k \le [\frac{n-3}{2}]$.
The fundamental representation of $\cg$, of dimension $4n-2$, decomposes under
the superprincipal $OSp(1|2)$ of $\wt{\cg}$ as
\be
\und{4n-2} = \rpi{k-1/2} \oplus \rpi{n-k-1/2}
\ee
The adjoint representation of $OSp(2n-2|2n)$ is given by
\be
\frac{{\bf Ad}[OSp(2n-2|2n)]}{OSp(2k-1|2k)\oplus OSp(2n-2k-1|2n-2k)}
 = \left.\left(\rpi{k-1/2} \oplus \rpi{n-k-1/2}\right)
\times \left(\rpi{k-1/2} \oplus \rpi{n-k-1/2}\right)\right|_S \nonumber
\ee
i.e.
\bea
&&\frac{{\bf Ad}[OSp(2n-2|2n)]}{OSp(2k-1|2k)\oplus OSp(2n-2k-1|2n-2k)}
 = \crr_{2n-2k-1} \oplus \crr_{2n-2k-3} \oplus ... \oplus \crr_1 \nonumber \\
&&\mb{} \oplus \crr_{2n-2k-5/2} \oplus \crr_{2n-2k-7/2}
\oplus ... \oplus \crr_{3/2}
\oplus \crr_{2k-1} \oplus \crr_{2k-3} \oplus ... \oplus \crr_1 \nonumber \\
&&\mb{} \oplus \crr_{2k-5/2} \oplus \crr_{2k-7/2} \oplus ... \oplus \crr_{3/2}
\oplus \crr_{n-1} \oplus \crr_{n-2} \oplus ... \oplus \crr_{n-2k} \nonumber \\
&&\mb{} \oplus\crr_{n-3/2} \oplus \crr_{n-5/2} \oplus ... \oplus
\crr_{n-2k+1/2}
\ena

\indent

Consider finally the superalgebra $\cg = OSp(2n|2n)$ with $\wt{\cg} =
OSp(2k+1|2k) \oplus OSp(2n-2k-1|2n-2k)$ and $1 \le k \le [\frac{n-1}{2}]$.
Under the superprincipal $OSp(1|2)$ of $\wt{\cg}$, the fundamental
representation of $\cg$, of dimension $4n$, decomposes as
\be
\und{4n} = \r{k} \oplus \rpi{n-k-1/2}
\ee
and we get for the $OSp(2n|2n)$ adjoint representation
\bea
\frac{{\bf Ad}[OSp(2n|2n)]}{OSp(2k+1|2k)\oplus OSp(2n-2k-1|2n-2k)}
&=& \left(\rule{.0mm}{5mm} \r{k} \times \r{k}\right)_A
\oplus \left(\rpi{n-k-1/2} \times \rpi{n-k-1/2}\right)_S \nonumber \\
&& \oplus \left(\r{k} \times \rpi{n-k-1/2}\right)
\ena
which leads to
\bea
&&\frac{{\bf Ad}[OSp(2n|2n)]}{OSp(2k+1|2k)\oplus OSp(2n-2k-1|2n-2k)}
=\crr_{2n-2k-1} \oplus \crr_{2n-2k-3} \oplus ... \oplus \crr_1 \nonumber \\
&&\mb{} \oplus \crr_{2n-2k-5/2} \oplus \crr_{2n-2k-7/2}
\oplus ... \oplus \crr_{3/2}
\oplus \crr_{2k-1} \oplus \crr_{2k-3} \oplus ... \oplus \crr_1 \nonumber \\
&&\mb{} \oplus \crr_{2k-1/2} \oplus \crr_{2k-3/2} \oplus ... \oplus \crr_{3/2}
\oplus \crr_{n-1} \oplus \crr_{n-2} \oplus ... \oplus \crr_{n-2k} \nonumber \\
&&\mb{} \oplus \crr_{n-1/2} \oplus \crr_{n-3/2}
\oplus ... \oplus \crr_{n-2k-1/2}
\ena

\indent

If $\wt{\cg} = OSp(2k-1|2k) \oplus OSp(2n-2k+1|2n-2k)$ with $1 \le k \le
[\frac{n-2}{2}]$, the fundamental representation of $\cg$, of dimension $4n$,
decomposes under the superprincipal $OSp(1|2)$ of $\wt{\cg}$ as
\be
\und{4n} = \r{n-k} \oplus \rpi{k-1/2}
\ee
and we have the following decomposition of the adjoint representation of
$OSp(2n|2n)$:
\bea
&&\frac{{\bf Ad}[OSp(2n|2n)]}{OSp(2k-1|2k)\oplus OSp(2n-2k+1|2n-2k)}
= \crr_{2n-2k-1} \oplus \crr_{2n-2k-3} \oplus ... \oplus \crr_1 \nonumber \\
&&\mb{} \oplus \crr_{2n-2k-1/2} \oplus \crr_{2n-2k-3/2}
\oplus ... \oplus \crr_{3/2}
\oplus \crr_{2k-1} \oplus \crr_{2k-3} \oplus ... \oplus \crr_1 \nonumber \\
&&\mb{} \oplus \crr_{2k-5/2} \oplus \crr_{2k-7/2} \oplus ... \oplus \crr_{3/2}
\oplus \crr_{n-1} \oplus \crr_{n-2} \oplus ... \oplus \crr_{n-2k+1} \nonumber
\\
&&\mb{} \oplus \crr_{n-1/2} \oplus \crr_{n-3/2}
\oplus ... \oplus \crr_{n-2k+1/2}
\ena

\subsubsection{The superalgebras $OSp(2|2n)$}

\indent

The superalgebra $OSp(2|2n)$ requires a special attention. Actually, the
regular SSAs of $\cg = OSp(2|2n)$ which admit a superprincipal
embedding are only $OSp(2|2)$ and $Sl(1|2)$.

\indent

Let $\wt{\cg} = OSp(2|2)$. Under the superprincipal $OSp(1|2)$ of $\wt{\cg}$,
the fundamental representation of $OSp(2|2n)$, of dimension $2n+2$, decomposes
as follows:
\be
\und{2n+2} = \rpi{1/2} \oplus \r{0} \oplus (2n-2) \rpi{0}
\ee
Therefore, the decomposition of the adjoint representation of $OSp(2|2n)$ under
the superprincipal $OSp(1|2)$ of $OSp(2|2) \subset OSp(2|2n)$ is
\bea
\frac{{\bf Ad}[OSp(2|2n)]}{OSp(2|2)} =
\crr_1 \oplus \crr_{1/2} \oplus (2n-2) \crr'_{1/2} \oplus (2n-1)(n-1) \crr_0
\oplus (2n-2) \crr'_0
\ena

\indent

Now, let $\wt{\cg} = Sl(1|2)$. Under the superprincipal $OSp(1|2)$ of
$\wt{\cg}$, the fundamental representation of $OSp(2|2n)$ decomposes as:
\be
\und{2n+2} = 2\rpi{1/2} \oplus (2n-4) \rpi{0}
\ee
In that case, the decomposition of the adjoint representation is
\bea
\frac{{\bf Ad}[OSp(2|2n)]}{Sl(1|2)} =
3\crr_1 \oplus \crr_{1/2} \oplus (4n-8)\crr'_{1/2} \oplus [(2n-3)(n-2) + 1]
\crr_0
\ena

\subsection{Summary of the results \label{soussect}}

\indent

The previous results can be easily extended to the case of sums of simple
Lie SSAs. The decomposition of the fundamental representation is
obtained by taking the corresponding $OSp(1|2)$ representation for each
factor of the sum, which can be read in the following tableau. Then, starting
from a decomposition of the fundamental representation of the form
\be
{\bf F} = \left(\bigoplus_i \r{i} \right) \oplus
\left(\bigoplus_j \rpi{j} \right)
\ee
the decomposition of the adjoint is given, in the orthosymplectic series, by
\be
{\bf Ad} =
\left.\left(\bigoplus_i \r{i} \right) \times \left(\bigoplus_i \r{i} \right)
\right|_A
\oplus
\left.\left(\bigoplus_j \rpi{j} \right) \times \left(\bigoplus_j \rpi{j}
\right) \right|_S \oplus
\left(\bigoplus_i \r{i} \right) \times \left(\bigoplus_j \rpi{j} \right)
\ee
and in the unitary series, by
\bea
&&{\bf Ad} =
\left(\bigoplus_i \r{i} \bigoplus_j \rpi{j} \right) \times
\left(\bigoplus_i \r{i} \bigoplus_j \rpi{j} \right) - \r{0} \mb{for} SL(m|n)
\ m\neq n \\
&&{\bf Ad} =
\left(\bigoplus_i \r{i} \bigoplus_j \rpi{j} \right) \times
\left(\bigoplus_i \r{i} \bigoplus_j \rpi{j} \right) - 2\r{0} \mb{for} PSl(m|m)
\ena
For explicit formulae, one has to apply the product rules given in
(\ref{luc6}) and (\ref{luc7}-\ref{luc9}).

\indent

$\begin{array}{ccc}
\cg & \wt{\cg} & \mbox{Fund. Rep. of \ } \cg \\ \hline && \\
Sl(m|n)
 &Sl(p+1|p) &\r{p/2} \oplus (m-p-1)\r{0} \oplus (n-p)\rpi{0} \\ && \\
 &Sl(p|p+1) &\rpi{p/2} \oplus (m-p)\r{0} \oplus (n-p-1)\rpi{0} \\ && \\
 \hline && \\
OSp(2m|2n)
 &OSp(2k|2k)
 &\begin{array}{c}\rpi{k-1/2} \oplus (2n-2k)\rpi{0} \\
  \oplus (2m-2k+1)\r{0}\end{array} \\ && \\
 &OSp(2k+2|2k)
 &\begin{array}{c}\r{k} \oplus(2m-2k-1)\r{0} \\
  \oplus (2n-2k)\rpi{0}\end{array} \\ && \\
 &Sl(p+1|p)
 &\begin{array}{c}2\r{p/2} \oplus 2(m-p-1)\r{0} \\
  \oplus 2(n-p)\rpi{0}\end{array} \\ && \\
 &Sl(p|p+1)
 &\begin{array}{c}2\rpi{p/2} \oplus 2(n-p-1)\rpi{0} \\
  \oplus 2(m-p)\r{0}\end{array} \\ && \\
 \hline && \\
OSp(2m+1|2n)
 &\left.\begin{array}{c} OSp(2k|2k) \\ OSp(2k-1|2k) \end{array} \right\}
 &\begin{array}{c}\rpi{k-1/2} \oplus (2n-2k)\rpi{0} \\
  \oplus (2m-2k+2)\r{0}\end{array} \\ && \\
 &\left.\begin{array}{c} OSp(2k+2|2k) \\ OSp(2k+1|2k) \end{array} \right\}
 &\begin{array}{c}\r{k} \oplus (2m-2k)\r{0} \\
  \oplus (2n-2k)\rpi{0}\end{array} \\ && \\
 &Sl(p+1|p)
 &\begin{array}{c}2\r{p/2} \oplus 2(m-p-1)\r{0} \\
  \oplus \r{0} \oplus 2(n-p)\rpi{0}\end{array} \\ && \\
 &Sl(p|p+1)
 &\begin{array}{c}2\rpi{p/2} \oplus 2(n-p-1)\rpi{0} \\
  \oplus \r{0} \oplus 2(m-p)\r{0}\end{array} \\ && \\
\hline && \\
OSp(2|2n)
 &OSp(2|2)
 &\rpi{1/2} \oplus \r{0} \oplus (2n-2)\rpi{0} \\ && \\
 &Sl(1|2)
 &2\rpi{1/2} \oplus (2n-4)\rpi{0} \\ && \\
\end{array}$

$\begin{array}{ccc}
\cg & \wt{\cg} & \mbox{Fund. Rep. of \ } \cg \\ \hline && \\
OSp(2n+2|2n)
 &\begin{array}{c}OSp(2k+1|2k) \oplus \\ OSp(2n-2k+1|2n-2k) \end{array}
 &\r{k} \oplus \r{n-k} \\ && \\
 \hline && \\
OSp(2n-2|2n)
 &\begin{array}{c}OSp(2k-1|2k) \oplus \\ OSp(2n-2k-1|2n-2k) \end{array}
 & \rpi{k-1/2} \oplus \rpi{n-k-1/2} \\ && \\
 \hline && \\
OSp(2n|2n)
 &\begin{array}{c}OSp(2k+1|2k) \oplus \\ OSp(2n-2k-1|2n-2k) \end{array}
 & \r{k} \oplus \rpi{n-k-1/2} \\ && \\
 &\begin{array}{c}OSp(2k-1|2k) \oplus \\ OSp(2n-2k+1|2n-2k) \end{array}
 & \r{n-k} \oplus \rpi{k-1/2}
\end{array}$

\subsection{The exceptional superalgebra G(3)}

\indent

The superalgebra $G(3)$ has dimension 31 and rank 3, with $\cg_B = G_2 \oplus
Sl(2)$ as bosonic part and the representation $(\und{7},\und{2})$ of $\cg_B$ as
fermionic part. The Dynkin diagrams of $G(3)$ are

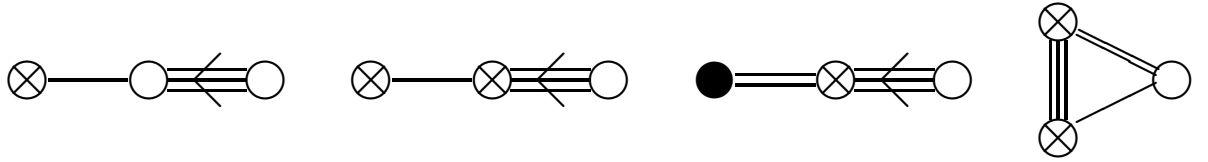
\begin{picture}(450,60)
\thicklines
\put(1,30){\circle{14}}
\put(-4,25){\line( 1, 1){10}}\put(-4,35){\line( 1, -1){10}}
\put(9,30){\line( 1, 0){ 30}}
\put(47,30){\circle{14}}
\put(54,34){\line( 1, 0){ 30}}
\put(54,30){\line( 1, 0){ 30}}
\put(54,26){\line( 1, 0){ 30}}
\put(64,30){\line( 1, 1){10}}\put(64,30){\line( 1, -1){10}}
\put(91,30){\circle{14}}

\put(131,30){\circle{14}}
\put(126,25){\line( 1, 1){10}}\put(126,35){\line( 1, -1){10}}
\put(139,30){\line( 1, 0){ 30}}
\put(177,30){\circle{14}}
\put(172,25){\line( 1, 1){10}}\put(172,35){\line( 1, -1){10}}
\put(184,34){\line( 1, 0){ 30}}
\put(184,30){\line( 1, 0){ 30}}
\put(184,26){\line( 1, 0){ 30}}
\put(194,30){\line( 1, 1){10}}\put(194,30){\line( 1, -1){10}}
\put(221,30){\circle{14}}

\put(261,30){\circle*{14}}
\put(269,32){\line( 1, 0){ 30}}\put(269,28){\line( 1, 0){ 30}}
\put(307,30){\circle{14}}
\put(302,25){\line( 1, 1){10}}\put(302,35){\line( 1, -1){10}}
\put(314,34){\line( 1, 0){ 30}}
\put(314,30){\line( 1, 0){ 30}}
\put(314,26){\line( 1, 0){ 30}}
\put(324,30){\line( 1, 1){10}}\put(324,30){\line( 1, -1){10}}
\put(351,30){\circle{14}}

\put(391,8){\circle{14}}
\put(386,3){\line( 1, 1){10}}\put(386,13){\line( 1, -1){10}}
\put(391,52){\circle{14}}
\put(386,47){\line( 1, 1){10}}\put(386,57){\line( 1, -1){10}}
\put(394,15){\line( 0, 1){ 30}}
\put(391,15){\line( 0, 1){ 30}}
\put(388,15){\line( 0, 1){ 30}}
\put(398,14){\line( 2, 1){ 30}}
\put(398,47){\line( 2, -1){ 30}}\put(399,49){\line( 2, -1){ 30}}
\put(434,30){\circle{14}}
\end{picture}

leading to the following regular sub(super)algebras:
\bea
&&G_2 \oplus A_1, G_2, A_2, A_1 \nonumber \\
&&B(1,1) \oplus A_1, B(1,1), C(2), B(0,1), A_2 \oplus B(0,1) \nonumber \\
&&A(0,2), A(0,1), A(1,0), D(2,1;3), G(3)
\ena
Only the superalgebras $B(0,1)$, $C(2)$, $B(1,1)$, $A(0,1)$, $A(1,0)$ and
$D(2,1;3)$ admit a superprincipal embedding. As an example, we will treat the
case of $B(1,1) \equiv OSp(3|2)$.
{}From the results of Section \ref{luc3}, the bosonic part $G_2 \oplus Sl(2)$
decomposes under the principal $Sl(2)$ of $SO(3) \oplus Sl(2)$ as
\be
{\bf Ad}[G_2 \oplus Sl(2)] = \cd_{3/2} \oplus \ov{\cd}_{3/2} \oplus 2\cd_1
\oplus 3\cd_0
\ee
and the fermionic part $(\und{7},\und{2})$ as
\be
(\und{7},\und{2}) = \cd_{3/2} \oplus 2\cd_1 \oplus \cd_{1/2} \oplus 2\cd_0
\ee
Putting together the $Sl(2)$ representations into $OSp(1|2)$ ones, one obtains
the following decomposition under the superprincipal $OSp(1|2)$ of $OSp(3|2)
\subset G(3)$:
\be
\frac{{\bf Ad}[G(3)]}{OSp(3|2)} =
\crr_{3/2} \oplus 2\crr'_{3/2} \oplus \crr_1 \oplus 3\crr_0
\oplus 2\crr'_0
\ee
The other cases are similar and are summarized in the table
\ref{lucT1}.

\subsection{The exceptional superalgebra F(4)}

\indent

The superalgebra $F(4)$ has dimension 40 and rank 4, with $\cg_B = Sl(2) \oplus
O(7)$ as bosonic part and the representation $(\und{2},\und{8})$ of $\cg_B$ as
fermionic part. Its Dynkin diagrams are:

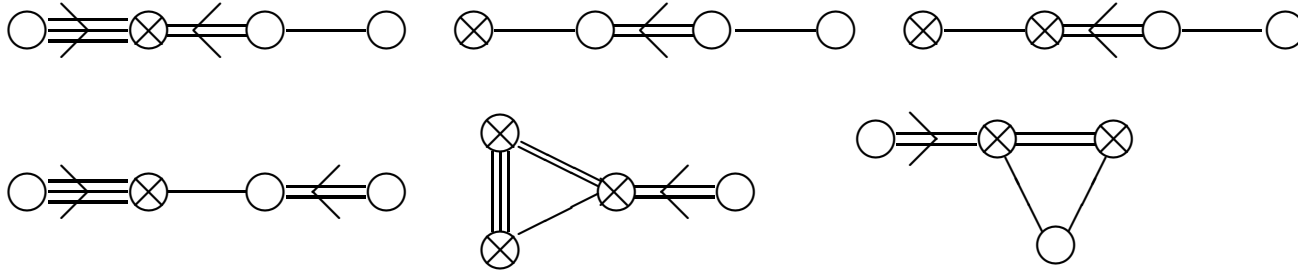
\begin{picture}(450,60)
\thicklines
\put(1,30){\circle{14}}
\put(9,34){\line( 1, 0){ 30}}
\put(9,30){\line( 1, 0){ 30}}
\put(9,26){\line( 1, 0){ 30}}
\put(24,30){\line( -1, 1){10}}\put(24,30){\line( -1, -1){10}}
\put(47,30){\circle{14}}
\put(42,25){\line( 1, 1){10}}\put(42,35){\line( 1, -1){10}}
\put(54,32){\line( 1, 0){ 30}}
\put(54,28){\line( 1, 0){ 30}}
\put(64,30){\line( 1, 1){10}}\put(64,30){\line( 1, -1){10}}
\put(91,30){\circle{14}}
\put(99,30){\line( 1, 0){ 30}}
\put(137,30){\circle{14}}

\put(170,30){\circle{14}}
\put(178,30){\line( 1, 0){ 30}}
\put(165,25){\line( 1, 1){10}}\put(165,35){\line( 1, -1){10}}
\put(216,30){\circle{14}}
\put(223,32){\line( 1, 0){ 30}}
\put(223,28){\line( 1, 0){ 30}}
\put(233,30){\line( 1, 1){10}}\put(233,30){\line( 1, -1){10}}
\put(260,30){\circle{14}}
\put(269,30){\line( 1, 0){ 30}}
\put(307,30){\circle{14}}

\put(340,30){\circle{14}}
\put(348,30){\line( 1, 0){ 30}}
\put(335,25){\line( 1, 1){10}}\put(335,35){\line( 1, -1){10}}
\put(386,30){\circle{14}}
\put(393,32){\line( 1, 0){ 30}}
\put(393,28){\line( 1, 0){ 30}}
\put(403,30){\line( 1, 1){10}}\put(403,30){\line( 1, -1){10}}
\put(381,25){\line( 1, 1){10}}\put(381,35){\line( 1, -1){10}}
\put(430,30){\circle{14}}
\put(438,30){\line( 1, 0){ 30}}
\put(477,30){\circle{14}}
\end{picture}

\begin{picture}(450,60)
\thicklines
\put(1,30){\circle{14}}
\put(9,34){\line( 1, 0){ 30}}
\put(9,30){\line( 1, 0){ 30}}
\put(9,26){\line( 1, 0){ 30}}
\put(24,30){\line( -1, 1){10}}\put(24,30){\line( -1, -1){10}}
\put(47,30){\circle{14}}
\put(42,25){\line( 1, 1){10}}\put(42,35){\line( 1, -1){10}}
\put(54,30){\line( 1, 0){ 30}}
\put(91,30){\circle{14}}
\put(99,32){\line( 1, 0){ 30}}
\put(99,28){\line( 1, 0){ 30}}
\put(109,30){\line( 1, 1){10}}\put(109,30){\line( 1, -1){10}}
\put(137,30){\circle{14}}

\put(180,8){\circle{14}}
\put(175,3){\line( 1, 1){10}}\put(175,13){\line( 1, -1){10}}
\put(180,52){\circle{14}}
\put(175,47){\line( 1, 1){10}}\put(175,57){\line( 1, -1){10}}
\put(183,15){\line( 0, 1){ 30}}
\put(180,15){\line( 0, 1){ 30}}
\put(177,15){\line( 0, 1){ 30}}
\put(187,14){\line( 2, 1){ 30}}
\put(187,47){\line( 2, -1){ 30}}\put(188,49){\line( 2, -1){ 30}}
\put(224,30){\circle{14}}
\put(218,25){\line( 1, 1){10}}\put(218,35){\line( 1, -1){10}}
\put(231,32){\line( 1, 0){ 30}}
\put(231,28){\line( 1, 0){ 30}}
\put(241,30){\line( 1, 1){10}}\put(241,30){\line( 1, -1){10}}
\put(269,30){\circle{14}}

\put(322,50){\circle{14}}
\put(330,52){\line( 1, 0){ 30}}
\put(330,48){\line( 1, 0){ 30}}
\put(345,50){\line( -1, 1){10}}\put(345,50){\line( -1, -1){10}}
\put(368,50){\circle{14}}
\put(363,45){\line( 1, 1){10}}\put(363,55){\line( 1, -1){10}}
\put(375,52){\line( 1, 0){ 30}}
\put(375,48){\line( 1, 0){ 30}}
\put(412,50){\circle{14}}
\put(407,45){\line( 1, 1){10}}\put(407,55){\line( 1, -1){10}}
\put(390,10){\circle{14}}
\put(385,15){\line( -1, 2){ 14}}
\put(395,15){\line( 1, 2){ 14}}
\end{picture}

The SSAs of $F(4)$ which admit a superprincipal embedding are
$A(0,1)$, $A(1,0)$, $C(2)$ and $D(2,1;2)$ (the extended
Dynkin diagrams of $F(4)$ can be found in \cite{FSS}). As an example, we will
treat the case of $C(2) \equiv OSp(2|2)$. The bosonic part $Sl(2) \oplus O(7)$
decomposes then as
\be
{\bf Ad}[Sl(2) \oplus O(7)] = 5\cd_1 \oplus 9\cd_0
\ee
and the fermionic part $(\und{2},\und{8})$ as
\be
(\und{2},\und{8}) = 8\cd_{1/2}
\ee
Putting together the $Sl(2)$ representations into $OSp(1|2)$ ones, one obtains
the following decomposition under the superprincipal $OSp(1|2)$ of $OSp(2|2)
\subset F(4)$:
\be
\frac{{\bf Ad}[F(4)]}{OSp(2|2)} =
5\crr_1 \oplus 3\crr_{1/2} \oplus 6\crr_0
\ee
The other cases are analogous and are summarized in the table \ref{lucT2}.

\sect{$OSp(1|2) \oplus  U(1)$ decompositions of simple Lie superalgebras}

\subsection{Introduction of the $U(1)$}

\indent

Now, we are in position to introduce the $U(1)$ factor. In the case of the
unitary superalgebras, since the formulae for $Sl(p+1|p)$ are completely
analogous to those of $Sl(n)$ (the $\crr_j$ representations replacing the
$\cd_j$ ones), one can write the following statement.

\indent

A decomposition of the fundamental representation ${\bf F}$ of $\cg = Sl(m|n)$
under the superprincipal $OSp(1|2)$ of $\wt{\cg} \subset \cg$ being given,
\be
{\bf F} = \left(\bigoplus_i n_i \r{i} \right) \oplus \left(\bigoplus_j
n_j \rpi{j} \right)
\ee
the corresponding decomposition under $OSp(1|2) \oplus U(1)$ has the form
\be
{\bf F} = \left(\bigoplus_i n_i \r{i}(y_i) \right) \oplus
\left(\bigoplus_j n_j \rpi{j}(y_j) \right)
\ee
identical representations (i.e. labelled by the same index $i$ or $j$) having
the same value of $y$.
Moreover, one has to impose the supertraceless condition
\be
\sum_i n_i y_i - \sum_j n_j y_j = 0
\ee
Then the decomposition of the adjoint is given by
\bea
{\bf Ad} = \left(\bigoplus_i n_i \r{i}(y_i) \bigoplus_j n_j \rpi{j}(y_j)
\right) \times \left(\bigoplus_i n_i \r{i}(-y_i) \bigoplus_j n_j \rpi{j}(-y_j)
\right) - \r{0}(0)
\ena
For an explicit calculation of this expression, one uses the fact that
\be
\left(\rule{.0mm}{5mm} n_i \r{i}(y_i)\right) \times
\left(\rule{.0mm}{5mm} n_j \r{j}(y_j)\right) =
n_i n_j \bigoplus_{k=|i-j|}^{i+j}
\r{k}(y_i+y_j) \mb{with $k$ integer and half-integer}
\ee
and the same formula for $\rpi{}$ representations.

\indent

In the case of the orthosymplectic superalgebras, one considers the following
decomposition of the $OSp(M|2n)$ fundamental representation:
\be
{\bf F} = \left(\bigoplus_i n_i \r{i} \right) \oplus
\left(\bigoplus_j n_j \rpi{j} \right)
\label{luc10}
\ee
which implies for the fundamental representations of $SO(M)$ and $Sp(2n)$:
\bea
\und{M} &=& \left(\bigoplus_i n_i \cd_i \right) \oplus \left(\bigoplus_j
n_j \cd_{j-1/2} \right) \nonumber \\
\und{2n} &=& \left(\bigoplus_i n_i \cd_{i-1/2} \right) \oplus
\left(\bigoplus_j n_j \cd_j \right)
\ena
For the $SO(M)$ part, one can introduce a non zero $U(1)$ eigenvalue $y_i$
only for representations $\cd_i$ with $i$ integer, which appear twice and only
twice. For the $Sp(2n)$ part, a non zero $U(1)$ eigenvalue $y_i$ is allowed
only for representations $\cd_i$ with $i$ half-integer, which appear twice and
only twice.

\noindent
For the superalgebra $\cg$ itself, one has to group the $Sl(2) \oplus U(1)$
representations $\cd_i(y_i)$ into $OSp(1|2) \oplus U(1)$ representations
$\crr_j(y_j) = \cd_j(y_j)\oplus\cd_{j-1/2}(y_j)$. Therefore, if the
decomposition of the $OSp(M|2n)$ fundamental representation ${\bf F}$ under
a certain $OSp(1|2)$ is given by (\ref{luc10}), non zero values $y$ of
the $U(1)$ factor are allowed for the following combinations:

\noindent
- the representation $\r{i}$ appears twice and only twice ($n_i = 2$), and
$i$ is integer,

\noindent
- the representation $\rpi{i}$ appears twice and only twice ($n_i = 2$), and
$i$ is half-integer.

\noindent
Moreover, $y$ can only take the values 0, 1/4 or 1/2 if $i \ne 0$ (which lead
to the values 0, $\pm 1/2$ or $\pm 1$ for the $U(1)$ factor in
the adjoint representation of $\cg$).
Finally, starting from a decomposition of the fundamental representation of
$OSp(M|2n)$ under $OSp(1|2) \oplus U(1)$ of the form
\be
{\bf F} = \left(\bigoplus_i \r{i}(y_i) \oplus \r{i}(-y_i) \right) \oplus
\left(\bigoplus_j \rpi{j}(y_j) \oplus \rpi{j}(-y_j) \right) \oplus
\left(\bigoplus_{i,n_i \ne 2} n_i \r{i}(0) \right) \oplus
\left(\bigoplus_{j,n_j \ne 2} n_j \rpi{j}(0) \right)
\ee
the decomposition of the adjoint is given by
\bea
{\bf Ad} &=&
\left.\left(\bigoplus_i \r{i}(y_i) \oplus \r{i}(-y_i) \bigoplus_{i,n_i \ne 2}
n_i \r{i}(0) \right) \times
\left(\bigoplus_i \r{i}(y_i) \oplus \r{i}(-y_i) \bigoplus_{i,n_i \ne 2}
n_i \r{i}(0) \right) \right|_A \nonumber \\
&\oplus & \left.
\left(\bigoplus_j \rpi{j}(y_j) \oplus \rpi{j}(-y_j) \bigoplus_{j,n_j \ne 2}
n_j \rpi{j}(0) \right) \times
\left(\bigoplus_j \rpi{j}(y_j) \oplus \rpi{j}(-y_j) \bigoplus_{j,n_j \ne 2}
n_j \rpi{j}(0) \right) \right|_S \nonumber \\
&\oplus &
\left(\bigoplus_i \r{i}(y_i) \oplus \r{i}(-y_i) \bigoplus_{i,n_i \ne 2}
n_i \r{i}(0)\right) \times \left(\bigoplus_j \rpi{j}(y_j) \oplus \rpi{j}(-y_j)
\bigoplus_{j,n_j \ne 2} n_j \rpi{j}(0)\right) \nonumber \\
&&
\ena

The (anti)symmetric products of $\r{}$ representations are given by the
formulae \ref{luc6} and (\ref{luc7}-\ref{luc9}) modulo the following
modifications due to the $U(1)$ eigenvalue:
\bea
\left(\rule{.0mm}{5mm} \r{i}(y_i) \oplus \r{i}(-y_i)\right) &\times &
\left(\rule{.0mm}{5mm} \r{i}(y_i) \oplus \r{i}(-y_i)\right) \left.\right|_A
= \left(\rule{.0mm}{5mm} \r{i} \times \r{i}\right)_A(2y_i) \oplus
\left(\rule{.0mm}{5mm} \r{i} \times \r{i}\right)_A(-2y_i)
\nonumber \\
&\oplus& (\r{i} \times \r{i})(0)
\ena
and
\bea
\left(\rpi{i}(y_i) \oplus \rpi{i} (-y_i)\right) &\times &
\left(\rpi{i}(y_i) \oplus \rpi{i}(-y_i)\right) \left.\right|_S
= \left(\rpi{i} \times \rpi{i}\right)_S(2y_i) \oplus
\left(\rpi{i} \times \rpi{i}\right)_S(-2y_i)
\nonumber \\
&\oplus& (\rpi{i} \times \rpi{i})(0)
\ena

\subsection{Superdefining vector}

\indent

The determination of the grading $H$ from the $OSp(1|2) \oplus U(1)$
decomposition of the fundamental representation is strictly the same as for the
algebras case. One just has to "double the calculation" since the bosonic part
of $\cg$ is in general the direct sum of two simple algebras. Using the same
basis for the Cartan
algebras (see section \ref{defV}), we will denote the defining vector as
\be
f=(f_1,...,f_n \ ; \ f'_1,...,f'_n)
\ee
where $f_i$ refers to the first simple algebra and the $f'_i$ to the second.
For example, for the case of $Sl(m|n)$ superalgebras, the contribution
of a representation is:
\bea
\crr_j(y) &\rightarrow&
(j+y,j-1+y,..,-j+y,0,..,0\ ;\ j-\half+y,j-\frac{3}{2}+y,..,-j+\half+y,0,..,0)
\nonumber\\
\crr^\pi_j(y) &\rightarrow&
(j-\half+y,j-\frac{3}{2}+y,..,-j+\half+y,0,..,0\ ;\ j+y,j-1+y,..,-j+y,0,..,0)
\nonumber
\ena
The other cases are analogous.

\sect{$W$ superalgebras from Lie superalgebras of rank up to 4}

\indent

In the following tables, we present an exhaustive classification of super
$W$ algebras arising from super Toda models based on classical superalgebras
of rank up to 4. The classification is listed in tables \ref{lucT0} to
\ref{lucT2}.

\indent

For the infinite series $\cg = A(m,n) = Sl(m+1|n+1)$ with $m \ne n$, $A(n,n) =
Sl(n+1|n+1)/U(1)$, $B(m,n) = OSp(2m+1|2n)$, $C(n+1) = OSp(2|2n)$ and $D(m,n) =
OSp(2m|2n)$, we give
the decomposition of the fundamental representation of $\cg$ with respect to
$OSp(1|2) \oplus U(1)$, the minimal (i.e. the lowest dimensional) regular
SSAs containing the $OSp(1|2)$ or (for the irregular cases) the
corresponding singular embedding. Then, we give the superspin content
with the same convention as for the bosonic tables. We recall that to a
$W_s$ superfield correspond two fields $w_s$ and $w_{s+1/2}$. When the
superspin is marked with a prime ($'$), the corresponding superfield $W_s$
has the "wrong" statistics (commuting fermions and anticommuting bosons).
In the same column, we give under the
superspin $s$ the hypercharge(s) $y$ when they exist.

\indent

For the two exceptional superalgebras $\cg = G(3)$ and $F(4)$, we give the
minimal regular SSA containing the $OSp(1|2)$ embedding, the
decomposition of the adjoint representation of $\cg$, and the superspin
content.

\begin{table}[p]
\begin{tabular}{|c|c|c|c|} \hline
$\cg$ &SSA &Decomposition of the &Superconformal spin \\
 &in $\cg$ &fundamental of $\cg$ &of the $W$ superfields \\
 &&&(Hypercharge) \\ \hline
 &&& \\
$A(0,1)$
 &$A(0,1)$ &$\rpi{1/2}$ &$\sm{3}{2}, 1$ \\
 &&& \\ \hline &&& \\
$A(0,2)$
 &$A(0,1)$ &$\rpi{1/2}(y) \oplus \rpi{0}(-y)$
 &$\begin{array}{c} \sm{3}{2}, 1, 1', 1', \sm{1}{2} \\ (0,0,2y,-2y,0)
 \end{array}$ \\
 &&& \\ \hline &&& \\
$A(1,1)$
 &$A(0,1)$ &$\rpi{1/2} \oplus \r{0}$ &$\sm{3}{2}, 1, 1, 1$ \\
 &&& \\ \hline &&& \\
$A(0,3)$
 &$A(0,1)$ &$\rpi{1/2}(y) \oplus 2\rpi{0}(-y/2)$
 &$\begin{array}{c} \sm{3}{2}, 1, 4\po1', 4\po\sm{1}{2} \\
 (0,0,\sm{3$y$}{2},\sm{3$y$}{2},\sm{-3$y$}{2},\sm{-3$y$}{2},0,0,0,0)
\end{array}$ \\
 &&& \\ \hline &&& \\
$A(1,2)$
 &$A(1,2)$ &$\rpi{1}$ &$\sm{5}{2}, 2, \sm{3}{2}, 1$ \\
 &&& \\
 &$A(0,1)$
 &$\begin{array}{c} \rpi{1/2}(0) \\ \oplus \r{0}(y) \oplus \rpi{0}(y)
 \end{array}$
 &$\begin{array}{c} \sm{3}{2}, 1, 1, 1, 1', 1', \sm{1}{2}, \sm{1}{2},
 \sm{1}{2}', \sm{1}{2}' \\
 (0,0,\sm{$y$}{2},\sm{-$y$}{2},\sm{$y$}{2},\sm{-$y$}{2},
 0,0,0,0) \end{array}$ \\
 &&& \\
 &$A(1,0)$ &$\r{1/2}(y) \oplus 2\rpi{0}(y/2)$
 &$\begin{array}{c} \sm{3}{2}, 5\po1, 4\po\sm{1}{2} \\
 (0,0,\sm{$y$}{2},\sm{$y$}{2},\sm{-$y$}{2},\sm{-$y$}{2},0,0,0,0) \end{array}$
\\
 &&& \\ \hline
\end{tabular}
\caption{ $A(m,n)$ superalgebras up to rank 4. \label{lucT0}}
\end{table}

\clearpage
\begin{table}[p]
\begin{tabular}{|c|c|c|c|} \hline
$\cg$ &SSA &Decomposition of the &Superconformal spin \\
 &in $\cg$ &fundamental of $\cg$ &of the $W$ superfields \\
 &&&(Hypercharge) \\ \hline
 &&& \\
$B(0,2)$
 &$B(0,1)$ &$\rpi{1/2} \oplus 2\rpi{0}$
 &$\sm{3}{2}, 1', 1', \sm{1}{2}, \sm{1}{2}, \sm{1}{2}$ \\
 &&& \\ \hline &&& \\
$B(1,1)$
 &$B(1,1)$ &$\r{1}$ &$2, \sm{3}{2}$ \\
 &&& \\
 &$\left.\begin{array}{c} C(2) \\ B(0,1) \end{array} \right\}$
 &$\rpi{1/2} \oplus \r{0}(y) \oplus \r{0}(-y)$
 &$\begin{array}{c} \sm{3}{2}, 1, 1, \sm{1}{2} \\ (0,y,-y,0) \end{array}$ \\
 &&& \\ \hline &&& \\
$B(0,3)$
 &$B(0,1)$ &$\rpi{1/2} \oplus 4\rpi{0}$
 &$\sm{3}{2}, 1', 1', 1', 1', 10 \po \sm{1}{2}$ \\
 &&& \\ \hline &&& \\
$B(1,2)$
 &$B(1,2)$
 &$\rpi{3/2}$ &$\sm{7}{2}, 2, \sm{3}{2}$ \\
 &&& \\
 &$B(1,1)$
 &$\r{1} \oplus 2\rpi{0}$ &$2, \sm{3}{2}, \sm{3}{2}', \sm{3}{2}', \sm{1}{2},
 \sm{1}{2}, \sm{1}{2}$ \\
 &&& \\
 &$\left.\begin{array}{c} C(2) \\ B(0,1) \end{array} \right\}$
 &$\begin{array}{c} \rpi{1/2} \oplus \r{0}(y) \oplus \r{0}(-y) \\ \oplus
 2\rpi{0} \end{array}$
 &$\begin{array}{c} \sm{3}{2}, 1, 1, 1', 1', 4 \po \sm{1}{2}, 4 \po \sm{1}{2}'
 \\ (0,y,-y,6 \po 0,y,y,-y,-y) \end{array}$ \\
 &&& \\
 &$\left.\begin{array}{c} C(2) \oplus B(0,1) \\ A(0,1) \end{array} \right\}$
 &$\rpi{1/2}(y) \oplus \rpi{1/2}(-y) \oplus \r{0}$
 &$\begin{array}{c} \sm{3}{2},\sm{3}{2}, \sm{3}{2}, 1, 1, 1, \sm{1}{2}
 \\ (2y,-2y,0,y,-y,0,0) \end{array}$ \\
 &&& \\ \hline &&& \\
$B(2,1)$
 &$\left.\begin{array}{c} D(2,1) \\ B(1,1) \end{array} \right\}$
 &$\r{1} \oplus \r{0}(y) \oplus \r{0}(-y)$
 &$\begin{array}{c} 2, \sm{3}{2}, \sm{3}{2}, \sm{3}{2}, \sm{1}{2} \\
 (0,2y,-2y,0,0) \end{array}$ \\
 &&& \\
 &$\left.\begin{array}{c} C(2) \\ B(0,1) \end{array} \right\}$
 &$\rpi{1/2} \oplus 4\r{0}$ &$\sm{3}{2}, 4 \po 1, 6 \po \sm{1}{2}$ \\
 &&& \\
 &$A(1,0)$ &$2\r{1/2} \oplus \r{0}$
 &$\sm{3}{2}, 1, 1, 1, 1', 1', \sm{1}{2}, \sm{1}{2}, \sm{1}{2}$ \\
 &&& \\ \hline
\end{tabular}
\caption{ $B(m,n)$ superalgebras of rank 2 and 3.}
\end{table}

\clearpage
\pagestyle{empty}
\begin{table}[p]
\begin{tabular}{|c|c|c|c|} \hline
$\cg$ &SSA &Decomposition of the &Superconformal spin
\\
 &in $\cg$ &fundamental of $\cg$ &of the $W$ superfields \\
 &&&(Hypercharge) \\ \hline
 &&& \\
$B(0,4)$
 &$B(0,1)$ &$\rpi{1/2} \oplus 6\rpi{0}$
 &$\sm{3}{2}, 6 \po 1', 21 \po \sm{1}{2}$ \\
 &&& \\ \hline &&& \\
$B(1,3)$
 &$B(1,2)$ &$\rpi{3/2} \oplus 2\rpi{0}$
 &$\sm{7}{2}, 2, 2', 2', \sm{3}{2}, \sm{1}{2}, \sm{1}{2}, \sm{1}{2}$ \\
 &&& \\
 &$B(1,1)$ &$\r{1} \oplus 4\rpi{0}$ &$2, \sm{3}{2}, \sm{3}{2}', \sm{3}{2}',
 \sm{3}{2}', \sm{3}{2}', 10 \po \sm{1}{2}$ \\
 &&& \\
 &$\left.\begin{array}{c} C(2) \oplus B(0,1) \\ A(0,1) \end{array} \right\}$
 &$\begin{array}{c} \rpi{1/2}(y) \oplus \rpi{1/2}(-y) \\ \oplus \r{0}
 \oplus 2\rpi{0} \end{array}$
 &$\begin{array}{c} 3 \po \sm{3}{2}, 1, 1', 1', 1, 1', 1', 1, 4 \po \sm{1}{2},
 2 \po \sm{1}{2}' \\ (2y,-2y,0,3 \po y, 3 \po -y, 7 \po 0) \end{array}$ \\
 &&& \\
 &$\left.\begin{array}{c} C(2) \\ B(0,1) \end{array} \right\}$
 &$\begin{array}{c} \rpi{1/2} \oplus 4\rpi{0} \\ \oplus \r{0}(y) \oplus
 \r{0}(-y) \end{array}$
 &$\begin{array}{c} \sm{3}{2}, 2 \po 1, 4 \po 1', 11 \po \sm{1}{2},
 9 \po \sm{1}{2}' \\ (0,y,-y,15 \po 0,4 \po y,4 \po -y,0) \end{array}$ \\
 &&& \\ \hline &&& \\
$B(2,2)$
 &$B(2,2)$ &$\r{2}$ &$4, \sm{7}{2}, 2, \sm{3}{2}$ \\
 &&& \\
 &$\left.\begin{array}{c} D(2,2) \\ B(1,2) \end{array} \right\}$
 &$\rpi{3/2} \oplus \r{0}(y) \oplus \r{0}(-y)$
 &$\begin{array}{c} \sm{7}{2}, 2, 2, 2, \sm{3}{2}, \sm{1}{2} \\
 (0,y,-y,0,0,0) \end{array}$ \\
 &&& \\
 &$\left.\begin{array}{c} D(2,1) \\ B(1,1) \end{array} \right\}$
 &$\begin{array}{c} \r{1} \oplus 2 \rpi{0} \\ \oplus \r{0}(y) \oplus \r{0}(-y)
 \end{array}$
 &$\begin{array}{c} 2, \sm{3}{2}, \sm{3}{2}, \sm{3}{2}, \sm{3}{2}',
 \sm{3}{2}', 4 \po \sm{1}{2}, 4 \po \sm{1}{2}' \\
 (0,2y,-2y,0,6 \po 0,y,y,-y,-y) \end{array}$ \\
 &&& \\
 &$\left.\begin{array}{c} D(2,1) \oplus B(0,1) \\ B(1,1) \oplus C(2)
 \end{array} \right\}$
 &$\r{1} \oplus \rpi{1/2} \oplus \r{0}$ &$2, 2, \sm{3}{2}, \sm{3}{2},
 \sm{3}{2}, \sm{3}{2}, 1, 1$ \\
 &&& \\
 &$\left.\begin{array}{c} C(2) \oplus C(2) \\ A(0,1) \end{array}
 \right\}$
 &$\begin{array}{c} \rpi{1/2}(y) \oplus \rpi{1/2}(-y) \\ \oplus 3\r{0}
 \end{array}$
 &$\begin{array}{c} \sm{3}{2}, \sm{3}{2}, \sm{3}{2}, 7 \po 1, \sm{1}{2},
 \sm{1}{2}, \sm{1}{2}, \sm{1}{2} \\ (2y,-2y,0,3 \po y,3 \po -y,0,4 \po 0)
 \end{array}$ \\
 &&& \\
 &$\left.\begin{array}{c} C(2) \\ B(0,1) \end{array} \right\}$
 &$\rpi{1/2} \oplus 4\r{0} \oplus 2\rpi{0}$ &$\sm{3}{2}, 4 \po 1, 2 \po 1',
 9 \po \sm{1}{2}, 8 \po \sm{1}{2}'$ \\
 &&& \\
 &$A(1,0)$ &$2\r{1/2} \oplus \r{0} \oplus 2\rpi{0}$ &$\sm{3}{2}, 7\po 1,
 1', 1', 6\po \sm{1}{2}, 2 \po \sm{1}{2}'$ \\
 &&& \\ \hline &&& \\
$B(3,1)$
 &$\left.\begin{array}{c} D(2,1) \\ B(1,1) \end{array} \right\}$
 &$\r{1} \oplus 4\r{0}$ &$2, 5 \po \sm{3}{2}, 6 \po \sm{1}{2}$ \\
 &&& \\
 &$\left.\begin{array}{c} C(2) \\ B(0,1) \end{array} \right\}$
 &$\rpi{1/2} \oplus 6\r{0}$ &$\sm{3}{2}, 6 \po 1, 15 \po \sm{1}{2}$ \\
 &&& \\
 &$A(1,0)$ &$2\r{1/2} \oplus 3\r{0}$
 &$\sm{3}{2}, 3 \po 1, 6 \po 1', 6 \po \sm{1}{2}$ \\
 &&& \\ \hline
\end{tabular}
\caption{ $B(m,n)$ superalgebras of rank 4.}
\end{table}

\clearpage
\pagestyle{plain}
\begin{table}[p]
\begin{tabular}{|c|c|c|c|} \hline
$\cg$ &SSA &Decomposition of the &Superconformal spin
\\
 &in $\cg$ &fundamental of $\cg$ &of the $W$ superfields \\
 &&&(Hypercharge) \\ \hline
 &&& \\
$D(2,1)$
 &$D(2,1)$ &$\r{1} \oplus \r{0}$ &$2, \sm{3}{2}, \sm{3}{2}$ \\
 &&& \\
 &$C(2)$ &$\rpi{1/2} \oplus 3\r{0}$
 &$\sm{3}{2}, 1, 1, 1, \sm{1}{2}, \sm{1}{2}, \sm{1}{2}$ \\
 &&& \\
 &$A(1,0)$ &$2\r{1/2}$
 &$\sm{3}{2}, 1, 1, 1, \sm{1}{2}, \sm{1}{2}, \sm{1}{2}$ \\
 &&& \\ \hline &&& \\
$D(2,2)$
 &$D(2,2)$ &$\rpi{3/2} \oplus \r{0}$ &$\sm{7}{2}, 2, 2, \sm{3}{2}$ \\
 &&& \\
 &$D(2,1)$ &$\r{1} \oplus \r{0} \oplus 2\rpi{0}$
 &$2, \sm{3}{2}, \sm{3}{2}, \sm{3}{2}', \sm{3}{2}', 5\po\sm{1}{2}$ \\
 &&& \\
 &$C(2)$ &$\rpi{1/2} \oplus 3\r{0} \oplus 2\rpi{0}$
 &$\sm{3}{2}, 1, 1, 1, 1', 1', 6 \po \sm{1}{2}, 6 \po \sm{1}{2}'$ \\
 &&& \\
 &$\left.\begin{array}{c} C(2) \oplus C(2) \\ A(0,1) \end{array} \right\}$
 &$\rpi{1/2} \oplus \r{0}(y) \oplus \r{0}(-y)$
 &$\begin{array}{c} \sm{3}{2}, \sm{3}{2}, \sm{3}{2}, 5\po1, \sm{1}{2},
 \sm{1}{2} \\ (3 \po 0,y,-y,5 \po 0) \end{array}$ \\
 &&& \\
 &$B(1,1) \oplus B(0,1)$
 &$\r{1} \oplus \rpi{1/2}$ &$2, 2, \sm{3}{2}, \sm{3}{2}, \sm{3}{2}, 1$ \\
 &&& \\
 &$A(1,0)$ &$2\r{1/2} \oplus 2\rpi{0}$ &$\sm{3}{2}, 7 \po 1, 6 \po \sm{1}{2}$
\\
 &&& \\ \hline &&& \\
$D(3,1)$
 &$D(2,1)$ &$\r{1} \oplus 3\r{0}$
 &$2, 4\po\sm{3}{2}, \sm{1}{2}, \sm{1}{2}, \sm{1}{2}$ \\
 &&& \\
 &$C(2)$ &$\rpi{1/2} \oplus 5\r{0}$ &$\sm{3}{2}, 5 \po 1, 10 \po \sm{1}{2}$ \\
 &&& \\
 &$A(1,0)$ &$2\r{1/2} \oplus 2\r{0}$
 &$\sm{3}{2}, 1, 1, 1, 4\po1', 4 \po \sm{1}{2}$ \\
 &&& \\ \hline
\end{tabular}
\caption{ $D(m,n)$ superalgebras up to rank 4.}
\end{table}

\clearpage
\begin{table}[p]
\begin{tabular}{|c|c|c|c|} \hline
$\cg$ &SSA &Decomposition of the &Superconformal spin
\\
 &in $\cg$ &fundamental of $\cg$ &of the $W$ superfields \\
 &&&(Hypercharge) \\ \hline
 &&& \\
$C(3)$
 &$A(0,1)$ &$\rpi{1/2}(y) \oplus \rpi{1/2}(-y)$
 &$\begin{array}{c} \sm{3}{2}, \sm{3}{2}, \sm{3}{2}, 1, \sm{1}{2} \\
 (0,2y,-2y,0,0) \end{array}$ \\
 &&& \\
 &$C(2)$ &$\rpi{1/2} \oplus \r{0} \oplus 2\rpi{0}$
 &$\sm{3}{2}, 1, 1', 1', \sm{1}{2}, \sm{1}{2}, \sm{1}{2}, \sm{1}{2}',
 \sm{1}{2}'$ \\
 &&& \\ \hline &&& \\
$C(4)$
 &$A(0,1)$
 &$\rpi{1/2}(y) \oplus \rpi{1/2}(-y) \oplus 2\rpi{0}$
 &$\begin{array}{c} 3 \po \sm{3}{2}, 1, 4\po1', 4 \po \sm{1}{2} \\
 (0,2y,-2y,0,y,y,-y,-y,4 \po 0) \end{array}$ \\
 &&& \\
 &$C(2)$ &$\rpi{1/2} \oplus \r{0} \oplus 4\rpi{0}$
 &$\sm{3}{2}, 1, 4\po1', 10 \po \sm{1}{2}, 4 \po \sm{1}{2}'$ \\
 &&& \\ \hline
\end{tabular}
\caption{ $C(n+1)$ superalgebras up to rank 4.}
\end{table}

\begin{table}[p]
\begin{tabular}{|c|c|c|} \hline
SSA & $OSp(1|2)$ decomposition & Superconformal spin \\
 &of $G(3)$ & of the $W$ superfields \\ \hline
 && \\
$A(1,0)$ &$\crr_1 \oplus 3\crr_{1/2} \oplus 4\crr'_{1/2} \oplus 3\crr_0
 \oplus 2\crr'_0$
 &$\sm{3}{2}, 1, 1, 1, 4\po1', \sm{1}{2}, \sm{1}{2}, \sm{1}{2},
 \sm{1}{2}', \sm{1}{2}'$ \\
 && \\ \hline && \\
$A(1,0)'$ &$2\crr_{3/2}' \oplus \crr_1 \oplus 3\crr_{1/2} \oplus 3\crr_0$
 &$2', 2', \sm{3}{2}, 1, 1, 1, \sm{1}{2}, \sm{1}{2}, \sm{1}{2}$ \\
 && \\ \hline && \\
$B(0,1)$ &$\crr_1 \oplus 6\crr_{1/2} \oplus 8\crr_0$
 &$\sm{3}{2}, 6 \po 1, 8 \po \sm{1}{2}$ \\
 && \\ \hline && \\
$B(1,1)$ &$\crr_{3/2} \oplus 2\crr'_{3/2} \oplus \crr_1 \oplus 3\crr_0 \oplus
 2\crr'_0$
 &$2, 2', 2', \sm{3}{2}, \sm{1}{2}, \sm{1}{2}, \sm{1}{2}, \sm{1}{2}',
 \sm{1}{2}'$
 \\ && \\ \hline && \\
$D(2,1;3)$ &$\crr_2 \oplus \crr_{3/2} \oplus 3\crr_1$
 &$\sm{5}{2}, 2, \sm{3}{2}, \sm{3}{2}, \sm{3}{2}$ \\ && \\ \hline
\end{tabular}
\caption{ The exceptional superalgebra $G(3)$. \label{lucT1}}
\end{table}

\clearpage
\begin{table}[p]
\begin{tabular}{|c|c|c|} \hline
SSA & $OSp(1|2)$ decomposition & Superconformal spin \\
 &of $F(4)$ & of the $W$ superfields \\ \hline
 && \\
$A(1,0)$ &$\crr_1 \oplus 7\crr_{1/2} \oplus 14\crr_0$
 &$\sm{3}{2}, 7 \po 1, 14 \po \sm{1}{2}$ \\
 && \\ \hline && \\
$A(0,1)$ &$\crr_1 \oplus 3\crr_{1/2} \oplus 6\crr'_{1/2} \oplus 6 \crr_0
 \oplus 2\crr'_0$
 &$\sm{3}{2}, 3 \po 1, 6 \po 1', 6 \po \sm{1}{2}, \sm{1}{2}', \sm{1}{2}'$ \\
 && \\ \hline && \\
$C(2)$ &$5\crr_1 \oplus 3\crr_{1/2} \oplus 6\crr_0$
 &$5 \po \sm{3}{2}, 3 \po 1, 6 \po \sm{1}{2}$ \\
 && \\ \hline && \\
$D(2,1;2)$ &$\crr_{3/2} \oplus 2\crr'_{3/2} \oplus 2\crr_1 \oplus 2\crr'_{1/2}
 \oplus 3\crr_0$
 &$2, 2', 2', \sm{3}{2}, \sm{3}{2}, 1', 1', \sm{1}{2}, \sm{1}{2}, \sm{1}{2}$
 \\ && \\ \hline
\end{tabular}
\caption{ The exceptional superalgebra $F(4)$.}
\end{table}

\begin{table}[p]
\begin{tabular}{|c|c|c|} \hline
SSA &Decomposition of the &Superconformal spin \\
 &fundamental of $D(2,1;\alpha)$ &of the $W$ superfields \\
 && \\ \hline && \\
$D(2,1)$ &$\r{1} \oplus \r{0}$ &$2, \sm{3}{2}, \sm{3}{2}$ \\
 && \\ \hline && \\
$C(2)$ &$\rpi{1/2} \oplus 3\r{0}$
 &$\sm{3}{2}, 1, 1, 1, \sm{1}{2}, \sm{1}{2}, \sm{1}{2}$ \\
 && \\ \hline && \\
$A(1,0)$ &$2\r{1/2}$
 &$\sm{3}{2}, 1, 1, 1, \sm{1}{2}, \sm{1}{2}, \sm{1}{2}$ \\
 && \\ \hline
\end{tabular}
\caption{ The exceptional superalgebra $D(2,1;\alpha)$. \label{lucT2}}
\end{table}

\clearpage
\sect{Quadratic-, Quasi- and $\Z_2\times\Z_2$- Superconformal
algebras}

\indent

We have a natural framework to study superconformal algebras. Let us
first recall that a quadratic-superconformal algebra is a Zamolodchikov
superalgebra made of one spin 2 field corresponding to $T(x)$ (and forming a
Virasoro algebra), $N$ fermionic supersymmetry charges $G^\alpha (x)$ which
are spin $\sm{3}{2}$ primary fields with respect to $T(x)$, and a Kac-Moody
(KM) algebra (i.e. spin 1 primary fields). The spin $\sm{3}{2}$ generators are
required to form a representation of the KM algebra, but the
quadratic-superconformal superalgebra is
not (in general) a Lie superalgebra in the sense that the PB $\{ G^\alpha (x),
G^\beta (x') \}_{PB}$ contains quadratic terms in the KM currents
\cite{quasic,sca1}.

The "usual" superconformal algebras, i.e. the Ademollo et al. algebras
\cite{Adem} and the one parameter algebra found in \cite{Troost},
are the only closed Lie superconformal algebras we know.
We will refer to them as {\em Lie} superconformal algebras and called the
corresponding supersymmetries "true" supersymmetries.

The same definition holds for a quasi-superconformal algebra \cite{quasic},
except that its spin $\sm{3}{2}$ fields $G^\alpha (x)$ are bosonic ("wrong"
statistics). As an example, the algebra explicited in section \ref{Wber},
possessing two spin-$\sm{3}{2}$ and one spin-1 fields, is quasi-superconformal.

An algebra with both bosonic and fermionic spin $\sm{3}{2}$ currents is called
$\Z_2\times \Z_2$ superconformal algebra. In that case, spin 1 fermions may
also appear.

\indent

It should be clear to the reader that the part \ref{part1} contains all the
tools necessary for the determination of the quasi-superconformal algebras,
whereas the quadratic and $\Z_2 \times \Z_2$
superconformal algebras can be obtained from
the part \ref{part2}. Note however, that the supersymmetric treatment we have
used (and which makes naturally appear a $N=1$ Lie superconformal algebra)
leads to the emergence of spin $\half$ fields. As it is now well-known, to
avoid these fermions, one can factorize them \cite{ferm}. These algebras
(without spin $\half$ fermions) have already been classified at the quantum
level in \cite{quasic}. We show hereafter that all the algebras of
\cite{quasic} can be realized at the classical level as symmetries of Toda
models. Moreover, two new (with respect to \cite{quasic}) $\Z_2 \times \Z_2$
superconformal algebras can be identified from the study of $G(3)$ and $F(4)$.

\subsection{Quasi-superconformal algebras}

\indent

{}From the study of part \ref{part1}, we can see that such algebras, with only
one spin 2 and no spin $s > 2$, are obtained when the fundamental
representation of $Sl(n)$ and $Sp(2n)$ (resp. $SO(n)$) algebras contains only
one (resp. two) $\cd_{1/2}$ representation(s). This means that we are reducing
these Lie
algebras with respect to a regular $A_1$. Using the results of part \ref{part1}
and \cite{slan} for the exceptional algebras $E_{6,7,8}$, we obtain the
classification of table \ref{d1}.

\subsection{Quadratic-superconformal algebras}

\indent

They are obtained from the reduction of a
superalgebra with respect to an $OSp(1|2)$ SSA. Note that "wrong"
statistic superfields may appear and lead to $\Z_2 \times \Z_2$ superconformal
algebras. From
the rules given in section \ref{pouf}, relating $\crr'$ representations of the
adjoint, to $\r{}$ and $\rpi{}$ representations of the fundamental, it is easy
to compute the allowed reductions. As an example, let us study the $Sl(m|n)$
algebras: the reduction with respect to $Sl(1|2)$ reads $\underline{n+m} =
\crr^\pi_{1/2} + (m-1)\crr_0 + (n-2) \crr^\pi_0$,
so that we must set $n=2$ to avoid
"wrong" statistics. Thus, only the $Sl(n|2)$ (or $Sl(2|n)$) algebra leads to
quadratic-superconformal algebras. The same calculation leads to the list:
\be
Sl(n|2) \mb{,} OSp(4|2n)\mb{,} OSp(n|2) \mb{,} F(4)\mb{,} G(3)
\ee

We summarize the results in the table \ref{aux1}. Note that the regular
superalgebra which characterizes the $OSp(1|2)$, provides the number $N_0$
of "true"
supersymmetries of the $W$ algebra: $N_0=1$ for a regular $OSp(1|2)$,
$N_0=2$ for the superprincipal $OSp(1|2)$ of $Sl(1|2)$ and $OSp(2|2)$,
$N_0=3$ if the previous $Sl(1|2)$ or $OSp(2|2)$ can be
embedded in an $OSp(3|2)$
SSA, and $N_0=4$ if the $Sl(2|1)$ or $OSp(2|2)$ is contained in
$OSp(4|2)$ or $D(2,1;\alpha)$ SSAs.

\subsection{$\Z_2 \times \Z_2$ superconformal algebras}

\indent

Their classification is easily deduced from the previous section.
We begin with the $\Z_2\times\Z_2$ superconformal algebras that do not
contain superspin $\half$ bosons, so that we can define a (right statistic)
super-KM algebra. These algebras are listed in table \ref{d2}.

\indent

If now one introduces the superspin $\half$ bosons, the number of allowed
superalgebras is much larger. In fact, in accordance with \cite{quasic}, we
find one (resp. two) $\Z_2\times\Z_2$ superconformal algebras from each
$A(m,n)$
and $C(n+1)$ (resp. $B(m,n)$ and $D(m,n)$) superalgebras. However, for $F(4)$
and $G(3)$, we find two new $\Z_2\times\Z_2$ superconformal algebras, different
from the two quadratic-superconformal algebras of \cite{quasic},
already listed in
table \ref{aux1}. This seems to indicate that these two algebras exist only at
the classical level. The results are summarized in table \ref{d3}.

\clearpage

\begin{table}[p]
\begin{tabular}{|c|c|c|c|} \hline
Algebra &Decomposition of &Conformal spin &Residual Kac-Moody \\
$\cg$ &the fundamental of $\cg$ &of the $W$ generators &algebra \\ \hline
 &&& \\
$Sl(n)$ &$\und{n} = \cd_{1/2} \oplus (n-2) \cd_0$
 &$\begin{array}{c} 2,\ 2(n-2) \po \sm{3}{2}, \\ (n-2)^2 \po 1 \end{array}$
 &$Sl(n-2) \oplus U(1)$ \\ &&& \\ \hline &&& \\
$SO(n)$ &$\und{n} = 2\cd_{1/2} + (n-4)\cd_0$
 &$\begin{array}{c} 2,\ 2(n-4) \po \sm{3}{2}, \\
{[\sm{$(n-4)(n-5)$}{2}+3]} \po 1
 \end{array}$
 &$SO(n-4) \oplus Sl(2)$ \\ &&& \\ \hline &&& \\
$Sp(2n)$ &$\und{2n} = \cd_{1/2} + (2n-2)$
 &$\begin{array}{c} 2,\ (2n-4) \po \sm{3}{2}, \\ (n-2)(2n-3) \po 1 \end{array}$
 &$Sp(2n-2)$ \\ &&& \\ \hline &&& \\
$G_2$ &$\und{7} = 2\cd_{1/2} + 3\cd_0$
 &$2, \sm{3}{2}, \sm{3}{2}, \sm{3}{2}, \sm{3}{2}, 1, 1, 1$
 &$Sl(2)$ \\ &&& \\ \hline &&& \\
$F_4$ &$\und{26} = 10\cd_{1/2} + 6\cd_0$
 &$2, 18 \po \sm{3}{2}, 13 \po 1$ &$Sp(6)$ \\ &&& \\ \hline &&& \\
$E_6$ &$\und{27} = 6\cd_{1/2} + 15\cd_0$
 &$2, 20 \po \sm{3}{2}, 35 \po 1$ &$Sl(6)$ \\ &&& \\ \hline &&& \\
$E_7$ &$\und{56} = 12\cd_{1/2} + 32\cd_0$
 &$2, 32 \po \sm{3}{2}, 66 \po 1$ &$SO(12)$ \\ &&& \\ \hline &&& \\
$E_8$ &$\und{248} = \cd_1 + 56\cd_{1/2} + 133\cd_0$
 &$2, 56 \po \sm{3}{2}, 133 \po 1$ &$E_7$ \\ &&& \\ \hline
\end{tabular}
\caption{ Classification of quasi-superconformal algebras.\label{d1}}
\end{table}

\clearpage

\begin{table}[p]
\begin{tabular}{|c|c|c|c|c|} \hline
&Min. includ. &$N_0$ &Superconformal spin
&Super KM \\
$\cg$ &regular SSA &&of the $W$ generators
&algebra \\ \hline
 &&&& \\
$A(1,n)$ &$A(1,0)$ &2
 &$\sm{3}{2}, (2n+1) \po 1, n^2 \po \sm{1}{2}$ &$A_{n-1} \oplus U(1)$ \\
 &&&& \\ \hline &&&& \\
$D(2,n)$ &$A(1,0)$ &4
 &$\sm{3}{2}, (4n-1) \po 1, [(n-1)(2n-1)+3] \po \sm{1}{2}$
 &$C_{n-1}\oplus3U(1)$ \\
 &&&& \\ \hline &&&& \\
$D(m,1)$ &$C(2)$ & 4
 &$\sm{3}{2}, (2m-1) \po 1, (m-1)(2m-1) \po \sm{1}{2}$ &$B_{m-1}$ \\
 &&&& \\ \hline &&&& \\
$B(m,1)$ &$C(2)$ & $\left\{\begin{array}{c} 3\ (m=1) \\ 4\
(m>1)\end{array}\right.$
 &$\sm{3}{2}, 2m \po 1, m(2m-1) \po \sm{1}{2}$ &$D_{m}$ \\
 &&&& \\ \hline &&&& \\
$G(3)$ &$B(0,1)$ &1
 &$\sm{3}{2}, 6 \po 1, 8 \po \sm{1}{2}$ &$A_2$ \\
 &&&& \\ \hline &&&& \\
$F(4)$ &$A(1,0)$ &2
 &$\sm{3}{2}, 7 \po 1, 14 \po \sm{1}{2}$ &$G_2$
 \\ &&&& \\ \hline
\end{tabular}
\caption{ Quadratic-superconformal algebras.\label{aux1}}
\end{table}

\begin{table}[p]
\begin{tabular}{|c|c|c|c|c|} \hline
&Min. includ. &$N_0$ &Superconformal spin & Super KM\\
$\cg$ &regular SSA &&of the $W$ generators & algebra \\ \hline
 &&&& \\
$A(1,n)$ &$A(1,0)$ &2
 &$\sm{3}{2}, 1, 2n \po 1', n^2 \po \sm{1}{2}$ & $A_{n-1}\oplus U(1)$
 \\ &&&& \\ \hline &&&& \\
$D(m,1)$ &$A(1,0)$ &4
 &$\sm{3}{2}, 3\po 1, 4(m-2) \po 1', [(m-2)(2m-5)+3] \po \sm{1}{2}$ &
$D_m\oplus3U(1)$
 \\ &&&& \\ \hline &&&& \\
$B(m,1)$ &$A(1,0)$ & 4
 &$\sm{3}{2},3\po1, 2(2m-3) \po 1', [(m-2)(2m-3)+3] \po \sm{1}{2}$ &
$B_{m-2}\oplus3U(1)$
 \\ &&&& \\ \hline &&&& \\
$B(0,n)$ &$B(0,1)$ &1
 &$\sm{3}{2}, (2n-2) \po 1', (n-1)(2n-1) \po \sm{1}{2}$ & $B_{n-1}$
 \\ &&&& \\ \hline
\end{tabular}
\caption{ $\Z_2 \times \Z_2$ superconformal algebras (no superspin $\half$
bosonic superfield).\label{d2}}
\end{table}

\begin{table}[p]
\begin{tabular}{|c|c|c|c|} \hline
&Min. includ. &$N_0$ &Superconformal spin \\
$\cg$ &regular SSA &&of the $W$ generators \\ \hline
 &&& \\
$A(m,n)$ &$A(1,0)$ &2
 &$\begin{array}{c} \sm{3}{2},\ (2m-1) \po 1,\ 2n \po 1', \\
 {[(m-1)^2 + n^2]} \po \sm{1}{2}, 2(m-1)n \po \sm{1}{2}' \end{array}$
 \\ &&& \\ \hline &&& \\
$D(m,n)$ &$C(2)$ & 4
 &$\begin{array}{c} \sm{3}{2},\ (2m-1) \po 1,\ (2n-2) \po 1', \\
 {[(m-1)(2m-1)+(n-1)(2n-1)]} \po \sm{1}{2}, \\ (2m-1)(2n-2) \po \sm{1}{2}'
 \end{array}$
 \\ &&& \\
 &$A(1,0)$ & 4
 &$\begin{array}{c} \sm{3}{2},\ (4n-1) \po 1,\ 4(m-2) \po 1',
 \\ {[(m-2)(2m-5) + (n-1)(2n-1) + 3]} \po \sm{1}{2},
 \\ 4(m-2)(n-1) \po \sm{1}{2}' \end{array}$
 \\ &&& \\ \hline &&& \\
$B(m,n)$ &$C(2)$ & $\left\{\begin{array}{c} 3\ (m=1) \\ 4\
(m>1)\end{array}\right.$
 &$\begin{array}{c} \sm{3}{2},\ 2m \po 1,\ (2n-2) \po 1', \\
 {[m(2m-1)+(n-1)(2n-1)]} \po \sm{1}{2}, \\ 4m(n-1) \po \sm{1}{2}' \end{array}$
 \\ &&& \\
 &$A(1,0)$ & 4
 &$\begin{array}{c} \sm{3}{2},\ (4n-1) \po 1,\ 2(2m-3) \po 1',
 \\ {[(m-2)(2m-3) + (n-1)(2n-1) + 3]} \po \sm{1}{2},
 \\ 2(2m-3)(n-1) \po \sm{1}{2}' \end{array}$
 \\ &&& \\ \hline &&& \\
$C(n+1)$ &$C(2)$ &2
 &$\begin{array}{c} \sm{3}{2},\ 1,\ (2n-2) \po 1', \\
 (n-1)(2n-1) \po \sm{1}{2}, (2n-2) \po \sm{1}{2}' \end{array}$
 \\ &&& \\ \hline &&& \\
$G(3)$ &$A(1,0)$ & 4
 &$\sm{3}{2}, 3\po1, 4\po1', 3\po\sm{1}{2}, 2\po\sm{1}{2}'$
 \\ &&& \\ \hline &&& \\
$F(4)$ &$A(0,1)$ & 4
 &$\sm{3}{2}, 3 \po 1, 6 \po 1', 6 \po \sm{1}{2}, 2 \po \sm{1}{2}'$
 \\ &&& \\ \hline
\end{tabular}
\caption{ $\Z_2 \times \Z_2$ superalgebras (with superspin $\half$ bosons).
\label{d3}}
\end{table}

\clearpage

\sect{Conclusion}

\indent

In the classification we have obtained, each $W$ (super)algebra is
characterized by its (super)conformal spin content and the couple $(Sl(2),\cg)$
if $\cg$ is a simple Lie algebra, respectively $(OSp(1|2),\cg)$ if $\cg$ is a
Lie superalgebra. The PB of the corresponding $W$ (super)algebra can
then be determined via the general method reminded in section \ref{sec2.1}.
However,
rather important simplifications occur when the $U(1)$ factor commuting with
$Sl(2)$, resp. $OSp(1|2)$, exists: the admitted $Y$ values are also provided
in our tables.

\indent

It has seemed to us necessary to reconsider in a first step the problem of the
$Sl(2)$ subalgebras in a simple Lie algebra $\cg$, in order to make explicit
our results in the algebraic case, and also to propose the generalization we
have obtained for the supersymmetric one.
We hope that the tables in which are gathered our results are presented in an
enough convenient way to allow a direct use. This has been at least the case
for us to easily recognize the superconformal algebras of \cite{quasic}.

\indent

Among the different problems one can immediately think of, an urgent one is of
course the quantum case. Some interesting works \cite{sca1,sca2,sca3,sca4,sca5}
already exist, but a general
treatment would be necessary. Another question we wish we could answer is how
large is the class of $W$ (super)algebras which are symmetries of Toda
theories, in the complete set of $W$ algebras.

\indent

{\large{\bf Acknowledgements}}

\indent

It is a pleasure to thank A. Deckmyn, F. Delduc, K. Hornfeck and A. Saveliev
for fruitful discussions.

\end{document}